\documentclass[10pt,journal,comsoc]{IEEEtran}
\usepackage{url}
\usepackage{caption}
\usepackage[noadjust]{cite}
\usepackage{graphicx}
\usepackage{multirow}
\usepackage{caption}
\usepackage{subcaption}
\usepackage{comment}
\usepackage{xspace}
\usepackage{caption}
\usepackage{subcaption}
\usepackage{makecell}
\usepackage{ragged2e}
\usepackage{rotating}
\usepackage{lscape}
\usepackage[table]{xcolor} 
\usepackage{nomencl}
\usepackage{mwe}
\makenomenclature
\usepackage{wrapfig}
\usepackage{makecell}
\usepackage{balance}
\usepackage[utf8]{inputenc}
\usepackage{ragged2e}
\usepackage{booktabs}

\usepackage{pifont}
\usepackage{tabulary}
\usepackage{comment}
\usepackage{lipsum}
\usepackage{array, tabularx, multirow, booktabs, diagbox}
\usepackage{adjustbox}

\newcommand{\cmark}{\ding{51}}%
\newcommand{\xmark}{\ding{55}}%
\usepackage{amsmath,amssymb,amsfonts}
\usepackage{algorithmic}
\usepackage{graphicx}
\usepackage{textcomp}
\def\BibTeX{{\rm B\kern-.05em{\sc i\kern-.025em b}\kern-.08em
		T\kern-.1667em\lower.7ex\hbox{E}\kern-.125emX}}

\begin{document}
	\title{Metaverse for Wireless Systems: Architecture, Advances, Standardization, and Open Challenges}
	
	\author{Latif~U.~Khan,~Mohsen~Guizani,~\IEEEmembership{Fellow,~IEEE},~Dusit~Niyato,~\IEEEmembership{Fellow,~IEEE},~Ala~Al-Fuqaha,~\IEEEmembership{Senior Member,~IEEE},~Mérouane~Debbah,~\IEEEmembership{Fellow,~IEEE}
		
		
		\IEEEcompsocitemizethanks{
			\IEEEcompsocthanksitem L.~U.~Khan and M.~Guizani are with the Department of Machine Learning, Mohamed Bin Zayed University of Artificial Intelligence, United Arab Emirates.
			\IEEEcompsocthanksitem D. Niyato is with the School of Computer Science and Engineering, Nanyang Technological University, Singapore.
			\IEEEcompsocthanksitem A. Al-Fuqaha is with the College of Science and Engineering, Hamad Bin Khalifa University, Qatar.
			\IEEEcompsocthanksitem M.~Debbah is with the Technology Innovation Institute, United Arab Emirates, and also with the Department of Machine Learning, Mohamed Bin Zayed University of Artificial Intelligence, United Arab Emirates.

			

	}}
	
	\markboth{ }{}%

	\maketitle
	



\begin{abstract}
The growing landscape of emerging wireless applications is a key driver toward the development of novel wireless system designs. Such a design can be based on the metaverse that uses a virtual model of the physical world systems along with other schemes/technologies (e.g., optimization theory, machine learning, and blockchain). A metaverse using a virtual model performs proactive intelligent analytics prior to a user request for efficient management of the wireless system resources. Additionally, a metaverse will enable self-sustainability to operate wireless systems with the least possible intervention from network operators. Although the metaverse can offer many benefits, it faces some challenges as well. Therefore, in this tutorial, we discuss the role of a metaverse in enabling wireless applications. We present an overview, key enablers, design aspects (i.e., metaverse for wireless and wireless for metaverse), and high-level architecture of metaverse-based wireless systems. Then, we discuss network management, reliability, and security of the metaverse-based system. Finally, we outline open challenges and present possible solutions. 
\end{abstract}




\maketitle

\section{Introduction}
\label{Introduction}
\setlength{\parindent}{0.7cm}The landscape of wireless systems incurred significant growth during the last few decades. Emerging wireless system applications have diverse requirements. The diverse requirements are in terms of user-defined metrics (e.g., quality of physical experience) and quality of service (QoS) requirements (e.g., strict latency and ultra-high reliability). Fulfilling these diverse requirements is difficult for existing wireless system infrastructures (e.g., 5G). Many recent works proposed the use of $6$G for such applications \cite{khan20206g,khan2021socially,khan2022digital}. $6$G is still in its infancy and many milestones are needed to realize its true implementation. The work in \cite{khan2022digital} proposed digital twin-based architecture for $6$G that consists of three layers, such as the physical interaction layer, twin layer, and service layer. A twin-based architecture tried to follow the trends of self-sustaining wireless systems and proactive-online learning-based systems. Self-sustaining wireless systems require the minimum possible intervention from the network operators and users for their operation to operate autonomously. On the other hand, proactive online-based wireless systems \footnote{Note that in this work we use the term proactive to refer to learning ML models before users request a service. } are needed because of the strict latency requirements of many emerging applications (e.g., healthcare and intelligent transportation systems). To manage wireless and computing resources for wireless applications with strict latency, there is a need to proactively analyze the wireless system. \par
\begin{figure}[!t]
	\centering
	\includegraphics[width=7cm, height=10cm]{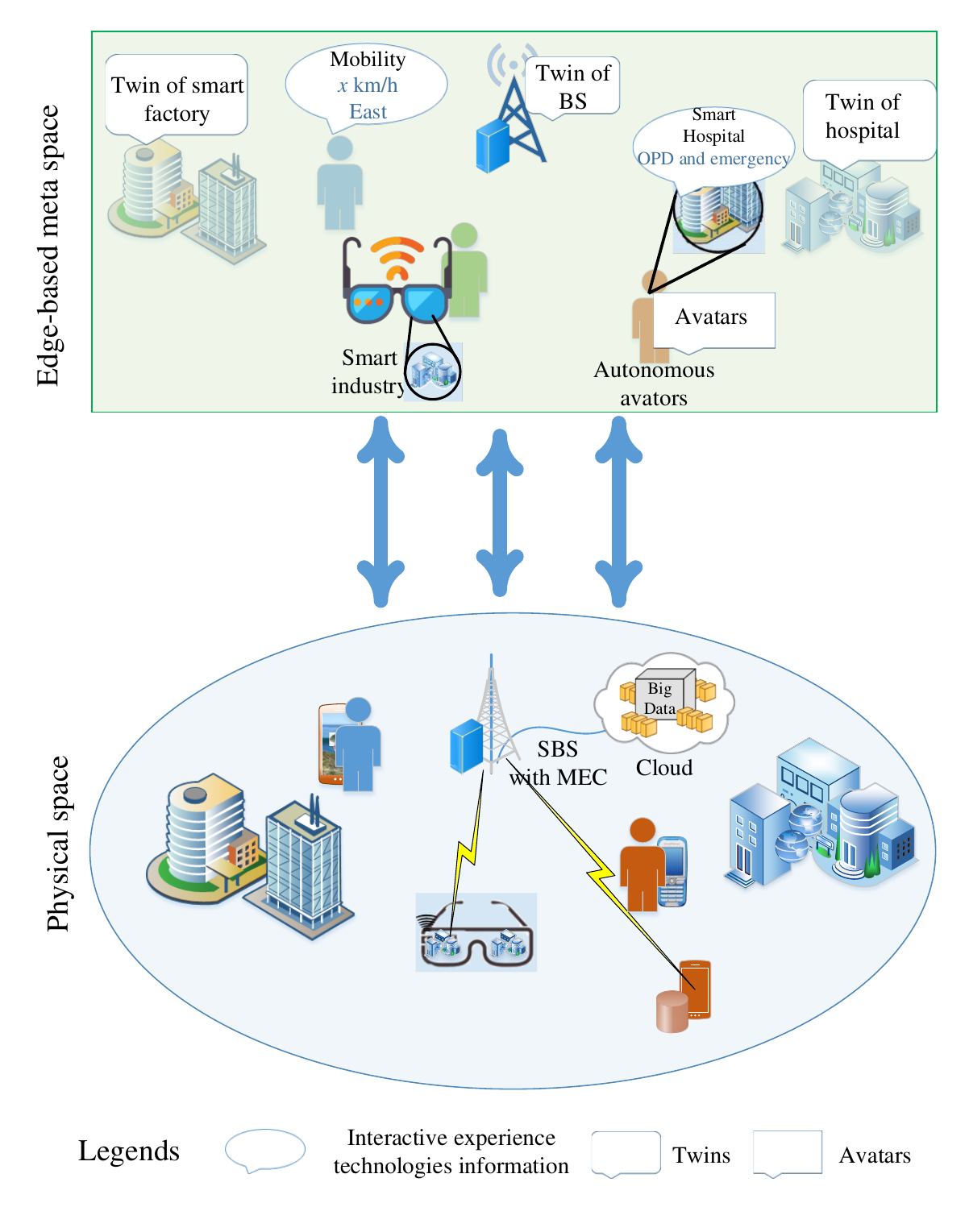}
	\caption{Example of a metaverse-based wireless system.}
	\label{fig:example}
\end{figure}

\begin{figure*}[!t]
	\centering
	\includegraphics[width=17cm, height=13cm] {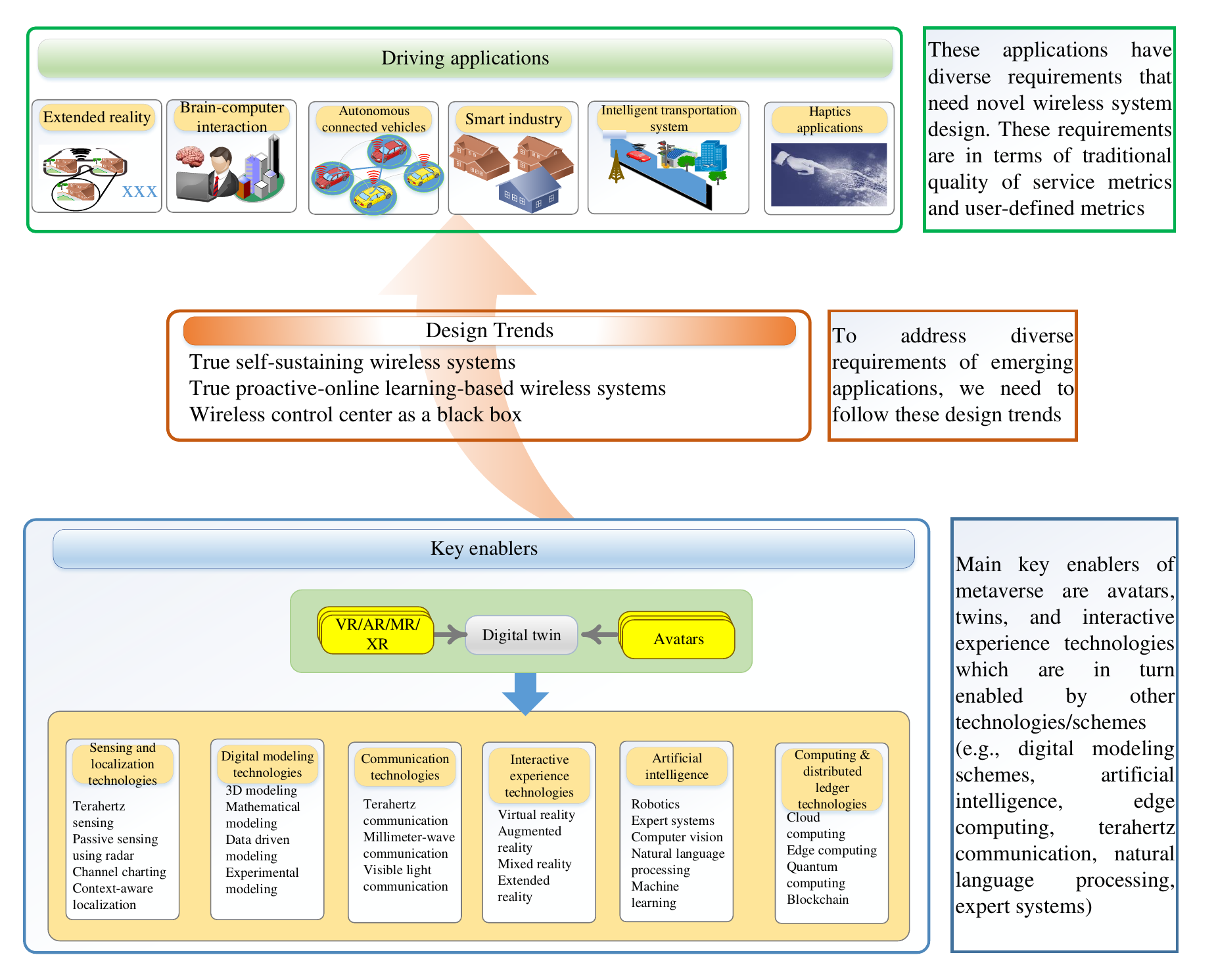}
	\caption{Metaverse for wireless systems: Applications, design trends, and key enablers.}
	\label{fig:intro}
\end{figure*}

Although a digital twin-based system can offer benefits, it seems difficult to truly meet the diverse requirements of wireless systems \cite{khan20206g,khan2022digital}. For instance, digital twin does not effectively consider the users/devices mobility which significantly affects the performance of wireless systems. Consider a terahertz (THz) communication system that is significantly affected (e.g., loss in line of sight (LOS) path) by the human body. Similarly, the mobility of devices/users significantly affects the performance of wireless systems. Therefore, we must effectively take into account the effect of user/device mobility. To do so, the work in \cite{khan2022metaverse} introduced the concept of metaverse that uses digital avatars to account for mobile devices/users. A metaverse-based system can be divided into two spaces, such as (a) meta space and (b) physical space \cite{khan2022machine}. A meta space is a logical space implemented either using edge or cloud or both edge and cloud. On the other hand, the physical space contains all the physical entities (e.g., edge/cloud servers and devices) that are required for wireless systems. An example of a metaverse-based wireless system is given in Fig.~\ref{fig:example}. The static entities are represented by twins in the meta space, whereas the mobile entities are represented by digital avatars. We will discuss more regarding the architecture of a metaverse-based wireless system in Section~\ref{High-Level Architecture}. An overview of emerging applications, design trends, and key enablers is given Fig.~\ref{fig:intro}. Emerging applications are characterized by diverse requirements that must be fulfilled. These requirements can be fulfilled by following the design trends of self-sustainability and proactive online learning analytics. These design trends are met using a metaverse-enabled design that predominantly uses avatars, digital twins, and interactive experience technologies. Next, we discuss the research statistics and research trends of the metaverse and Internet of Everything (IoE). \par

\begin{figure}[!t]
	\centering
	\includegraphics[width=8cm, height=6cm]{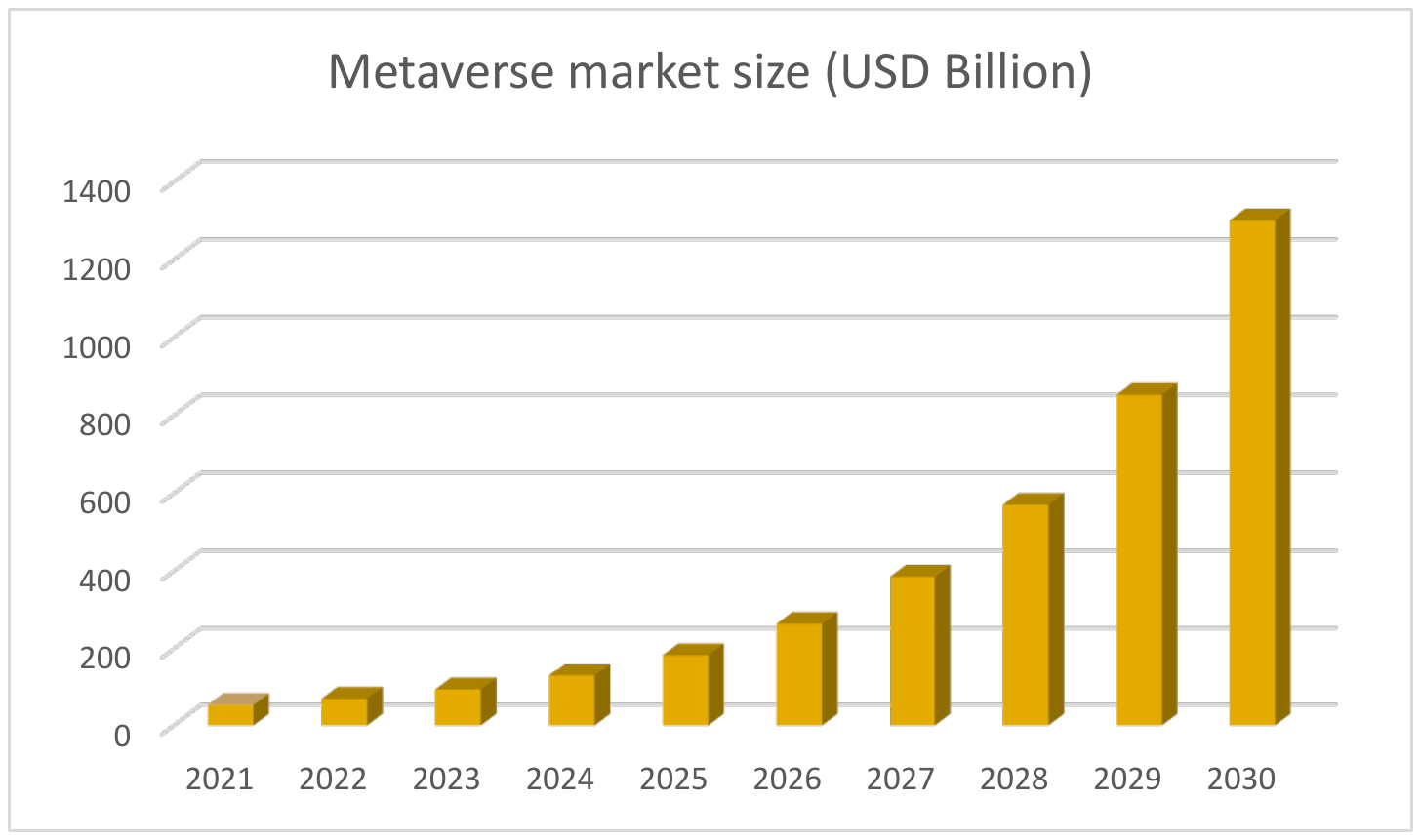}
	\caption{Market value of metaverse.}
	\label{meta_statistics}
\end{figure}

\begin{table*}[]
	\rowcolors{2}{gray!25}{white}
\caption {Summary of existing surveys and tutorials and their primary focus} \label{tab:surveyssummaries} 
\begin{center}
\begin{tabular}{p{2cm}p{2cm}p{2cm}p{1.8cm}p{2cm}p{2cm}p{3.2cm}}
\toprule
    \textbf{Reference}   & \textbf{Wireless for metaverse}& \textbf{Metaverse for wireless}  & \textbf{Recent advances} & \textbf{Standardization} & \textbf{Focus} & \textbf{Remark} \\ \midrule
    Ning~\textit{et al.}~\cite{ning2021survey} &\xmark &\xmark & \xmark & \xmark & Surveyed the concept and current activities in various countries for realizing metaverse. & N/A\\ \midrule
    Wang~\textit{et al.}~\cite{wang2022survey} &\xmark &\xmark & \xmark & \cmark (only for communication between virtual and real world) & Surveyed concept, security, and privacy of metaverse. & Our work will present standardization for ML-enable wireless metaverse\\ \midrule
    Gadekallu~\textit{et al.}~\cite{gadekallu2022blockchain} &\xmark &\xmark & \xmark & \xmark & Surveyed the role of blockchain for metaverse & N/A\\ \midrule
    Khan~\textit{et al.}~\cite{khan2022metaverse} &\xmark &\cmark & \xmark & \xmark & Presented vision of metaverse for wireless networks & N/A\\ \midrule
    Khan~\textit{et al.}~\cite{khan2022machine} &\xmark &\xmark & \xmark & \xmark & Presented role of ML in enabling metaverse-based wireless system & N/A\\ \midrule
    Xu~\textit{et al.}~\cite{xu2022full} &\cmark (considered network edge for enabling metaverse) &\xmark & \xmark & \xmark & Surveyed concept, enablers, computing, and communication for edge-based metaverse. & Our work present a novel architecture with meta space  (i.e., twins and digital avatars based on virtual machines) and physical space. We will discuss how to deploy them using edge, cloud, and devices. Moreover, we will present novel challenges compared to existing surveys\\ \midrule
    Our Tutorial & \cmark  & \cmark   &\cmark   &  \cmark  & \cmark  & N/A\\ 
\bottomrule 
\end{tabular}
\end{center}
\end{table*}

\subsection{Research Trends and Statistics}
According to statistics of \cite{meta_market_1}, the market value of metaverse in $2021$ was USD $51.69$ billion and it is expected to increase at a compound annual growth rate (CAGR) of $44.5$\% to reach USD $1.3$ trillion by $2030$, as shown in Fig.~\ref{meta_statistics}. In $2021$, the market share of North America was $46\%$ which made it the highest contributor among all regions in the world. Among regions, Asia Pacific will expect the highest growth among all regions. On the other hand, META, NVIDIA Corporation, Epic Games, Microsoft, Snap Inc., Nextech AR Solutions Inc., The Sandbox, Decentraland, Roblox Corporation, and Qualcomm Technologies, Inc. will be major players in the metaverse market. Among applications (e.g., gaming, social media, and virtual reality), the gaming sector seems to have the highest share in $2021$.\par
Other than metaverse, the IoE market share is expected to reach USD 3,335.2 Billion, globally, by 2027 at 15.1\% CAGR \cite{IoE_market_1}. The key drivers of this increase are the increased implementation of M2M systems and the emergence of numerous disruptive technologies. On the other hand, the key players are Cisco System Inc. (US), Nokia Corporation (Finland), Samsung (South Korea), Huawei Technologies Co Ltd. (China), Amazon Web Services (US), Qualcomm (US), AT\& T Inc. (US), Koninklijke Philips (Netherlands), Mesh systems LLC (US), and Robert Bosch AG (Germany). Additionally, among the regions, North America is the dominating region. From the aforementioned discussion, it is clear that metaverse and IoE will be key research technologies in the foreseeable future due to their increasing demand and market shares. \par

\subsection{Existing Surveys and Tutorials}
Various works \cite{ning2021survey,wang2022survey,gadekallu2022blockchain,khan2022metaverse,khan2022machine,xu2022full} considered metaverse. The work in \cite{ning2021survey} surveyed metaverse applications and their recent advances. Moreover, they studied various initiatives for realizing the metaverse in various countries. Another work \cite{wang2022survey} discussed the metaverse fundamentals and security aspects. Specifically, the authors discussed the security threats and solutions to various components (e.g., physical space and data management) of the metaverse. The work in \cite{gadekallu2022blockchain} surveyed the role of blockchain in the metaverse. The authors in \cite{khan2022metaverse} presented the vision of metaverse for enabling wireless applications. Specifically, they presented key requirements, general architecture, and open challenges. Another work \cite{khan2022machine} discussed the role of machine learning in enabling metaverse-enabled wireless systems. The work in \cite{xu2022full} surveyed key enablers, computing, and communication technologies for the metaverse. Different from the existing works \cite{ning2021survey,wang2022survey,gadekallu2022blockchain,khan2022metaverse,khan2022machine,xu2022full}, our work (as given in Table~\ref{tab:surveyssummaries}) presents the fundamentals, key enablers, and recent advances. Additionally, we present the network management, security, and reliability of meta space and physical space. Finally, present standardization of the machine learning-enabled metaverse.      

\begin{figure*}
	\centering
	\includegraphics[width=18cm, height=10cm]{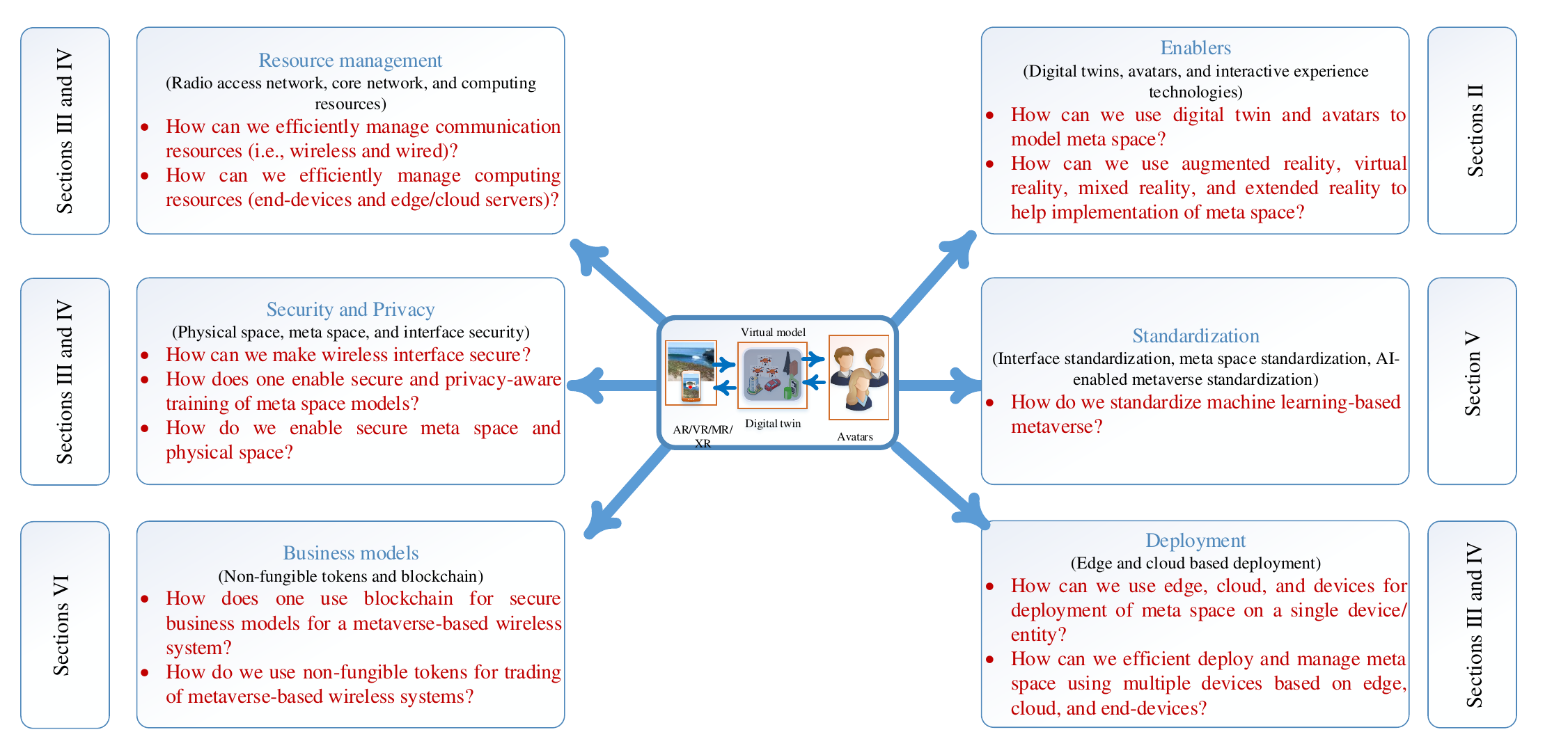}
	\caption{Overview of research questions and relevant sections.}
	\label{fig:questions}
\end{figure*}

\begin{figure}[!t]
	\centering
	\includegraphics[width=8cm, height=18cm]{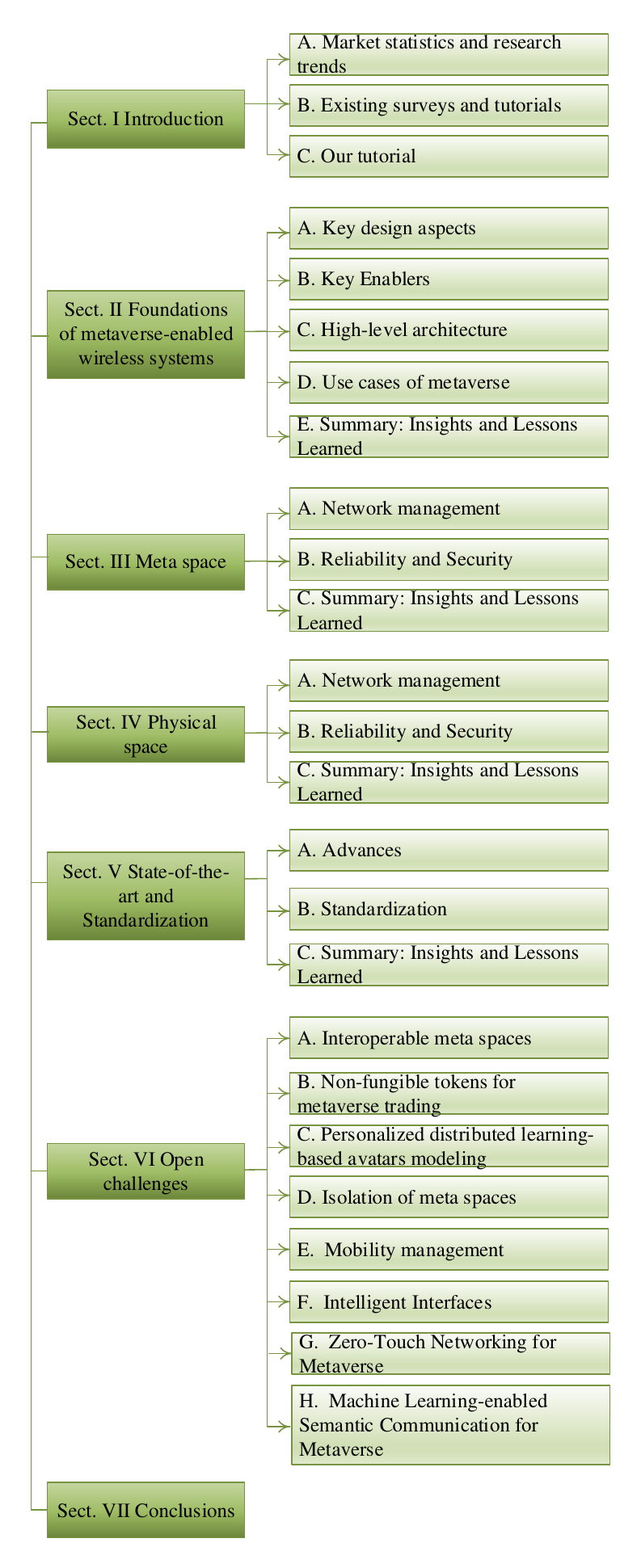}
	\caption{Road map of the tutorial.}
	\label{fig:roadmap_paper}
\end{figure}

\subsection{Our Tutorial}
Our tutorial is different from the existing works \cite{ning2021survey,wang2022survey,gadekallu2022blockchain,khan2022metaverse,khan2022machine,xu2022full} and will present an overview, design aspects (i.e., metaverse for wireless and wireless for metaverse), and an architecture consisting of meta space and physical space along with interfaces. We also review the networking, reliability, and security and present the novel aspects compared to existing works. Furthermore, we discuss the standardization of machine learning-based metaverse and present recent advances of enabling wireless applications by metaverse. Finally, we present open challenges. Specifically, our tutorial will answer the following questions:
\begin{itemize}
    \item How do we use a metaverse to enable IoE applications by performing efficient resource management?
    \item How do we use emerging wireless and computing technologies (e.g., $6$G, edge, and cloud computing) to enable metaverse?
    \item How do we standardize machine learning-based metaverse?
    \item How does one efficiently implement metaverse by using edge/cloud?
    \item How do we efficiently enable metaverse by effectively performing resource management and reliable operation? 
\end{itemize}
More detailed discussions about research questions and their relevant sections are shown in Fig.~\ref{fig:questions}. The summary of our tutorial is shown in Fig.~\ref{fig:roadmap_paper} and our contributions are summarized as follows.
\begin{itemize}
    \item We present an overview, key design aspects, key enablers, and architecture for the metaverse of wireless systems.
    \item We discuss networking management, reliability, and security for enabling meta space and physical space. 
    \item We present the recent of metaverse towards enabling emerging IoE applications. Additionally, we discuss standardization of machine learning-based metaverse.
    \item Finally, we present novel open challenges and their possible solutions.
\end{itemize}

\section{Foundations of Metaverse-Based Wireless Systems}

\subsection{Design Aspects}
\begin{figure}[!t]
	\centering
	\includegraphics[width=8cm, height=8cm]{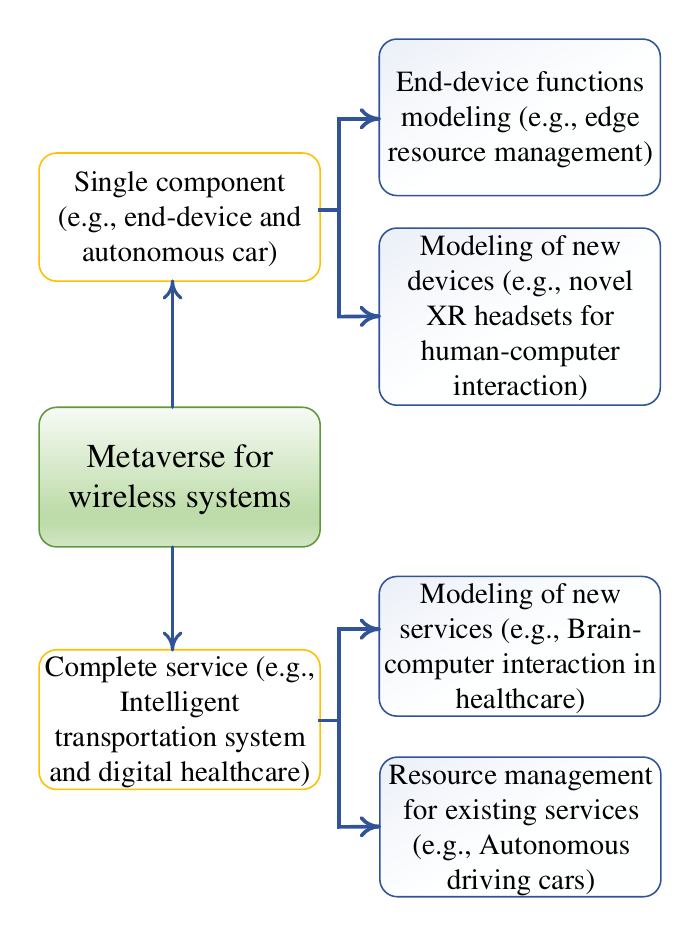}
	\caption{Possible uses of metaverse in wireless systems.}
	\label{fig:metverse_definition}
\end{figure}

\begin{figure}[!t]
	\centering
	\includegraphics[width=8cm, height=6cm]{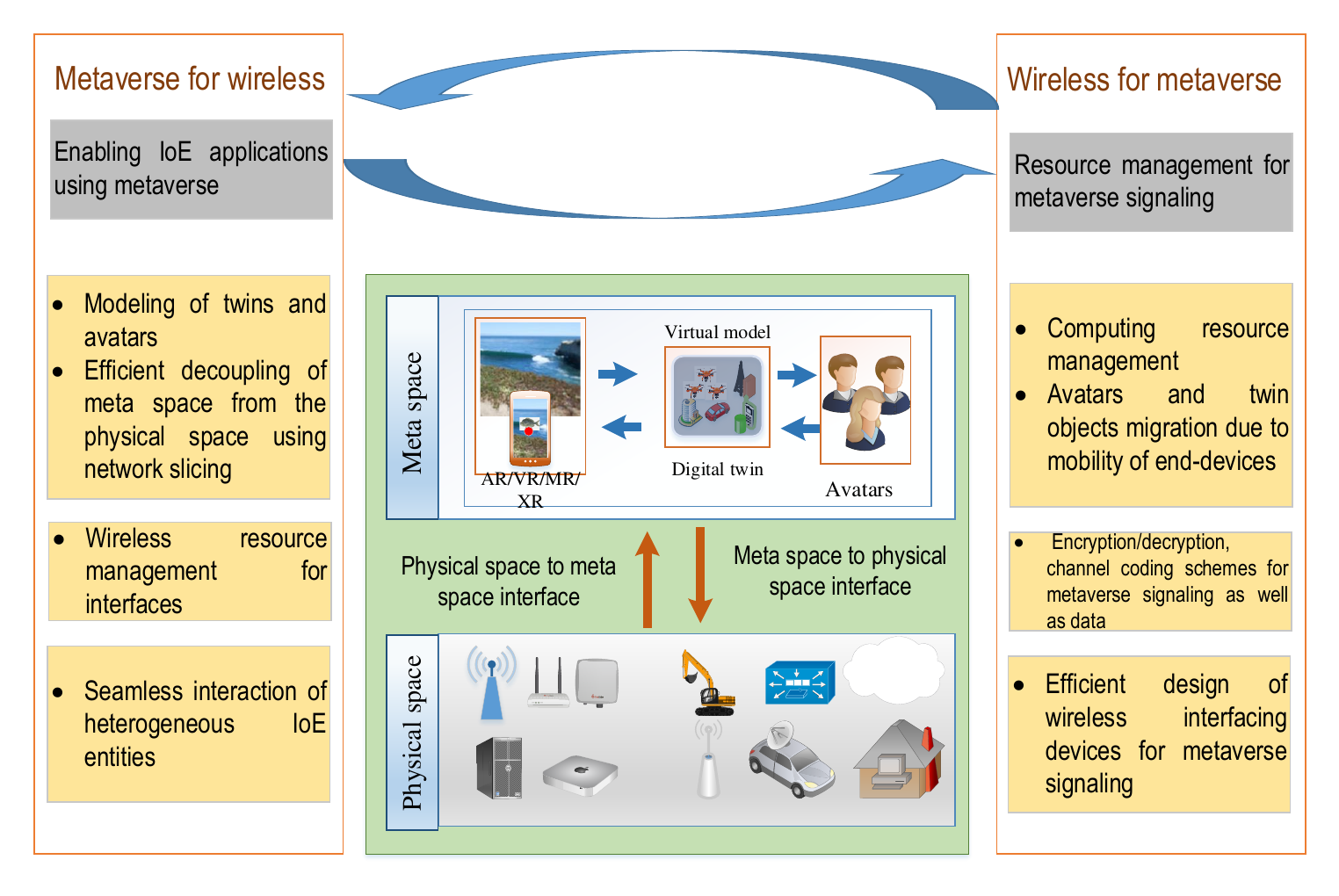}
	\caption{Metaverse and wireless system design aspects.}
	\label{fig:design aspects}
\end{figure}

A metaverse in the perspective of a wireless network can be used to represent various network entities, as shown in Fig.~\ref{fig:metverse_definition}. There are two main design aspects associated with the metaverse and wireless systems, as shown in Fig.~\ref{fig:design aspects}. For every aspect, the role is shown for meta space, interfaces, and physical space. Note that a more detailed discussion about the architecture will be given in Section~\ref{High-Level Architecture}. The design aspects are metaverse for the wireless and wireless for the metaverse. A metaverse for wireless systems deals with resource optimization of computing and communication resources for effectively enabling various wireless applications. On the other hand, wireless for metaverse deals with carrying out signaling for metaverse-enabled wireless system operation. Such a signaling will be used for the efficient placement of meta space objects (i.e., twins and digital avatars). For instance, consider the deployment of meta space using multiple edge and cloud servers (a more detailed discussion about the deployment will be given in Section~\ref{Efficient Deployment of Twins and Avatars}). To do so, there is a need to efficiently deploy meta space in such a way as to minimize the cost. Such a cost can be possibly transmission latency and energy consumption. To minimize this, one must choose a set of edge servers that will result in low latency and low transmission energy. Additionally, to enable proactive analytics, we must perform training to obtain pre-trained models. Such pre-trained models can be based on either centralized learning or distributed learning. Centralized learning will require the migration of device data to the cloud for training; however, it has an inherent issue of privacy leakage. To address this, one can use distributed learning that is based on iterative interaction between the devices and edge/cloud. To enable such interaction for getting pre-trained models, there is a need for effective computing and communication resources management. Other than wireless for metaverse, metaverse for wireless will require efficient management of resources for various applications. For instance, consider infotainment in autonomous driving cars that require computing resources at both cars and edge servers installed at the roadside units (RSUs). Due to the presence of many computing tasks in autonomous cars, there is a need for offloading these tasks to the RSUs. How to perform such offloading and management of computing as well as communication resources? A metaverse will effectively enable such offloading by making offloading decisions and management of computing as well as communication resources.\par      

\begin{table*}[]
\caption {Focus of existing surveys and tutorials for metaverse key enablers} \label{tab:keyenablers} 
\begin{center}
\begin{tabular}{p{3cm}p{5cm}p{6cm}}
\toprule
    \textbf{Existing surveys/tutorial}   & \textbf{Mentioned enabler}& \textbf{Focus}  \\ \midrule
    Khan~\textit{et al.}~\cite{khan2022metaverse} & \begin{itemize} \item Digital twins \item Avatars \item Interactive experience technologies \end{itemize} & This work introduced the key enablers along with role of ML in modeling them without primarily focusing on mobility modeling of avatars in meta space.  \\
    \midrule
    Khan~\textit{et al.}~\cite{khan2022machine} & \begin{itemize} \item Digital twins \item Avatars \item Interactive experience technologies \end{itemize}  & This work focused on role of ML for modeling twins and avatars without discussing mobility as well as pose estimation for modeling avatars. \\
    \midrule
    Xu~\textit{et al.}~\cite{xu2022full} & \begin{itemize} \item Digital twins \item Avatars \item Interactive experience technologies \end{itemize} & The authors discussed in detail about interactive experience technologies without providing in depth information about modeling of avatars/twins. \\
    \midrule
    Our work & \begin{itemize} \item Digital twins \item Avatars \item Interactive experience technologies  \end{itemize} & Our work discusses in detail about the twins/avatars modeling as well as interactive experience technologies. Additionally, we discuss in detail about mobility modeling and pose estimation for modeling avatars in meta space.  \\
\bottomrule
\end{tabular}
\end{center}
\end{table*}

\subsection{Key Enablers}
\subsubsection{Digital Twins}
A digital twin is a virtual model of the physical wireless system (example scenario is shown in Fig.~\ref{fig:example}) \cite{khan2022machine}. Such a virtual model will be used for analysis and help in controlling network components/devices. One can model a digital twin using experimental modeling, data-driven modeling, and mathematical modeling. In mathematical modeling, various assumptions are made during modeling of the real world systems. For instance, non-linear functions of robotic systems are generally modeled using linear assumptions, and thus might not reflect more effective modeling. To handle this issue, we can employ experimental modeling. In experimental modeling, a series of experiments are carried out for modeling a physical system. Similar to mathematical modeling, experimental modeling also suffers from degradation due to experimental errors (i.e., human errors and machine errors). One can address the issues related to experimental and mathematical modeling by using data-driven modeling. Data-driven modeling will use machine learning. Machine learning can be based either on centralized training or distributed training. A centralized training-based machine learning trains a model at a centralized location. It requires moving device data to the centralized location and thus might have privacy loss due to the presence of a malicious third-party server used for running the machine learning model. To address this privacy leakage issue, one can use a distributed training-based machine learning (i.e., federated learning (FL)) \cite{khan2023resource,khan2023federateddd}. FL for modeling twins in the metaverse will involve frequent interaction between end-devices and twins deployed at edge/cloud. In FL, end-devices compute their local models and send them to the edge/cloud for aggregation. After aggregation, the global twin model is shared with devices for further improving their local models. Such iterative interaction requires a significant amount of communication resources. Although FL enables on-devices machine learning, there are many scenarios where there are significant limitations on the available computing at end-devices, and thus they might not be able to compute their local models within the deadline. To address this issue, a few works \cite{thapa2022splitfed,singh2019detailed,gao2020end} proposed split FL (SFL) that is based on computing partial local models by the end-devices and the remaining at the edge/cloud servers. Different from traditional FL, SFL needs to offload partial local model computing tasks to the edge/cloud servers, therefore, there must be an effective computing resource allocation scheme for SFL.

\subsubsection{Digital Avatars}
\begin{figure}[!t]
	\centering
	\includegraphics[width=8cm, height=5cm]{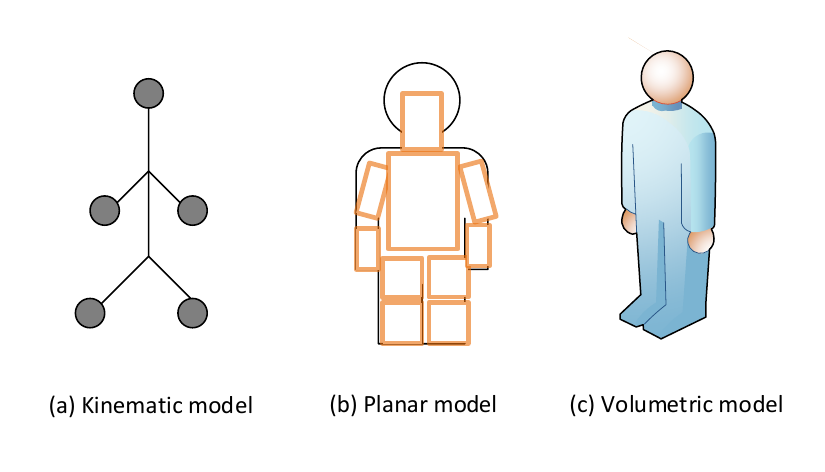}
	\caption{Categories of human modeling for pose detection.}
	\label{fig:human_model}
\end{figure}

A digital avatar is a digital replica of the humans/ mobile devices controlled by humans in the physical interaction world. One can use the digital twin to represent the virtual model of humans in meta space. Additionally, interactive experience technologies (e.g., AR, VR, MR, XR) can also represent humans in a virtual world. However, both interactive experience technologies and digital twins might not effectively represent humans in a meta space. A human body in physical wireless systems significantly impacts the quality of service (QoS). Different from traditional metaverse where avatars can be two-dimensional and three-dimensional ($3$D) \cite{avatar_1}, metaverse for a wireless system should consider $3$D avatars. Such a $3$D avatar effectively represents the behavior of humans in the metaverse, as shown in Fig.~\ref{fig:human_model}. From onward, we will use the digital twin avatars for $3$D avatar. To model $3$D avatars for the metaverse, one can propose novel tools. Existing software tools for $3$D modeling of humans are MakeHuman, Daz Studio, iClone, and Mixamo \cite{avatar_2}. These tools are used for illustration, animation, cinema, and video games. Although these tools can effectively model human in a computer simulation, animation, etc., there is a need to propose a novel design for metaverse-based wireless system.\par

\begin{figure}[!t]
	\centering
	\includegraphics[width=8cm, height=6cm]{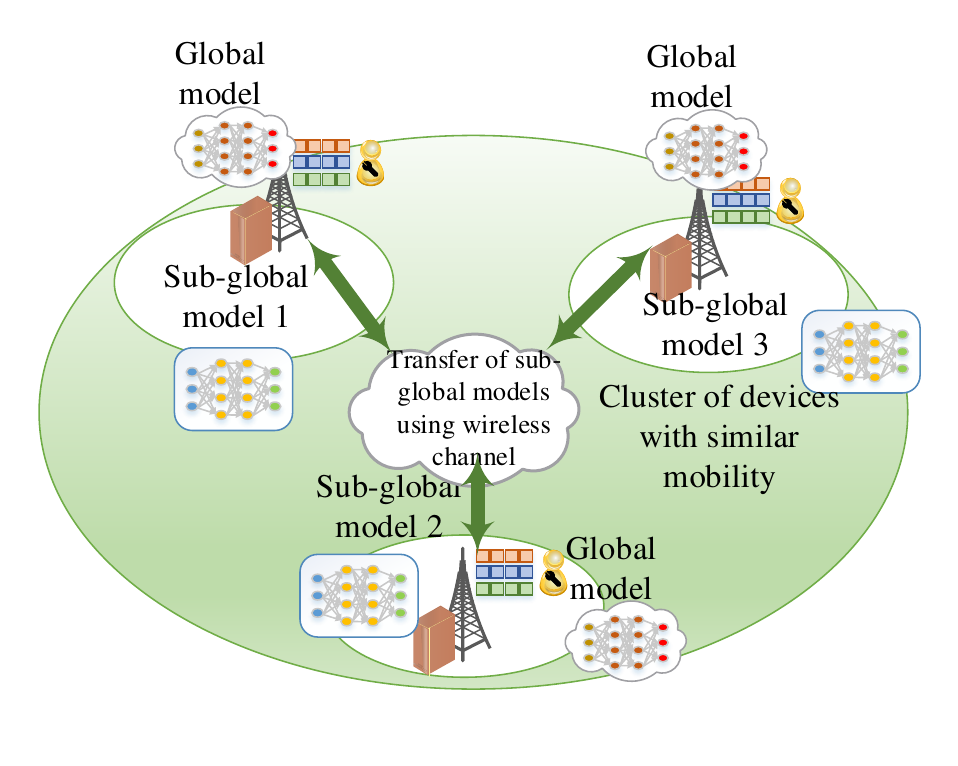}
	\caption{Dispersed federated learning for training meta space models \cite{khan2020dispersed}.}
	\label{fig:dfl}
\end{figure}

The nature of a $3$D model of a human in a metaverse-based wireless system will be different compared to other applications (e.g., animations). The difference lies in the incorporation of wireless system features in the avatar of the metaverse-based wireless system. For instance, Terahertz (THz) communication is significantly affected by the human body. A LOS path might be affected by humans. Additionally, THz communication is significantly affected by the concentration of red blood cells (RBCs) in the human blood. Other applications (e.g., $3$D printing, human-computer interaction) also require an effective model of avatars. An overview of avatars in a metaverse-based wireless system is shown in Fig.~\ref{fig:example}. First, there is a need to effectively create a $3$D model of a human. Next, we should add the effect (e.g., loss in LOS path for THz communication) of a wireless system in the avatars. We should also effectively model an avatar's mobility that can effectively follow human mobility patterns. Such mobility patterns can be modeled using various techniques (e.g., optimization theory and deep learning). Mobility management is of significant importance in traditional wireless networks and many works proposed various schemes. However, here, a metaverse-enabled wireless system is different compared to traditional wireless networks. A meta space deployed at the network edge must be migrated to the new edge depending on the device's mobility. Such migration of meta space can be either live or non-live. More detailed discussion about migration schemes will be provided in Sections~\ref{Devices Mobility Management} and \ref{Efficient Deployment of Twins and Avatars}. For mobility management, one can use prediction schemes based on deep learning for predicting the device's mobility. Based on the predicted mobility, one can perform migration of meta space. On the other hand, in a metaverse-enabled wireless system, one must manage the mobility of devices during the training of meta space models using distributed learning. It is desirable for devices to remain a range of edge servers performing aggregation for fast convergence. Therefore, one should manage device mobility during the training process of distributed learning models for meta space. Mostly, devices are mobile, therefore, one must manage such mobility. We can use dispersed federated learning (i.e., shown in Fig.~\ref{fig:dfl}) that is based on the clustering of devices. Note that devices that will remain within the vicinity of each other will be placed in a cluster and a sub-global model is learned, as shown in Fig~\ref{fig:dfl}. Next, to train sub-global models, one can share the sub-global models among different clusters to yield a global model.      \\
To enable interaction between avatars and the digital twin objects in a meta space for accurate operation and analysis prior to deployment, there is a need for efficient computer vision techniques, such as human-pose tracking, emotion, and expression recognition, and gesture recognition, among others \cite{CV_1}. Human pose tracking enables the estimation of multi-person human geometric and motion information. Such an estimation of human key points-trajectories is necessary for the performance optimization of metaverse-based wireless systems. For instance, the exact location and movement of human body parts can be used for accurate channel estimation of wireless signals. For instance, meta space having avatars and twins can be used for the analysis of wireless systems. As meta space is running a virtual world, therefore, accurate pose estimation of avatar is necessary for better analysis. Also, if the meta space interacts with physical space during the run-time control of physical objects, there is a need to accurately estimate human positions. Additionally, a human pose estimation will have a significant impact on various other applications (e.g., human-computer interaction, healthcare applications, AR, and VR). Therefore, human pose tracking in a metaverse-based wireless system has significant importance for wireless systems. To do pose estimation, the first step is to model humans. There are three categories, such as kinematic, planar, and volumetric, of human models shown in Fig~\ref{fig:human_model}. A human body has limbs and joints as well as it has body shape information and contains body kinematic structure. The kinematic model is based on a representation of the human body using limb orientation and joint positions. To represent humans in more detail, one can use planar models that use rectangle kind of representations to show different parts of the human body. Although we can represent humans using a kinematic model and planar model, there is a need for a more detailed model of humans, such as a volumetric model (i.e., 3D human reconstruction) for the metaverse.\par
\begin{figure}[!t]
	\centering
	\includegraphics[width=8cm, height=7cm]{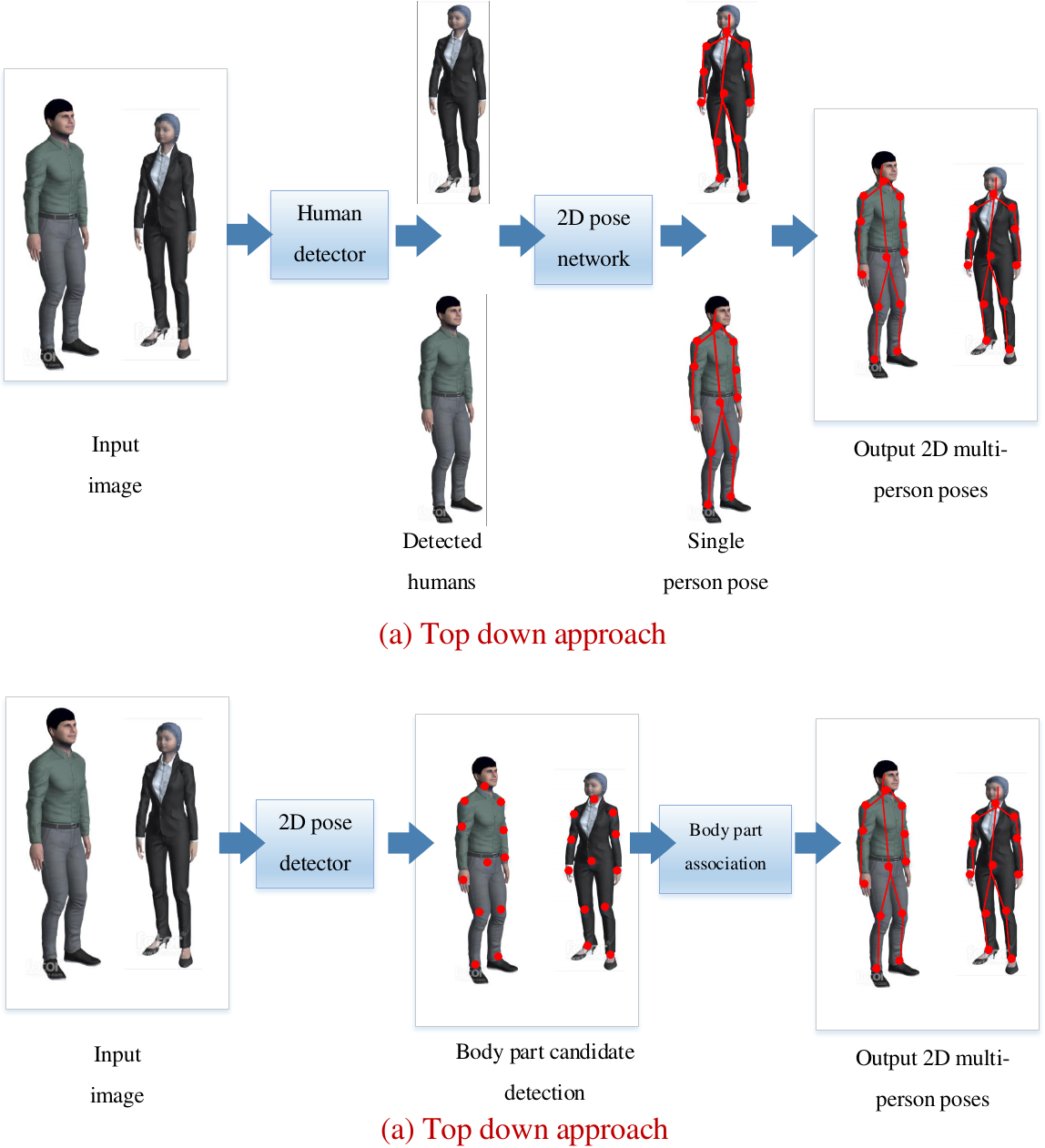}
	\caption{Example of 2D pose estimation \cite{dang2019deep}.}
	\label{fig:2Dpose}
\end{figure}
\begin{figure}[!t]
	\centering
	\includegraphics[width=8cm, height=7cm]{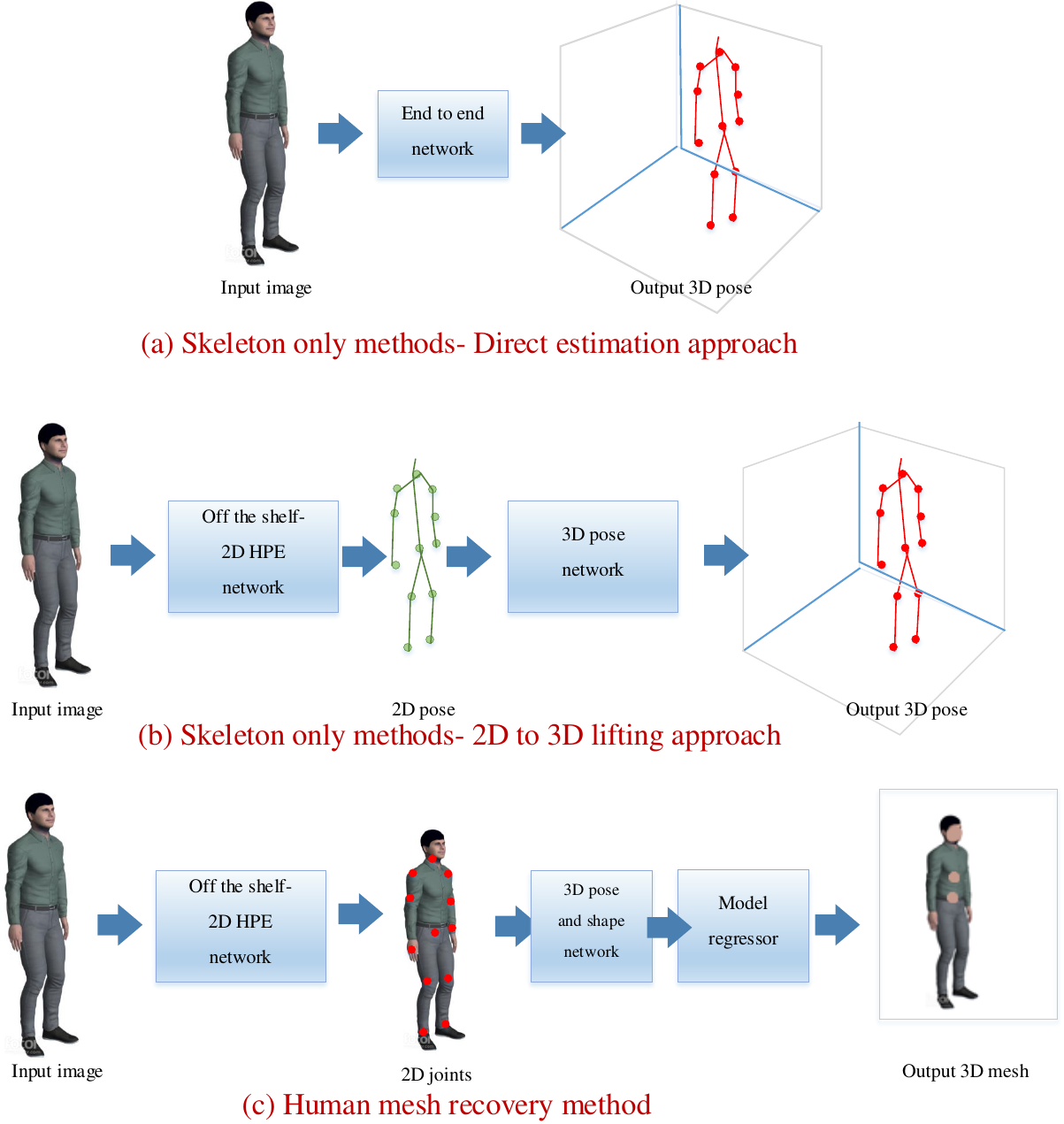}
	\caption{Example of 3D pose estimation \cite{dang2019deep}.}
	\label{fig:3Dpose}
\end{figure}

In addition to human modeling, there is a need for efficient human pose estimation. Human pose estimation can be of two types: two-dimensional ($2$D) and three-dimensional ($3$D), as shown in Figs.~\ref{fig:2Dpose} and \ref{fig:3Dpose}. In $2$D pose estimation, poses are estimated using images (in terms of pixel values), whereas $3$D pose estimation involves estimation results in the three-dimensional spatial arrangement of the human body. Recent works considered deep learning for estimating human poses and have shown promising results \cite{newell2016stacked, xiao2018simple, sun2019deep}. Mostly, these works are focused on transforming low-resolution images into high-resolution images for accurately estimating human poses. Sun \textit{et al.} in \cite{sun2019deep} proposed an architecture for 2D estimation, namely, \emph{HighResolution Net} for maintaining high resolution during the whole process for estimating 2D human pose using movements of joints, as shown in Fig.~\ref{fig:human_model}c. Although 2D pose estimation of humans can offer benefits, it has a few limitations. 2D pose estimation can estimate only joint movements, not exact human models and thus, might not be more desirable for metaverse-based systems. Therefore, there is a need for 3D pose tracking of humans while modeling digital avatars. In \cite{yang2022metafi}, the authors proposed a WiFi-based IoT-enabled human pose estimation system, namely, MetaFi for digital avatars. Specifically, the authors used Wi-Fi signals to estimate the human pose for the metaverse as motivated by the use of Wi-Fi for human activity recognition. The MetaFi system comprises two COTS WiFi routers (i.e., TP-Link N750) acting as a receiver and transmitter. Such data is sent to the server for AI model inference. Although MetFi AI model can be easily used for human pose estimation, it might not perform well in all scenarios. To do so, there is a need for considering large data sets from a wide variety of users. A certain group of people in an institution might not want to share such data outside their institution. To address this issue, one can use cross-silo federated learning that can train a human pose estimation model within one institution and then share only the learning model updates with the other institution. Such an approach will better preserve the end-user privacy but at the cost of slowing the convergence rate. Also, there will be many communication rounds between different institutions that will require careful design for cost-efficient communication.

\subsubsection{Interactive Experience Technologies}
A trend of a virtuality-reality continuum is followed by interactive experience technologies. VR is based on synthetic views along with additional information and is on the virtuality end. On the other hand, AR is on the reality end. AR is based on enhancing physical view by using additional information. In a mixed reality (MR), the virtual world and physical words are merged and they interact in real time. Extended reality (XR) merges all three interactive experience technologies, such as AR, VR, and MR. XR will enable fine-grained human-specific information perception. Therefore, one can say that XR head-mounted display, sensors, and embedded systems are the main source of entrance to metaverse-enabled wireless systems \cite{wang2022survey}. Although interactive experience technologies will effectively enable metaverse, there is a need for efficient design based on edge computing. In a metaverse-enabled wireless system, there will be a massive number of running interactive experience technologies. Such devices will require on-demand computing resources with low latency. To do so, one must efficiently manage edge computing resources for various metaverse devices. Note that interactive experience technologies are well studied in the literature, still, there is a need for more research. The behavior of interactive experience in a metaverse will be different and more complex compared to general applications \cite{khan2022metaverse}. Therefore, there is a need for careful design considerations regarding the integration of interactive experience technologies in metaverse-enabled wireless systems. There can be many cases where the need for interactive experience technologies in the metaverse will be crucial \cite{IXR_1}. Example use cases are metaverse-assisted remote expert, metaverse-based real-time collaboration, and metaverse-based industrial maintenance, among others. Consider a remote expert system for industrial maintenance based on metaverse. Cameras and sensors installed near the industrial machine can take images and add annotations using interactive experience technologies. These images are sent to the remote expert using emerging communication technologies. The remote expert after adding annotations and suggestions will be shared with the industrial machine operators for providing guidance to remove faults. Meanwhile, the metaverse can use the data of the faults to train/further train machine learning meta space models. Such pre-trained models will be stored using a blockchain network.\par

\begin{figure*}[!t]
	\centering
	\includegraphics[width=18cm, height=12cm]{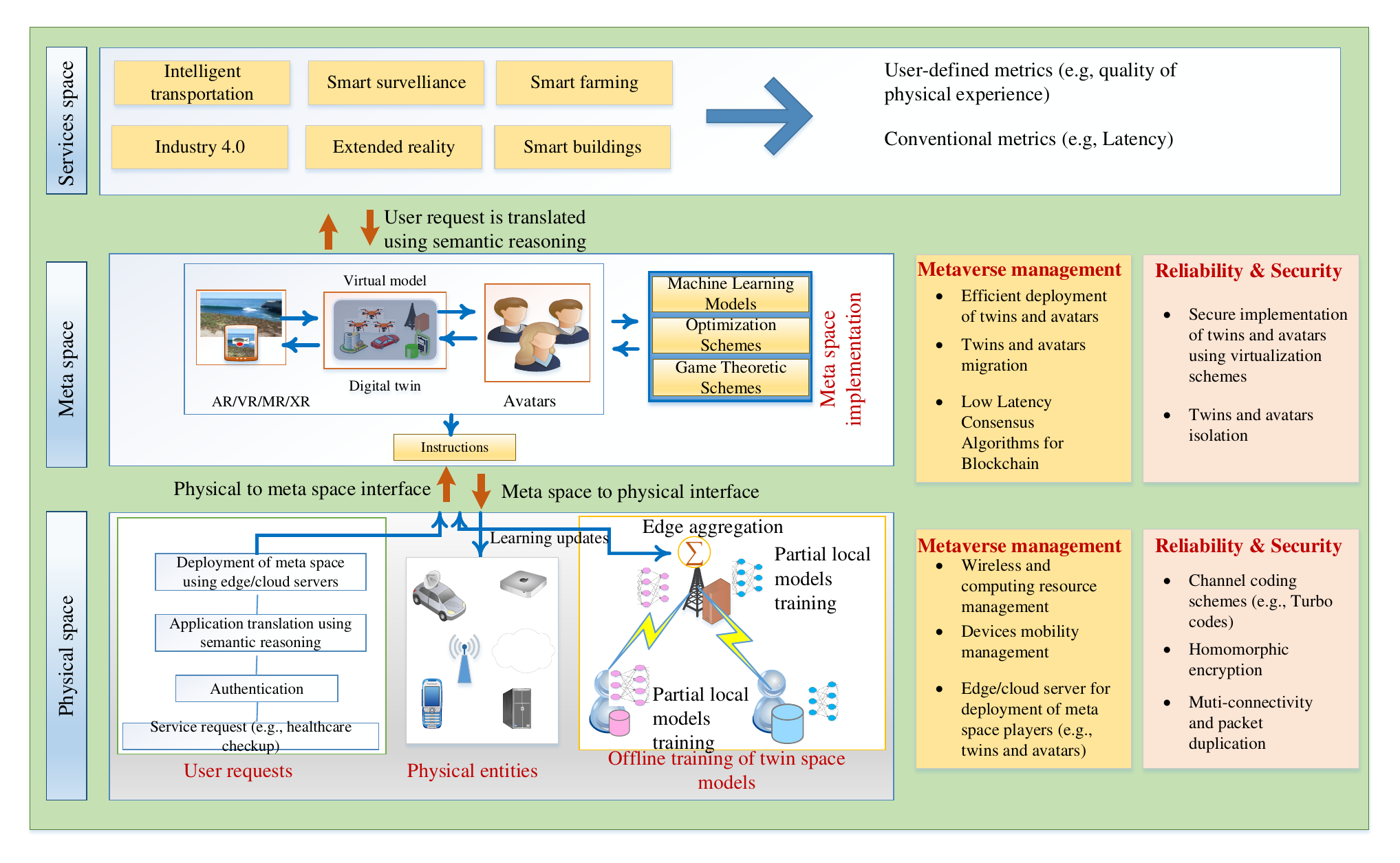}
	\caption{An overview of general architecture for metaverse.}
	\label{fig:architecture}
\end{figure*}

\begin{figure*}[!t]
	\centering
	\includegraphics[width=14cm, height=12cm]{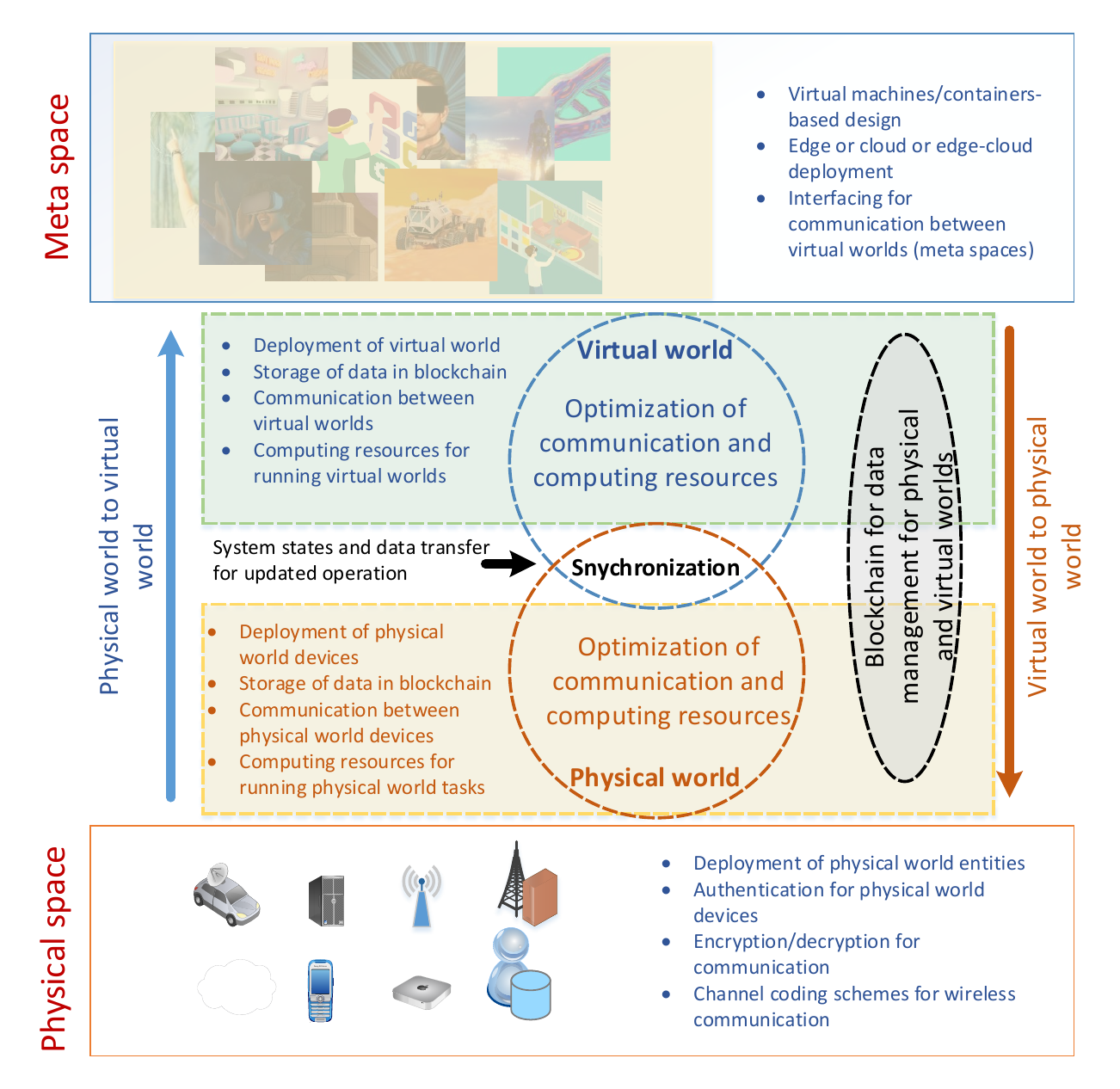}
	\caption{An overview of the interaction between physical and virtual worlds.}
	\label{fig:resarch_flow}
\end{figure*}

\subsection{High-Level Architecture}
\label{High-Level Architecture}
First, we discuss the general architecture of a metaverse-based wireless system, as shown in Fig.~\ref{fig:architecture} \cite{khan2022machine,khan2022metaverse}. The architecture mainly consists of three spaces: physical interaction space, meta space, and services space. A physical interaction space has all the physical devices, humans, edge/cloud servers, and other network switches necessary for establishing a wireless system. A meta space is a logical space that handles the interaction of a digital twin, digital avatars, and interactive experience technologies to analyze/control the physical wireless system. On the other hand, services space enables users to request services from a metaverse-enabled wireless system. An overview of the interaction between meta space and physical space is given in Fig.~\ref{fig:resarch_flow}. To deploy meta space, there is a need to first model digital twins and avatars. One can use various ways to model them, such as mathematical modeling, experimental modeling, and data-driven modeling. In mathematical modeling, we made a series of assumptions (e.g., a linear approximation to a non-linear model). Coping with this limitation, one can use experimental modeling. Experimental modeling consists of a series of experiments that may suffer from experimental errors and equipment malfunctioning. To address these limitations, one can use data-driven modeling. Data-driven modeling uses data generated by wireless applications to train machine learning models. An example of a metaverse-based wireless system is shown in Fig.~\ref{fig:example} in Section~\ref{Introduction}. Fig.~\ref{fig:example} shows the role of digital twins, avatars, and interactive experience technologies in wireless systems. Digital twins in wireless systems can be used to model the static entities in wireless systems. These static entities are buildings, base stations, and mountains, etc. Digital avatars refer to mobile devices and users. For instance, a user sitting inside an autonomous car can be modeled using an avatar for the autonomous car. Similarly, humans with wearables can be modeled using avatars. Modeling digital avatars in meta space might be more challenging compared to digital twins. Digital twins of static entities have no mobility, and thus easier to model compared to mobile avatars. Modeling the exact mobility of avatars is difficult. Additionally, mobile users significantly affect the performance of the wireless system. We must effectively model various effects (e.g., due to wireless signal attenuation, wireless signal reflection/refraction, and wireless signal energy absorption) of humans on wireless signals in meta space. For instance, Terahertz (THz) communication is significantly affected by the human body. LOS communication is significantly affected by humans for the THz band. Additionally, THz communication is affected by the concentration of red blood cells (RBCs). The molecular noise and path-loss decrease with a rise in RBCs, and vice versa. Therefore, there is a need to effectively model avatars in meta space. \par 
There are two main phases in a metaverse operation: offline training and operation. Offline training is used for the training of meta space models. Such training can be performed either using centralized machine learning or distributed machine learning. Although centralized learning can result in faster convergence, but at the cost of loss of users' privacy due to the transfer of data from devices/end-users to a centralized location for training. To remedy this, distributed learning can be used that is based on training local learning models at devices without moving their data to a centralized location. Then, local learning models will be shared with the meta space for aggregation, as shown in Fig.~\ref{fig:architecture}. On the other hand, during the operation phase, the devices will request services from the meta space. The meta space will in turn perform resource optimization to instantly serve the end-users using pre-trained models, optimization theory, and game theory. For both phases (i.e., offline training and operation), there is a need for wireless and computing resource optimization. Later, in a tutorial, we will discuss how can we perform resource optimization of computing and wireless resources. Additionally, we will discuss interfaces for communication between meta space and physical space, deployment of meta space, and meta space design later on in detail in this tutorial. \par
On the other hand, it is necessary there must be synchronization between the meta space and physical space, as shown in Fig.~\ref{fig:resarch_flow}. Changes in the physical space will significantly affect the meta space. Similarly, during the operation, changes in the meta space will have significant impact on the physical space. For instance, training of machine learning models in the meta space will use data or function of the data in the physical space. Therefore, the physical space data or function of data (i.e., in case of federated analytics) should be provided timely to the meta space. Note the function of data in case of federated learning is the locally trained model obtained by running iterative algorithm on device. Other than this, in the case of sensory measurements (e.g., images in autonomous cars and temperature data) should be wisely shared with the meta space. If we increase the frequency of sharing the sensory data, more communication resources will be needed and vice versa. Similarly, the sampling frequency (e.g., taking samples of the temperature data) should be properly adjusted. A high sampling frequency will utilize computing resources and more communication resources will be need to share them with the meta space. On the other hand, if we choose the low sampling frequency, the data (e.g., measured temperature) shared with the meta space will not well reflect the physical space and thus may result in less accurate results. Note here, for synchronization in metaverse, there is a need for special care due to the presence of immersive 3D streaming, real-time communication, and multi-sensory data \cite{wang2022survey}. Such communication will require more bandwidth in addition to different constraints compared to traditional wireless communication. For instance, if we consider multi-sensory data, the communication must be reliable to get the accurate results. On the other hand, there must be more bandwidth (e.g., similar to eMBB communication) for immersive 3D communication to enable immersive 3D experience to massive number of devices in the physical world. Therefore, there must be efficient and novel resource scheduling schemes for synchronization between the physical and the virtual world.       

\begin{table*}[]
\caption {Overview of existing projects of metaverse} \label{tab:existing_metaverse_projects} 
\begin{center}
\begin{tabular}{p{2.5cm}p{3cm}p{3cm}p{3cm}p{3cm}}
\toprule
    \textbf{Project}   & \textbf{Description}& \textbf{Developers} & \textbf{Objectives} & \textbf{Remarks}\\ \midrule
    Decentraland (MANA)   &  It is a 3D virtual world, browser based platform developed for user to experience a virtual world. Users can buy various entities (e.g., plots) using non-fungible tokens. &  Ari Meilich and Esteban Ordano & To realize virtual worlds of the various real world components.  & This platform is particularly developed for experience various virtual worlds without primarily focus on the emerging wireless applications.  \\ \midrule 
    The Sandbox & Game studio Pixowl &  The sandbox is a 2D game developed for mobile and windows users. & To realize gaming experience based on virtual world of the real word. & The focus of sandbox is gaming without effectively considering wireless applications. \\ \midrule 
    Axie Infinity & Sky Mavis & Axie infinity is actually non-fungible token-based online video game. It enables smart gaming with the freedom of enabling players to buy and sell items.  & Focus is on smart gaming & Axie infinity focuses on developing smart gaming experience for users without taking into account the actual wireless applications factors (e.g., fading and error rate). \\ \midrule
    Sorare & it is a football game that allows players to manage, buy, and sell virtual teams using cards. & Nicolas Julia and Adrien Montfort & To enable a fantasy football game & Sorare is based on enabling a virtual fantasy football with the purpose of promoting smart gaming.   \\

\bottomrule
\end{tabular}
\end{center}
\end{table*}

\subsection{Use Cases of Metaverse}
Here, we discuss several existing metaverse projects as summarized in Table~\ref{tab:existing_metaverse_projects}. 




\subsubsection{Decentraland (MANA)}
Decentraland developed by Ari Meilich and Esteban Ordano, runs on the Ethereum blockchain and is a 3D virtual world, browser based platform developed for the user to experience a virtual world \cite{case_study_1}. In this platform, users can buy various entities (e.g., plots) using non-fungible tokens. The first token of Decentraland was launched in 2017 at ICO boom that managed to reach \$26 million. Recently in the 2021 bull run, major brands are witnessed in Decentraland metaverse crypto projects \cite{DLND}. These brands include Sotheby’s, Miller Lite, PricewaterhouseCoopers, Atari, Adidas, and Samsung. Decentraland uses RC-20 token standard to deal with its cryptocurrency, namely, MANA. Moreover, decentraland has two tokens: LAND (i.e., ERC-721 token) and estate (i.e., ERC-721 token). Both of these tokens are used to depict parcels of land in decentraland. Using tokens, different players in a decentraland can buy virtual lands as a gamble at casino. \par   

\subsubsection{The Sandbox}
The sandbox is a 2D game developed by game studio Pixowl for mobile and windows users \cite{sandbox_1}. Players in a sandbox can build, own, and monetize the gaming experience in the Ethereum blockchain \cite{sandbox_2}. The two most important and known versions of the sandbox are The Sandbox Evolution (2016) and The Sandbox (2011). Both of them combine hits download of more than 40 million on iOS and Android. In 2018, developer/Publisher Pixowl brought the sandbox to the blockchain ecosystem with the goal of providing game manufacturers with true ownership using non-fungible tokens. Additionally, this will provide them with incentives in participating in a metaverse ecosystem. \par
\subsubsection{Axie Infinity}
Axie infinity was developed by Sky Mavis which is actually a non-fungible token-based online video game \cite{axieinfinity_1}. In Axie infinity, there are kingdoms of Axies and all players battle for the kingdoms. Moreover, within a game, all players can earn and sell their components. This game has been marketed by Sky Mavis using "play-to-earn" model that is based on the concept that every player should pay a starting cost. Sky Mavis estimation of the new player costs were $\$400$ and $\$307$ in $2020$ and $2022$, respectively. Although Axie infinity can enable a good gaming experience, there is a need for significant modifications in order to apply it for real world wireless applications that will require proactive analytics and online control.           

\subsubsection{Sorare}
It is a non-fungible tokens-enabled football game \cite{sorare_1}. To secure the ownership, sorare uses the Ethereum blockchain network for operation. It is located in Paris, Ile-de-France, France, and has approximately 39 investors that include Antoine Griezmann and aldeA Ventures. Sorare was founded in 2018 and its cards sold until 2020 were worth of \$1.8 million. In this virtual game, various players can own cards that are denoted by non-fungible tokens (i.e., ERC-721 tokens standard on Ethereum).

\subsection{Summary: Insights and Lessons Learned}
In this section, we discussed the fundamentals of the metaverse, key enablers, and general architecture of the metaverse-based wireless systems. Several lessons learned are as follows:
\begin{itemize}
    \item There is a need for effectively taking into account both design aspects, wireless for metaverse and metaverse for wireless. To account for the wireless for metaverse, there is a need for studying emerging wireless technologies (e.g., 6G) and computing technologies to enable metaverse signaling (e.g., metaverse management signals between meta space and physical space). On the other hand, to enable wireless applications using metaverse, there is a need for efficient design of metaverse using emerging technologies, such as ML, edge computing, blockchain, and optimization, among others.
    \item There are two main phases in a metaverse-based wireless system design, such as learning phase and the operation phase. In the learning phase, learning of meta space models takes place using distributed learning that involves iterative communication between the end-devices and the meta space. On the other hand, a user may request a service from the meta space. Therefore, there is a need for novel resource scheduling algorithms for a metaverse-based wireless system.  
    \item There is a need for efficient modeling of digital avatars and twins while designing a metaverse-based wireless system. Although one can use mathematical modeling and experimental modeling, they might not produce good results due to assumptions in mathematical modeling and experimental errors. To overcome this, there is a need for privacy preserving, distributed ML-based twin and avatar models. However, such modeling has many challenges, such as statistical and system heterogeneity, wireless channel impairments, and fairness. System heterogeneity refers to variations in the available computing power for learning tasks among all devices, whereas statistical heterogeneity refers to the non-independent and non-identical distribution of local datasets. On the other hand, fairness refers to the dominance of some of the devices on the global model more than the other devices and thus, the learned global model will be biased. Therefore, to efficiently design twins and avatars using distributed learning, we must propose novel schemes that take into account fairness, statistical and system heterogeneity, and wireless impairments.   
    \item We learned that most of the existing platforms of the metaverse focused on smart gaming and creating virtual worlds for the users without effectively focusing on real-time applications (e.g., smart homes, intelligent transportation systems, and Industry 4.0). Therefore, there is a need for novel metaverse engines that can effectively enable wireless applications by taking into account various tasks. These tasks are effective resource allocation of computing and wireless resources, proactive analysis of the wireless system prior to deployment, and online control of wireless devices.    
   
\end{itemize}


\begin{figure}[!t]
	\centering
	\includegraphics[width=8cm, height=8cm]{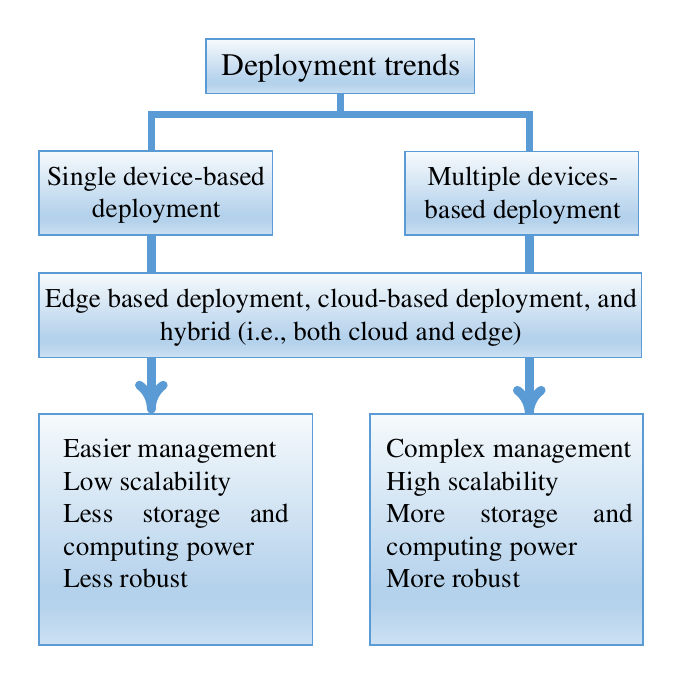}
	\caption{Meta space deployment schemes.}
	\label{fig:deployment trend}
\end{figure}

\section{Management of Meta Space}
\label{Meta Space}

\subsection{Network Management}

\subsubsection{Efficient Deployment of Twins and Avatars}
\label{Efficient Deployment of Twins and Avatars}
There can be two main deployment trends, such as single device-based deployment (e.g., edge server) and multiple devices-based deployments (e.g., multiple edge servers), as shown in Fig.~\ref{fig:deployment trend}. We first discuss the deployment using a single device. To deploy meta space, one can use edge servers located in close vicinity to the end-devices \cite{taleb2017mobile}. Such deployment of meta space will result in low latency, however, with low storage and computing power. Generally, we have less storage capacity and less computing power at the network edge compared to the remote cloud \cite{zahmatkesh2020fog,ning2019vehicular}. Therefore, one can deploy the meta space at a remote cloud when the latency requirements are not strict and high storage as well computing power is needed. Meta space based on edge servers has more context awareness compared to cloud-based meta space \cite{khan2020federated}. The reason for this context-awareness (e.g., devices location and mobility pattern) is due to the fact that edge servers are located close to the devices, and thus likely have more information about the network devices. Also, mobility management of devices for edge-based meta space is easier compared to cloud-based meta space because of the fact that edge servers are closer the devices, and thus more readily available knowledge about the devices' position and mobility patterns. \par   
Although the aforementioned discussion for implementing meta space on a single device can offer many benefits, it has a few limitations. The prominent one is scalability which is typically low for a single device-based implementation. A meta space located at a centralized location will suffer from high control signaling overhead, and thus suffer from high latency. Such a high latency might not be desirable for many strict latency applications (e.g., digital healthcare). To address this limitation, one can use meta space implementation in a hierarchical fashion using multiple devices \cite{peng2019hierarchical, smeliansky2018hierarchical, tong2016hierarchical}. One can have a root meta space and secondary meta spaces. The secondary meta spaces can be deployed to serve a part of whole devices. The primary meta space will coordinate among all the secondary meta spaces. The secondary meta space will handle all the signaling required for serving the devices within its vicinity. Such an approach of deploying meta spaces in a hierarchical fashion will enable scalable operation by serving a large number of users without significantly adding latency to the system. Other than scalability, robustness is also important. A meta space implemented on an edge server might stop working due to a malfunction or a security attack. Implementing meta space on distributed devices can offer more robustness compared to a single device implementation. However, this will be at the cost of management complexity. Therefore, one must make a tradeoff between robustness, latency, and management complexity during the deployment of meta space for a metaverse based wireless system.        

\begin{figure*}[!t]
	\centering
	\includegraphics[width=18cm, height=6cm]{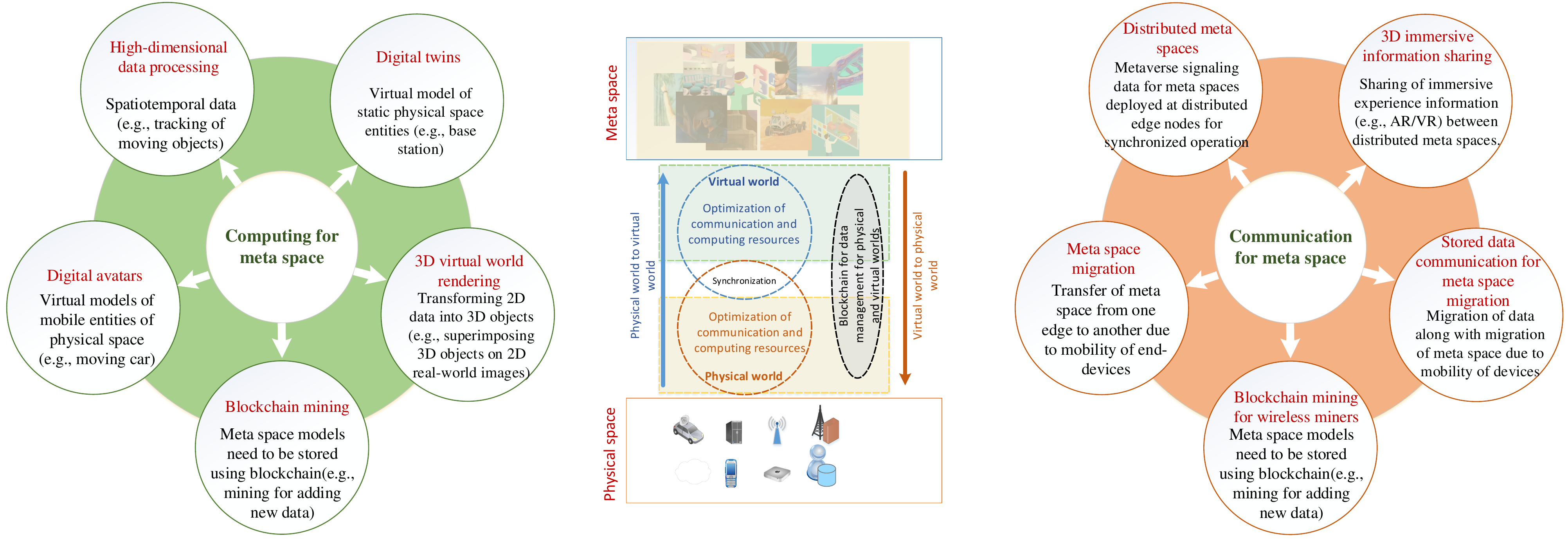}
	\caption{An overview of meta space computing and communication tasks.}
	\label{fig:meta_space_comp_comm}
\end{figure*}

\subsubsection{Computing and Communication Resource Management}
An overview of computing and communication management tasks for meta spaces is shown in Fig.~\ref{fig:meta_space_comp_comm}. The computing tasks in meta space can be high-dimensional processing of data, computing for digital twins modeling, computing for avatars modeling, 3D virtual world rendering, and blockchain mining. To perform computation for modeling of digital twins and avatars in a meta space, one can use various schemes requiring different computing power (i.e., CPU-cycles/sec). For instance, running simulations based on mathematical optimization to yield models of digital twins and avatars will have different requirements compared to data-driven (i.e., machine learning-based schemes) for twins and avatars modeling \cite{khan2020dispersed,khan2022metaverse}. Generally, mathematical optimization will use fewer computing resources, but at the cost of assumptions that might lead to less accurate modeling \cite{khan2022digital}. On the other hand, data-driven modeling has high computing complexity, but it will produce more accurate models. Other than modeling avatars and twins, one can encounter processing of high-dimensional data in a metaverse. Such high-dimensional data can be spatiotemporal data of autonomous driving based on metaverse \cite{yaqoob2020blockchain,zhang2023spatiotemporal,akhauri2021improving}. In addition, 3D virtual world rendering will require significant computing resources \cite{ ferrao2023environment,heinemann2022repix,jot2012interactive,cacciaguerra2004wireless,gan2017personalized,papaefthymiou2017gamified}. The first step in 3D virtual world rendering is 3D modeling that can be performed using mathematical modeling for representation. However, mathematical modeling might not be able to truly reflect the actual wireless scenario components (e.g., mobile cars). To address this challenge, one can use novel and emerging machine learning schemes. Note that 3D rendering in case of metaverse for wireless might be different. For instance, in the 3D rendering for AR/VR, one can focus on lighting effect. However, in the perspective of wireless systems, there is not a significant need to focus on the lighting effects. Such a lighting effect can enhance the 3D model illusion only without taking into the effect of twins and avatars on the wireless signals \cite{khan2022metaverse}. Instead, one should focus on the effect of 3D objects on the propagation of wireless signals (e.g., effect of users mobility on THz communication). Additionally, one must require computing resources for running blockchain consensus algorithm. Here, blockchain will be used to store meta space data in a transparent and immutable fashion. Blockchain consensus algorithms require significantly high computing power \cite{tasatanattakool2018blockchain}. Therefore, one must propose efficient schemes that will require less computing power while fulfilling the latency and energy constraints. \par

\subsubsection{Twins and Avatars Migration}
\begin{figure*}[!t]
	\centering
	\includegraphics[width=18cm, height=8cm]{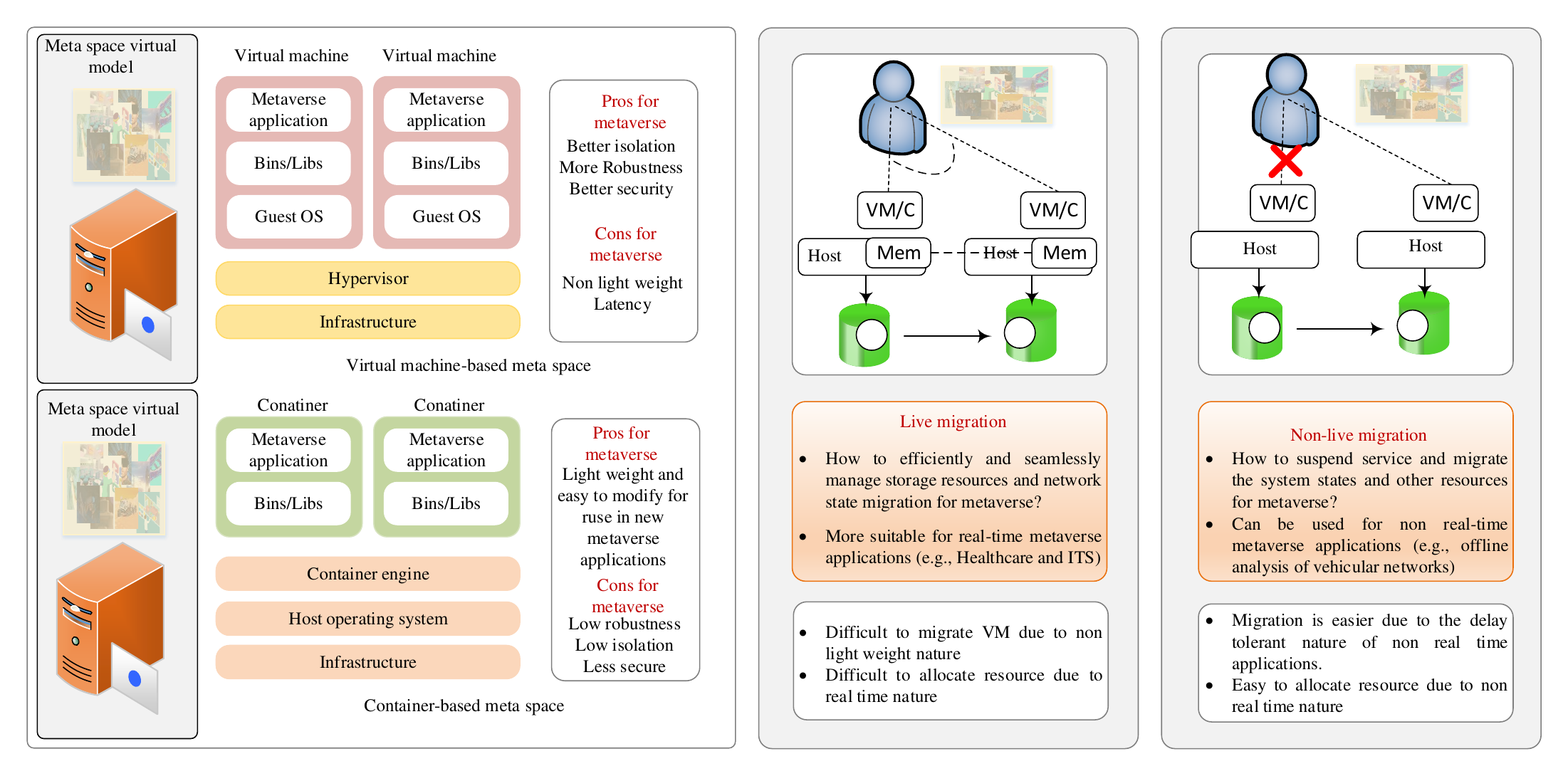}
	\caption{Overview of virtual machines, containers, and their migration schemes for metaverse.}
	\label{fig:VM_Containers}
\end{figure*}

Twins and avatars are deployed in the meta space upon the end-user request to serve them. To deploy, there is a need for using resources on-demand and then release these resources after use. To implement meta space (i.e., twins and avatars) at edge/cloud, one can use the concept of virtual machines and containers. We can use virtual machines/containers to create on-demand meta space and then release the computing resources after using it. Virtual machines are implemented using the virtualization of layers including hardware, whereas containers are implemented using software layers, as shown in Fig.~\ref{fig:VM_Containers}. Implementation of containers makes them easy because they only involve high software layers. This is the main reason why containers are lightweight and easy to modify for reuse in future metaverse applications. On the other hand, containers-based meta space will have low robustness due to having less isolation compared to virtual machines-based meta space that is implemented using virtualization of both hardware and software. On the other hand, virtual machines-based meta space will offer more isolation and robustness, but at the cost of high weight compared to containers-based meta space implementation. Therefore, one must make a tradeoff between robustness, reusability, and security.    \par

\begin{figure*}[!t]
	\centering
	\includegraphics[width=18cm, height=8.5cm]{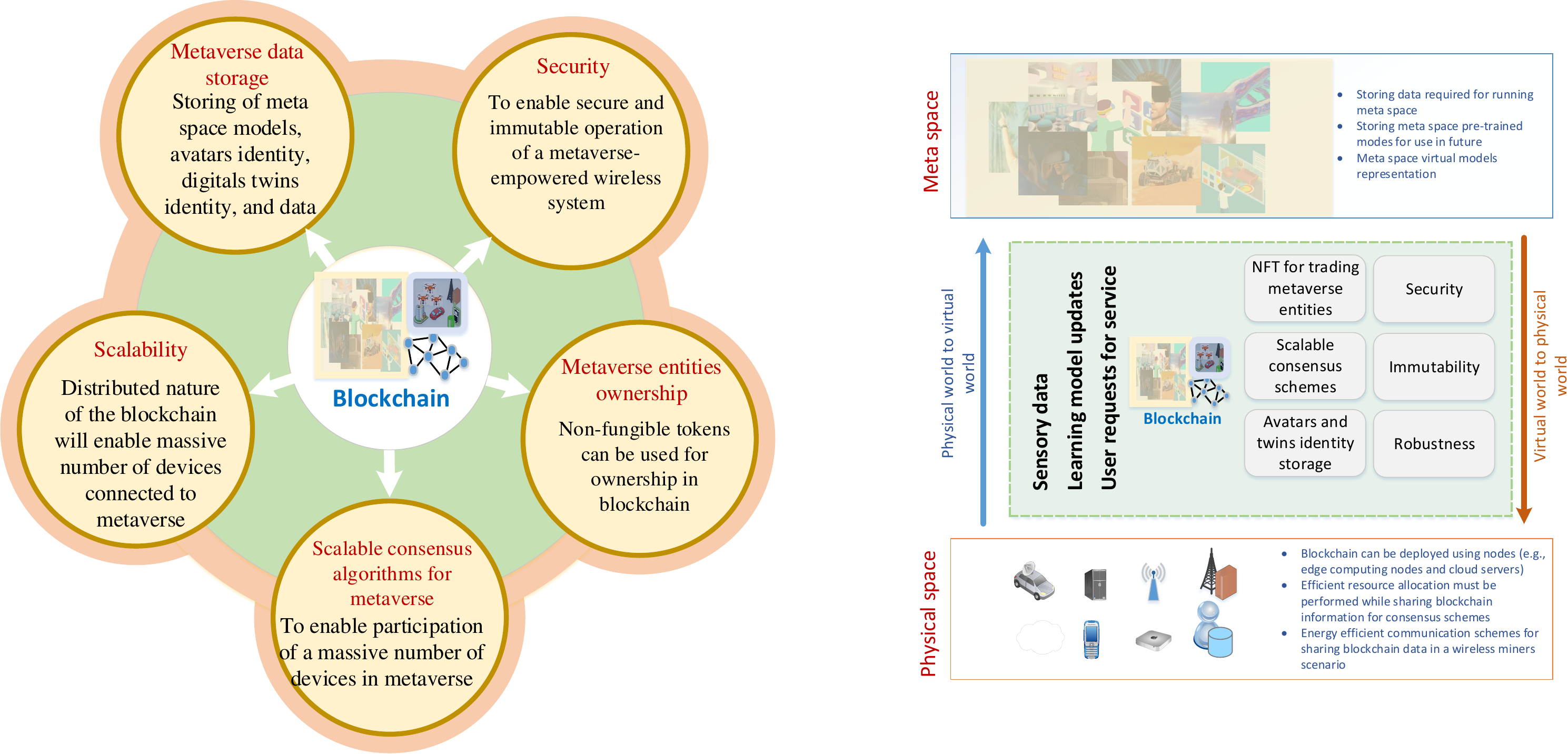}
	\caption{Overview of role of blockchain in metaverse.}
	\label{fig:blockchain}
\end{figure*}

Although virtual machine and containers based twins and avatars in a meta space can be used to serve end-users, the mobility of devices will cause many challenges in deployment. For instance, a meta space serving autonomous cars requires seamless communication during the serving period. However, autonomous cars have mobility, and thus they may go out of the range of their meta space deployed to serve them. To resolve this issue, there is a need for efficient migration of meta space to account for mobility. Other than mobility, hardware failure and imbalance loads can be tackled using meta space migration. Mainly, we can have two main types of meta space migration, such as live migration and non-live migration \cite{zhang2018survey}. In live migration, the meta space will be migrated towards the other supporting devices (i.e., edge or cloud server) without shutting down, whereas non-live migration first shut down or suspend before migrating the meta space to another facility. In non-real-time applications (e.g., training metaverse-based smart keyword suggestion in keyboard), one can use a non-live migration, and the states of meta space (based on virtual machines/containers) are transferred to the new running facility after suspending. Additionally, there is no need for transfer of the meta space states in case of shutting down. For real-time applications (e.g., infotainment and remote patient monitoring), there is a need for seamless service. Therefore, for such services, one should preferably use live migration. Although live migration can offer the benefit of the seamless running of applications, it has challenges in memory data migration and network connectivity. To tackle these issues, there is a need for managing the mobility of the devices. Based on the predicted mobility of devices, one can proactively live migrate the meta space to the new facility. Such kind of mobility can be predicted using various techniques. Most prominent is ML-based mobility prediction \cite{tang2019delay}. Although the mobility management scheme of \cite{tang2019delay} can be used for getting a mobility prediction model, it has privacy leakage issues. Users from some of the areas might not want to share their data with a centralized server. To address this one can use federated learning that will enable sending of only learning model updates instead of the whole data.  \par
Both for live and non-live migration schemes, there is a need for efficient resource allocation to carry out the migration process \cite{sharma2022energy, xiao2012dynamic, yang2020qos, ruan2019virtual}. Migration can be performed either using a wired or wireless network. A wired network has sufficient bandwidth, whereas a wireless network requires careful design due to communication resources (i.e., resource blocks) constraints. For instance, a meta space running on an edge server needs to migrate to the other edge server if the end-user moves to coverage of the new base station running that edge server. Such a migration can be performed wireless which will require efficient resource allocation with a variety of constraints. For instance, reusing wireless resource blocks of existing devices needs a resource allocation scheme such that interference caused due to reuse of wireless resources, to the existing devices should not exceed the maximum allowed limit. Other factors that should be taken are the efficient allocation of transmit power. Additionally, the migration delay should not exceed the maximum allowed latency. Therefore, one must perform resource allocation in such a way as to fulfill the latency constraints. Wireless resource allocation can be performed using various schemes. These schemes can be heuristic schemes, decomposition-relaxation schemes, game theory, matching theory, and convex optimization-based schemes \cite{ khan2022dispersed, khan2020federated2}. Typically, heuristic schemes (i.e., exhaustive search) check for all combinations and thus will give better results. However, checking all possibilities is significantly computationally expensive and may not be much desirable for practical strict latency applications. To overcome this, one can try a different scheme. A decomposition-relaxation scheme first decomposes the problem into sub-problems. Then, the decomposed sub-problems are solved separately by relaxing the binary resource allocation variables into continuous variables. Such approximation will lead to approximation error. One can use other schemes based on either game theory or matching theory to avoid approximation error and give faster convergence with less computational complexity. Yang \textit{et al.} in \cite{yang2020qos} proposed a heuristic scheme for migrating multiple virtual machines while fulfilling the latency constraints. Another work \cite{ruan2019virtual} also considered virtual machines migration for clouds. Although the works in \cite{yang2020qos} and \cite{ruan2019virtual} showed good performance, there is a need to propose novel schemes for meta space migration. For strict latency applications, the meta space will use edge servers that can be deployed either at base stations, unmanned aerial vehicles, and moving cars, among others. Migrating such a meta space will challenging due to mobility and wireless channel impairments as well communication resources constraints. Therefore, we should propose novel algorithms based on game theory/matching theory for meta space migration. \par                  

\subsubsection{Low Latency Consensus Algorithms for Blockchain}
An overview of blockchain for use in a metaverse is given in Fig.~\ref{fig:blockchain}. A blockchain in a metaverse can enable secure identity management, distributed metaverse data storage, metaverse entities ownership, scalability, and smart contracts \cite{belotti2019vademecum, berdik2021survey}. Typical in a metaverse, there will a wide variety of massive number of players. Therefore, there is a need for efficient unique identity management in an immutable manner. Additionally, a metaverse data can be stored using the distributed storage in a blockchain using immutable and transparent manner. Blockchain can be used to store metaverse data (e.g., pre-trained meta space models and authentication keys) in a distributed manner and thus more robust to failures. As the blockchain is based on the distributed concept, therefore, metaverse based on blockchain can enable a scalable operation. One of the possible ways to further increase the scalability of blockchain for metaverse can be sharding that uses the concept of dividing the whole network into many sub-networks \cite{yu2020survey,hafid2020scaling}. In a metaverse-enabled wireless system, there will be a variety of decentralized and distributed datasets. These datasets will be used for various purposes, such as training machine learning models and operations (e.g., cached data and security-related information). To enable these datasets in a transparent, efficient, and immutable manner, one can use blockchain \cite{wang2021blockchain}. One of the main advantages of blockchain is that no node can change the data without collusion. Note that blockchain can update the distributed datasets after running the consensus algorithm. A consensus algorithm is a fault-tolerant mechanism that enables an agreement on a set of rules agreed by decentralized nodes in contrast to a centralized authority. Therefore, one can use blockchain for various purposes in a metaverse-based wireless system. In a metaverse for a wireless system, a set of wireless devices used to train distributed learning leveraged blockchain to avoid a single point of failure issue \cite{kim2019blockchained}. Similarly, we can consider wireless miners that communicate with each other. Every miner can have a block with two parts: body and header. The block body can carry information about metaverse applications (e.g., control data and meta space pre-trained models). If some new update is to be added to the blockchain network, a miner generates a hash value after running the consensus algorithm. If the hash value is less than the target value, then the miner is allowed to update the distributed ledger by its generated block. The generated block is transmitted to all the miners. Some of the receiving nodes might be successful in solving the problem and broadcast their own block in the network before receiving the generated block of the other node. This event is called forking. There must be a measure to avoid this forking event by controlling the block generation rate. Another factor is the efficient allocation of wireless resources that can minimize the transmission latency, and thus forking. On the other hand, the consensus algorithm must be scalable with low latency. Additionally, the consensus algorithms must consider the privacy leakage issue as well because of their distributed nature. In a metaverse-based wireless system, there will be a massive number of nodes. To handle distributed datasets using blockchain, we must propose consensus algorithms that can work well without adding a significant delay to the system. Other than scalability, latency is another issue with running blockchain consensus algorithms. Therefore, we must propose novel blockchain consensus algorithms that are scalable and offer low latency along with better privacy preservation. \par
\begin{figure}[!t]
	\centering
	\includegraphics[width=6cm, height=9cm]{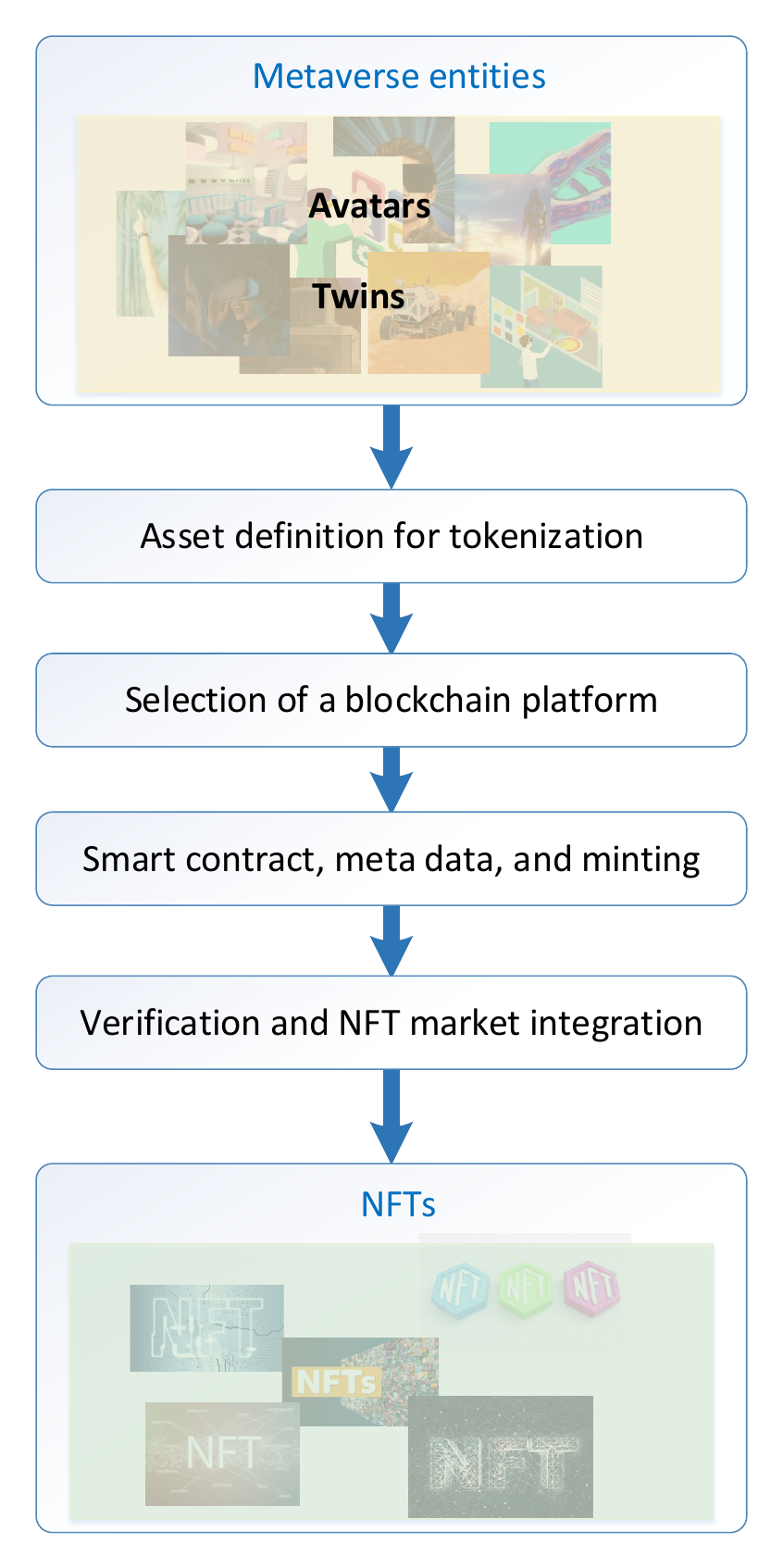}
	\caption{NFTs creation for metaverse.}
	\label{fig:NFT}
\end{figure}
Non-fungible tokens (NFT) will plays a significant role in a metaverse-empowered wireless system. In a metaverse, there are two main players, such as avatars (i.e., for representing mobile users and devices), digital twins (e.g., virtual model of a piece of land and buildings), from the perspective of digital ownership. To enable such ownership one can use NFTs that are based on utilizing standards such as ERC-721 or ERC-1155 on the Ethereum blockchain \cite{valeonti2021crypto}. NFTs are unique identities used to represent as well as trade metaverse entities. An overview of creating NFTs for a metaverse is given in Fig.~\ref{fig:NFT}. First of all, assets are defined to be represented using NFTs. The next step is to select a suitable platform. Currently, many works are using Ethereum with its ERC-1155 and ERC-721 tokens. After selecting a blockchain platform, there is a need to propose smart contracts that defines rules and functions of the NFT, such as royalty mechanism, metadata storage, and ownership, etc. Additionally, there is a need to define meta data (e.g., name, description, and image of the asset in meta space). Then, the next process is to mint a unique instance of the NFT on blockchain. The final step is verification and NFT market management.         

\begin{figure*}[!t]
	\centering
	\includegraphics[width=18cm, height=6.5cm]{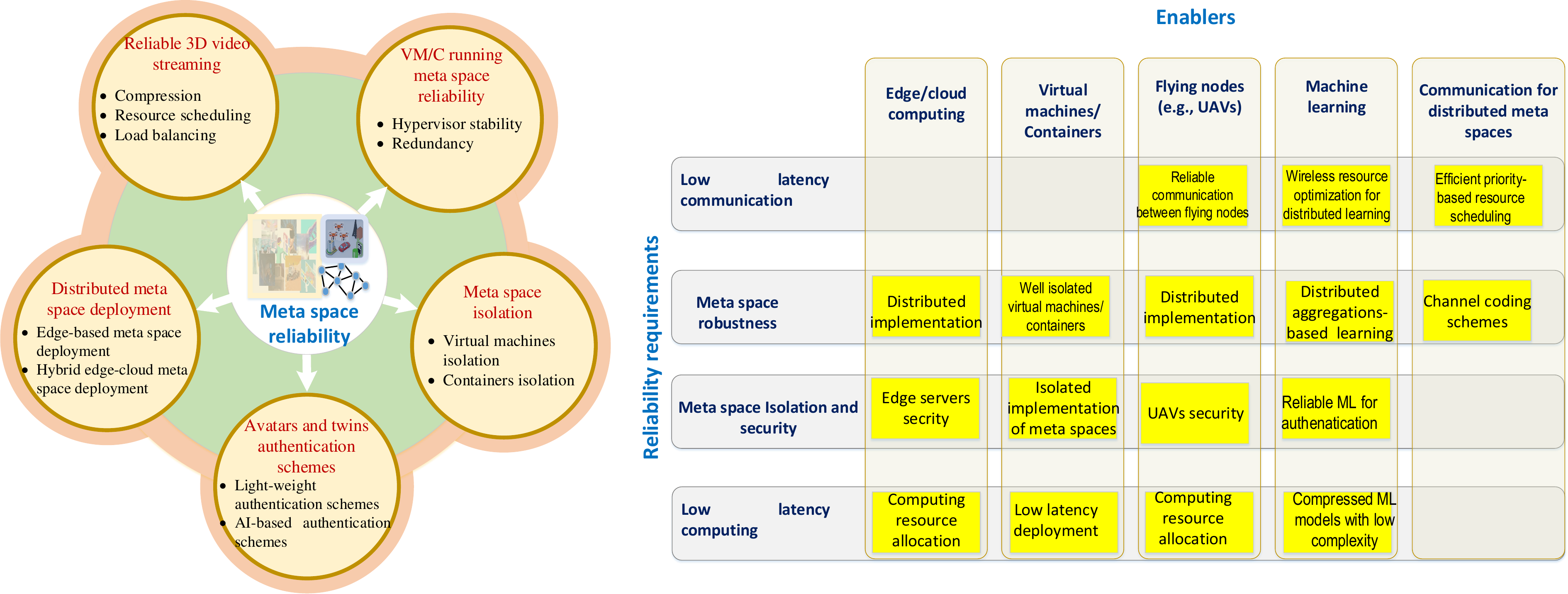}
	\caption{Overview of reliability of meta space.}
	\label{fig:meta_space_reliability}
\end{figure*}

\subsection{Reliability and Security}
Overview of reliability and security for meta space is given in Fig.~\ref{fig:meta_space_reliability}. The relationship between reliability requirements and key enablers of meta space is also given in Fig.~\ref{fig:meta_space_reliability}. The key reliability and security tasks in meta space are reliable $3$D video streaming, virtual machines/containers running meta space reliability, meta space isolation, distributed deployment of meta space, and avatars/twins authentication schemes. Similar to network slicing, there are two ways to implement meta space for various applications: (a) dedicated physical space hardware and (b) shared physical space hardware \cite{kazmi2019network, zhang2019overview,foukas2017network}. In the case of dedicated physical space hardware, one can deploy meta space to serve users. Such an approach will have to advantage of easier management and better performance but will cost high and is practically not feasible. To overcome this high issue, one can use shared physical space hardware that allows multiple meta spaces to operate. Although using shared physical space hardware for multiple meta spaces will be a good and feasible solution, it has a few implementation challenges, such as resource allocation, reliability, security, and isolation \cite{zhang2017network}. For instance, meta spaces deployed to service intelligent transportation and healthcare at the same network edge might suffer from security concerns if one of them is being attacked by a malicious user due to sharing of the same physical space hardware. Therefore, there is a need for efficient isolation of meta spaces. Note that isolation for meta spaces will be at various levels: access network isolation, computing resource isolation, and core network isolation \cite{ordonez2017network}. Effective isolation will result in a secure and reliable operation. For a dedicated model, an access network slice for meta space has a dedicated user and control plane traffic, spectrum, and MAC scheduler \cite{kazmi2019network}. This approach will ensure low latency, the isolated, secure, and reliable operation but will cost high and not allow elastic operation. Every meta space slice will have access to its own medium access control, radio link control, and radio resource control instances along with resource blocks. On the other hand, using dedicated physical space hardware, there is a need for sharing of spectrum, MAC scheduler, and control plane. Specifically, the resource blocks for serving different meta spaces will be managed by a single scheduler and thus, will face many management and isolation challenges. To resolve these challenges, one can modify the medium access control scheduler that can use resource blocks of different network operators and allocate them to various meta spaces in an efficient way. To do so, one can have an objective function that is based on maximizing the utility (i.e., overall throughput) while fulfilling the meta space user requirements (e.g., reliability and latency). On the other hand, for such an interaction between meta space schedulers, network operations, and end-users, there must be some efficient incentive mechanism. Such an incentive mechanism will enable buying of resources from network operators and selling them to meta space users with the aim to maximize the profit while improving users performance. Similar to access network, one must propose novel scheduling schemes for sharing of computing as well as core network resources. All of the above schemes will use optimization theory, game theory, deep reinforcement learning, and graph theory, among others \cite{leconte2018resource,kazmi2019network,khan2020network}. On the other hand, there are various ways of implementation of meta space. These possible ways can be meta space implementation using an edge server, cloud, edge-cloud, or devices \cite{khan2022metaverse}. Implementing meta space using devices can have easier management but at the cost of low reliability and security. An edge server running meta space might suffer from malfunction either because of a security attack or physical damage. To resolve this, one can deploy meta space using multiple edge/cloud servers. This approach will lead to more computing power and storage capacity as well as security and reliability but at the cost of management complexity. Based on the aforementioned facts, one can say that careful attention must be given to the implementation of meta spaces. Other than this, there must be secure interfaces for communication between meta space and physical space. For such security, one can use encryption/decryption schemes.               

\begin{figure*}[!t]
	\centering
	\includegraphics[width=18cm, height=6cm]{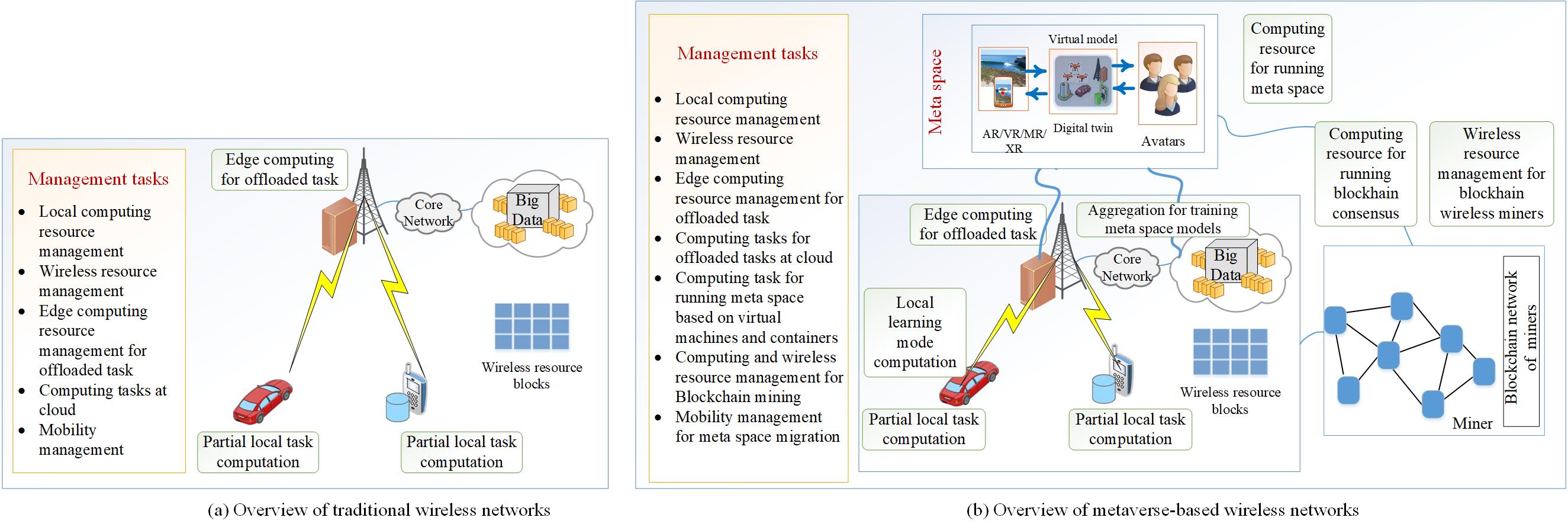}
	\caption{Overview of computing resource and communication resource management tasks for physical space.}
	\label{fig:resource_management}
\end{figure*}

\subsection{Summary: Lessons learned and Insights}
This section described how can we design and deploy meta space over the physical infrastructure. Specifically, efficient deployment of meta space, meta space migration, reliability, and security are discussed. Several lessons learned are as follows.
\begin{itemize}
    \item We learned that is a need for efficient deployment of meta space using edge and cloud. Deployment of the meta space requires storage and computing resources. For edge, there will be computing resource limitations, whereas, for cloud, latency is the problem. Therefore, deployment of meta space should be efficiently performed. Additionally, running the meta spaces for different applications on the same edge requires careful design for optimally allocating computing and storage resources. 
    \item Deployment of meta space (i.e., avatars and twins) on edge/cloud server must be performed intelligently and on-demand either using virtual machines or containers depending on the specifications. For instance, virtual machines are implemented using virtualization of layers including hardware as well as software layers, and thus gives better isolation and security. However, these features are at the cost of non-light weight nature compared to a container. Therefore, we must wisely choose containers and virtual machines for the implementation of on-demand meta space at edge/cloud.
    \item Mostly, the emerging wireless applications are real-time, therefore, we should use live migration of meta space from one edge to another depending on the mobility of end-devices. To do so, there is a need for an effective ML-based scheme for the prediction of device mobility. Based on the predicted mobility of devices, one can proactively start the migration of meta space to avoid latency in the service. For such kinds of predictions, one must propose effective algorithms based on emerging schemes of ML. To do so, one can use distributed learning with better convergence. Normally, distributed learning has a low convergence rate due to devices and statistical heterogeneity as well as fairness issues. Therefore, there is a need for efficient novel distributed learning schemes for predicting the mobility of devices.    
    \item To effectively isolate the meta space of one application from others, there is a need for novel isolation schemes that allow the operation of various twin spaces for different applications on the shared physical infrastructure. For wireless resources, one can use the concept of virtualization which can be achieved using a modification of the existing resource schedulers at the medium access control layer. To do so, there should be an efficient novel algorithm based on either contract theory, Stackelberg game, or matching theory that will enable buying of wireless resources from various network operators and selling them to the different meta spaces to increase an overall utility (i.e., that accounts for network operators profits and meta space users performance).     
\end{itemize}

\section{Physical Space}

\subsection{Network Management}

\begin{figure}[!t]
	\centering
	\includegraphics[width=8cm, height=8cm]{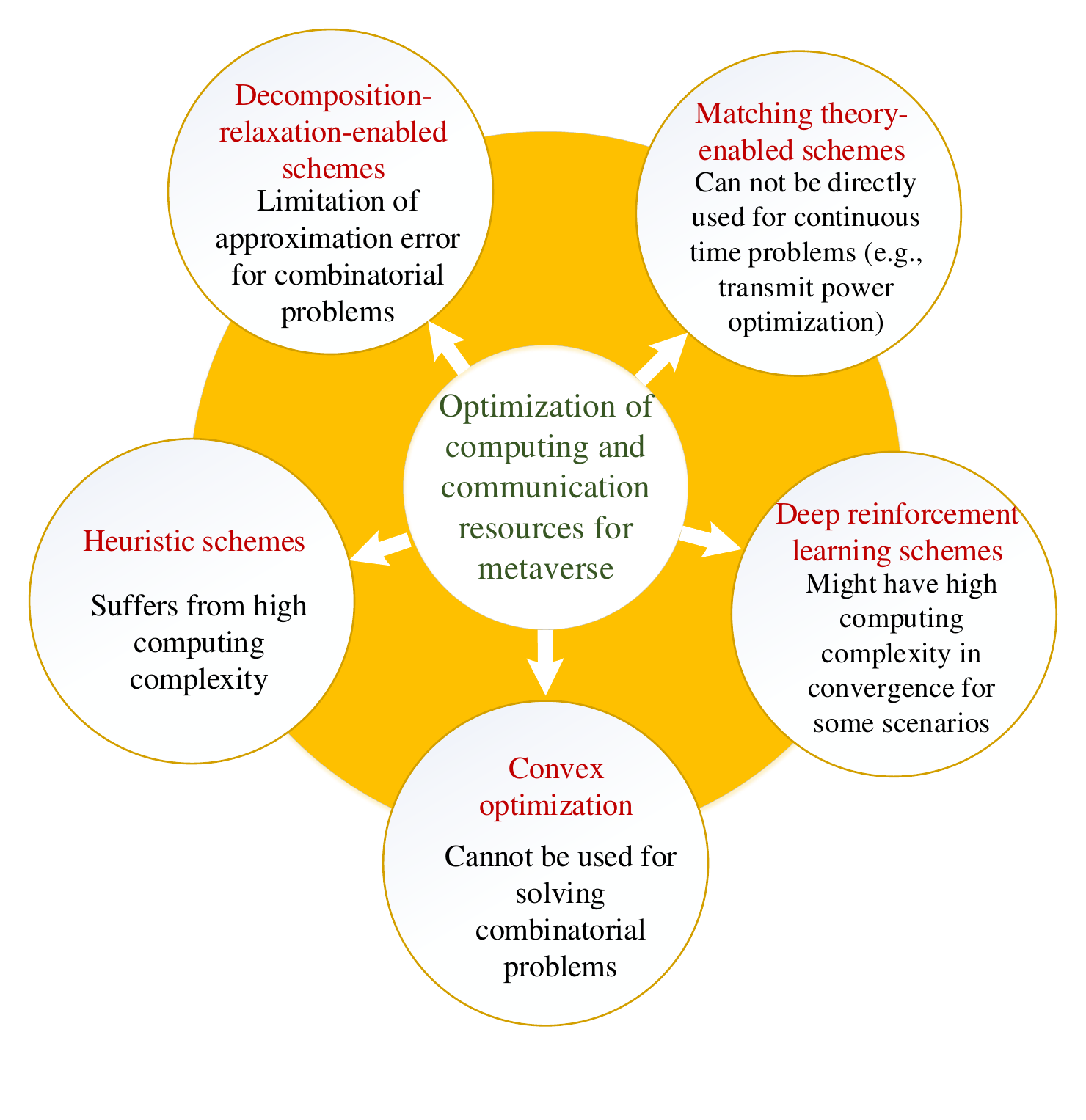}
	\caption{Overview of resource optimization schemes.}
	\label{fig:ph_resource_management}
\end{figure}

\subsubsection{Wireless and Computing Resource Management}
The physical space of the metaverse has a wide variety of players, such as edge/cloud servers, base stations, autonomous cars, moving devices, and unmanned aerial vehicles, among others \cite{khan2022metaverse}. For the successful enabling of metaverse-based wireless systems, there is a need for seamless interaction among devices. Additionally, the key element of the metaverse, namely, meta space will be run by the wireless system hardware. To do so, there is a need for efficient allocation of wireless and computing resources. An overview of computing and wireless resource optimization schemes is given in Fig.~\ref{fig:resource_management}. Typically, wireless resources of access networks need careful design of resource allocation schemes. The design of the wireless resource allocation scheme depends mainly on the access scheme (e.g., orthogonal multiple access (OFDMA) and non-orthogonal multiple access (NOMA) \cite{dai2015non,ding2017survey,ding2015application}. Typically, devices in wireless systems have computing tasks and they do not have sufficient resources. Therefore, they compute a part of their task and send the remaining to the nearby base stations enabled by edge servers. Additionally, edge and cloud servers should support virtual machines and containers running meta space. For interaction among devices and edge servers, one must efficiently allocate computing and wireless servers. Also, for efficient implementation of meta space, the computing resources must be efficiently managed. The works in \cite{kim2014joint, gao2019user, fooladivanda2019joint,trabelsi2017user,yemini2019optimal} considered the association and allocation of wireless resources in a cellular network. In \cite{kim2014joint}, the authors considered an association-interference problem. They formulated a problem to maximize the network utility. Due to the non-convex nature of the formulated problem, a dual decomposition is used. In another work, \cite{gao2019user}, Gao \textit{et al.} proposed an energy-efficient scheme for user association and on/off of the base stations. The problem is formulated as a non-convex nonlinear programming problem which is decomposed into two sub-problems for an easier solution. The work in \cite{fooladivanda2019joint} proposed a joint user association and resource allocation in heterogeneous cellular networks. Similarly, the works in \cite{trabelsi2017user} and \cite{yemini2019optimal} discussed resource allocation and association. \par
Other works \cite{saeik2021task,wang2022decentralized, luo2020game,kan2018task} considered task offloading in edge computing. In \cite{saeik2021task}, the authors surveyed different techniques used for offloading in edge and cloud computing. specifically, they studied various applications based on edge computing and then challenges related to edge offloading. Another work \cite{wang2022decentralized} proposed a decentralized scheme for task offloading in an edge computing system. The authors in \cite{luo2020game} proposed a game theoretic scheme for enabling efficient task offloading between multiple users and multiple base stations. The works in \cite{saeik2021task,wang2022decentralized, luo2020game} mainly performed association of devices to base stations; however, they did not consider the edge computing and offloading of resource-constrained end-devices. In a metaverse-based wireless system, there will be a need for local computational task offloading as well optimization of edge servers computing resources for performing various tasks, such as the running of meta space, offloaded computing task, aggregation of meta space models for distributed learning, computing partial local learning models for split distributed learning, and running blockchain miners, among others. On the other hand, the work in \cite{kan2018task} considered both tasks offloading and resource allocation for edge computing. Similarly, other works \cite{zhang2017joint,zhang2019joint} considered joint computing and wireless resources (i.e., transmit power allocation and resource block allocation). In \cite{zhang2017joint}, the authors proposed a game theoretic scheme for joint computational offloading and resource allocation in mobile edge computing. They formulated an objective function that accounts for energy consumption and
monetary cost. Due to the NP-hard nature of the formulated problem, a joint offloading and resource allocation optimization game was proposed to solve the formulated problem. Another work \cite{zhang2019joint} considered a system of multiple edge servers and users. They formulated an objective for minimizing the cost that considers the time and energy consumption of devices. To solve the formulated problem, the authors proposed a two-stage algorithm using alternating optimization and one-dimensional search. One-dimensional search performs offloading decisions, whereas the second stage, alternating optimization, performs resource optimization. 
Although the works in \cite{zhang2017joint,zhang2019joint,kan2018task} can be used for allocation for computing, offloading of tasks, and wireless resources in a typical wireless system, they will not perform well for a metaverse-based wireless system. In a metaverse-based wireless system (shown in Fig.~\ref{fig:ph_resource_management}), the scenario is different and there are a wide variety of players involved in resource management, such as end-devices computing resource, wireless resource blocks, offloaded task computation at the edge servers, edge computing resource for running meta space, and storage resource for blockchain miners. Note that for different services, one can deploy different meta spaces. To meet the aforementioned challenges of a metaverse-based wireless system, there is a need for novel frameworks that will joint perform computing resources (i.e., local devices computing resources for local model computing and local task computing, whereas edge computing resources for meta space running, computing offloaded tasks, computing partial local meta models for split learning case) and communication resource (i.e., for transmission of learning updates, user requests, offloaded task, and mining information). \par          

\begin{figure*}[!t]
	\centering
	\includegraphics[width=18cm, height=6cm]{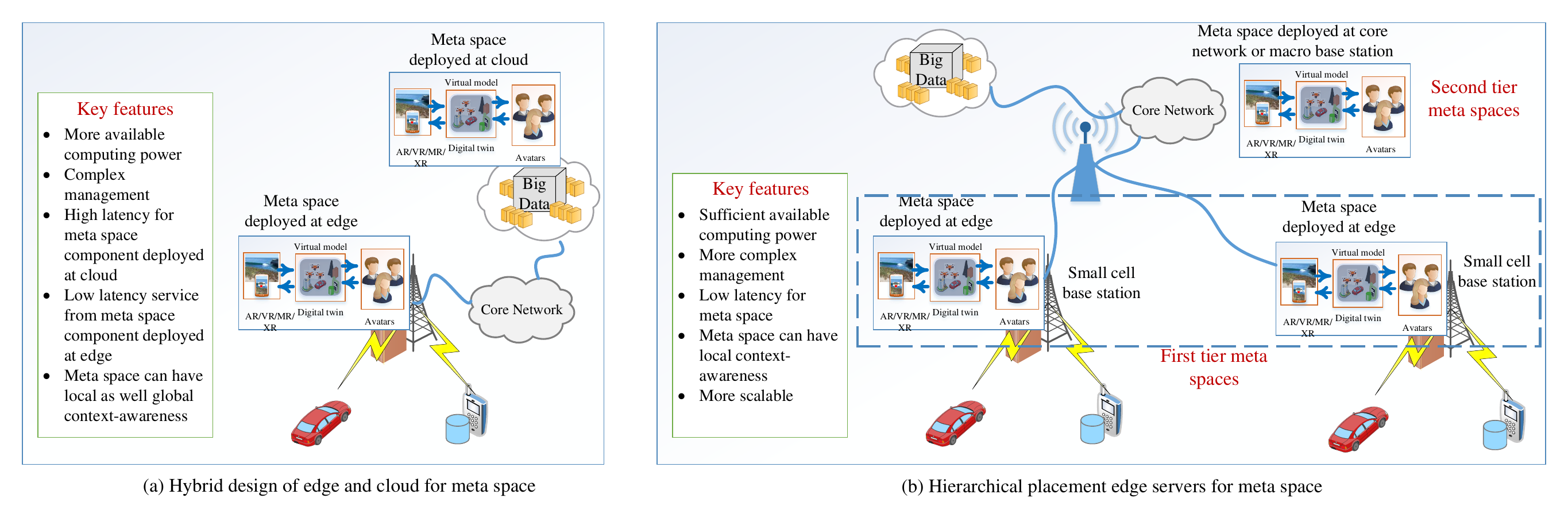}
	\caption{Overview of edge and cloud deployment for running meta space.}
	\label{fig:edge_cloud_deployment}
\end{figure*}

\subsubsection{Devices Mobility Management}
\label{Devices Mobility Management}
The mobility of devices poses different challenges in a metaverse compared to traditional wireless networks. For instance, a mobile device connected to one base station can easily be handed over to the new base station if it enters its coverage area. However, the case is different in the metaverse where simply handover will not work. There must be different and novel schemes to address the mobility of users. There are two main phases in the metaverse: (a) offline training of meta space models and (b) online operation \cite{khan2022metaverse,khan2022machine}. For training, one can use various schemes, such as centralized ML or distributed learning \cite{khan2022federated}. Distributed learning can offer many benefits over centralized ML. For distributed learning, frequent communication takes place between the meta space deployed at the network edge/cloud and end-devices. For such interaction, there must be seamless communication between devices and the edge/cloud server. It is desirable that the devices should remain in the coverage area of the edge-based base station. such a fashion will generally result in a faster convergence \cite{khan2020federated,khan2020dispersed}. Frequent changes in the devices for a typical edge server in case of multiple edge servers will result in changes in local datasets (i.e., of devices), and thus will suffer from a slow convergence rate. To resolve this issue, one can use a clustering approach that should be based on the clustering of devices that have more probability to remain within the coverage area of each other with one of the nodes as a central node acting for aggregation \cite{khan2020dispersed}. On the other hand, during serving the end-devices by a meta space deployed at edge/cloud, we should also tackle mobility. For serving the devices, the meta space will enable them with efficient resource management that will require seamless communication among devices and meta space. Therefore, during the training phase and operation phase, there is a need for efficient management of device mobility. \par
Mobility management of devices in wireless systems is considered by various works \cite{lee2011spectrum,smys2015self, fernandes2010vertical,siddiqui2006mobility,achour2015mobility}. Broadly, one can divide the management schemes into categories: (a) within a network of one network operator and (b) between different network operators. For instance, a device connected to meta space deployed on edge supported by one network operator can go under the coverage area. In this case, mobility management will be easy and will generally require less signaling information compared to the case when the device moves to the coverage of new network operators. Therefore, there is a need for novel schemes that can efficiently handle the mobility of devices served by meta space. To continue seamless operation, the meta space should also be migrated based on the mobility of devices. In \cite{lee2011spectrum}, the authors proposed a spectrum-aware mobility management scheme. Specifically, they presented an architecture for mitigating heterogeneous spectrum availability. Using this architecture, a unified mobility management framework is presented to cope with the issue of mobility events. Moreover, the authors proposed inter-cell resource allocation. Other works \cite{smys2015self,fernandes2010vertical,siddiqui2006mobility,achour2015mobility} surveyed and presented schemes for mobility management of devices in wireless networks. However, note that the nature of a metaverse-based wireless system is different. Along with the mobility of devices, there is a need to migrate corresponding meta space (i.e., those serving the mobile devices) as well. Therefore, traditional mobility management schemes will not work well for a metaverse-based wireless system, and we should propose novel schemes.\par

\subsubsection{Edge and Cloud Deployment}
\label{Edge and Cloud Deployment}
To deploy edge and cloud servers for serving wireless system users, there is a need for efficient deployment of edge and cloud servers \cite{khan2020edge}. Every edge and cloud server requires sufficient backup power for operation. They will require cooling especially for cloud servers \cite{patel2003smart,liu2012renewable,evans2012different}. Additionally, edge servers have limited computing power, and thus they should be deployed intelligently. Therefore, there is a need for efficient deployment of edge and cloud servers. Such efficient placement of edge/cloud servers is necessary for efficient running meta space based on virtual machines and containers. Various works \cite{vitello2021mobility,shao2022cost,cong2022coopedge} considered efficient deployment of edge servers. In \cite{vitello2021mobility},  Vitello \textit{et al.} proposed the efficient placement of edge data centers in urban environments. They focused on using user mobility along with spatial deployments to assist in the efficient deployment of data centers. Another work \cite{shao2022cost}, studied three key issues, such as capacity at edge locations, user association, and edge location, related to edge servers deployment. Cong \textit{et al.} in \cite{cong2022coopedge} proposed a scheme for cost-efficient deployment for cooperative edge computing. Specifically, the idea of the authors was to share the edge servers (i.e., overlapped) during the time of peak load in a cooperative manner. The advantage of this approach is to avoid using a large number of edge servers to fulfill the demands during peak hours. \par
On the other hand, there are computing limitations on the edge server running meta space/s. To address this, one can have a hybrid placement of meta space/s. Such a hybrid placement will allow us to use both edge and cloud for placing meta space. This approach will enable to use of edge computing resources by the meta space first and then the cloud computing resource if needed. Although this approach of hybrid deployment can offer the benefits of high computing power, performing a task by meta space deployed in the cloud will suffer from a high latency that is undesirable. To resolve this issue, one can use the concept of hierarchical edge deployment of meta space/s, as shown in Fig.~\ref{fig:edge_cloud_deployment}. Although hierarchical fashion Fig.~\ref{fig:edge_cloud_deployment}b can significantly improve the performance of a metaverse, it has a few limitations. Its context awareness (i.e., information about the surrounding network nodes) is local. The reason for this is the low converge area associated with the small cell base stations. On the other hand, context-awareness is global for hybrid due to the fact that the cloud is associated with a large coverage area, and thus might have information about the nodes located over a large geographical area. The hierarchical fashion of deploying edge servers for enabling a meta space can offer scalability as well. The reason for this is the number of devices associated with a meta space deployed at the edge will be less. Only the top tier will provide service if the computing power available in the bottom tier is insufficient. Additionally, a top-tier meta space will control the bottom-tier meta spaces. This approach will enable more scalable operations. Note that we can have multiple meta spaces for enabling a single service/application (e.g., infotainment in autonomous cars). For enabling infotainment, one should perform caching in addition to other schemes. Such caching based on meta space can be performed either at meta spaces deployed at the first tier and also at the second tier. Such a fashion of hierarchical caching has been considered by many works \cite{9417323}. Therefore, we must efficiently deploy meta space/s for metaverse-based wireless systems.

\subsection{Reliability and Security}
In the physical space of the metaverse-based wireless system, there are a wide variety of players, such as edge/cloud servers, end-devices, blockchain miners, software-defined networking switches, and unmanned aerial vehicles, among others \cite{khan2022metaverse}. Devices' physical access from a malicious user is very difficult because of their distributed nature. Therefore, one must deploy effective authentication schemes to avoid attacks due by malicious users. One must propose efficient and lightweight authentication schemes. Such a scheme can be based on tokens generated by a server (e.g., Auth2 protocol) or a non-tokens-based scheme that uses the user name and password \cite{el2019survey}. Authentication can be performed using various ways: one-way authentication, two-way authentication, and three-way authentication. One-way authentication involves authenticating only one party (e.g., client) without considering the other (e.g., server). This approach might have low complexity but might suffer from inefficiency in the case of the malicious second user for which authentication is not required. To address this limitation, one can have two-way authentication, both parties agree to authenticate with each other. To make the system more secure, one can use three-way authentication that involves a third party authenticating the two parties. Other than authentication schemes, there must be some mechanism for secure wireless communication. A malicious user might access the wireless signal and cause leakage/alteration of sensitive information. Additionally, during training of the meta space model using distributed learning model, a malicious user can access the wireless local learning model and infer the device-sensitive information \cite{el2019survey}. Therefore, there is a need for good encryption schemes before transmitting wireless signals for a metaverse. A data encryption scheme transforms plaintext data into encoded data, namely, ciphertext to avoid the man-in-the-middle attacks that can result in the leakage of devices' sensitive data. One can use various encryption/decryption schemes. One of the popular ones is homomorphic encryption \cite{acar2018survey}. An advantage of using homomorphic encryption is that there is no need of sharing a key between the two parties involved in communication to avoid privacy concerns. In homomorphic encryption, the receiving party can operate on the data without the need for decryption. For instance, training a meta space learning model using distributed learning, devices send their locally trained models to the meta space where aggregation has to take place. A malicious aggregation server can infer the devices' sensitive information using their learning model updates. Therefore, here we can use homomorphic encryption to encrypt the local model and at the aggregation server, one can perform aggregation without the need for decryption \cite{khan2021federatedl,fang2021privacy,ma2022privacy}. Homomorphic encryption can be divided into three types: (a) partial homomorphic encryption, (b) somewhat homomorphic encryption, and (c) fully homomorphic encryption. In partial homomorphic encryption, only one type of operation can be performed an unlimited number of times, whereas a somewhat homomorphic encryption scheme allows some types of operations that will be performed a limited number of times. However, fully homomorphic encryption allows an unlimited number of times of operations. Note that homomorphic encryption enables effective security, there is a need for efficient wireless resource allocation as it results in a significant overhead, especially fully homomorphic encryption. Therefore, there must be a tradeoff while selecting a homomorphic encryption scheme.\par
Other than security, there must be reliable communication between the devices of physical space. For reliable communication, one can use effective channel coding that enables encoding the input bits into a coded sequence of bits to make the system robust against channel errors. One can use linear block codes, convolutional codes, and Turbo codes \cite{zhan2018channel,arora2020survey,biglieri2000coding}. Although linear block codes have low computing complexity, they might not perform well in all scenarios. To overcome this, one can use convolution codes that may perform well but will be generally at the cost of the increase in computing and communication costs. On the other hand, one can use Turbo codes that are based on either parallel or series concatenation of linear block codes or convolutional codes. Generally, Turbo codes can outperform all other schemes, but they have high computing and communication cost. Therefore, one must make a tradeoff between performance and cost. For Ultra-Reliable Low Latency Communications, recent works proposed the use of Short Block-Length Codes \cite{shirvanimoghaddam2018short,sybis2016channel}. Bose, Chaudhuri, and Hocquenghem (BCH), Low-density parity-check (LDPC) codes, convolutional codes, and Turbo codes can be used for URLLC. Similarly, one can use these codes for requesting devices in a metaverse-based wireless system. BCH codes have shown good reliability under optimal decoding conditions among various codes (e.g., polar codes and convolutional codes) \cite{shirvanimoghaddam2018short}. From the aforementioned discussion, one can say that we must properly select a code with low overhead for metaverse-based wireless systems.

\subsection{Summary: Lessons Learned and Insights}
In this section, we discussed various management functions (e.g., resource management and deployment of edge and cloud servers) of the physical space. Moreover, we discussed the reliability and security of physical space. Several lessons learned from this section are as follows.
\begin{itemize}
    \item There must be efficient joint computing and wireless resource allocation schemes for a metaverse-based wireless system. Such a resource allocation in a metaverse-enabled wireless system is different compared to traditional resource allocation problems due to the presence of a wide variety of players. Such a problem will be a kind of mixed integer non-linear programming problem (MINLP) along with numerous constraints. To solve such kind of problem, there is a need for novel solutions based on decomposition-relaxation, game theory, deep reinforcement learning, and graph theory \cite{khan2020federated}. 
    \item It is evident that the deployment of edge servers for running meta spaces must be performed intelligently. One can deploy meta space at the network edge or cloud or both edge and cloud. Deployment at the network edge will result in more context awareness compared to cloud-based meta space but at the cost of low computing and storage resources. More context awareness (e.g., device location) will result in better mobility management and vice versa. On the other hand, one can use both cloud and edge for the deployment of meta space to offer benefits of both edges (i.e., low latency and more context awareness) and cloud (i.e., more storage and computing power). 
    \item Novel low overhead channel coding schemes should be proposed for a metaverse-enabled wireless system. These low overhead channel coding schemes can be comprised of the existing schemes (e.g., Turbo codes and linear block codes) or modified versions for further reducing the overhead while fulfilling the bit error rate requirements of the applications as well metaverse signaling. 
    \item Mobility of the devices must be given proper attention as it will significantly affect the performance of a metaverse-enabled wireless system. Both during training of meta space models and service request/operation, there is a need for effective mobility management. For mobility management, one can use novel schemes based on deep reinforcement learning or federated learning. 
\end{itemize}

\begin{table*}[htp!]
\fontsize{8}{9}\selectfont
\caption {State-of-the-art: key contributions, primary focused area, design aspect, and architecture/experimental model.} \label{tab:advances1} 
\begin{center}%
\begin{tabular}{p{2cm}p{7.5cm}p{1.5cm}p{1.5cm}p{1.5cm}p{1.5cm}}
\toprule
    \textbf{Reference}   & \textbf{Key contributions}  & \textbf{Primary focus} & \textbf{Design aspect} & \textbf{Framework or Experimental testbed} & \textbf{Remarks}\\ \midrule
    Zhang~\textit{et al.}~\cite{zhang2022multi} & \begin{itemize} \item Proposed a metaverse-enabled healthcare framework that diagnoses a patient. \item For ensuring the security and privacy of data owners (i.e., patients), an encryption framework is proposed. \end{itemize}& Healthcare & Wireless for metaverse & Framework & N.A\\ \midrule
   He~\textit{et al.}~\cite{he2023three} & \begin{itemize} \item Proposed a three-dimensional holographic communication system for metaverse \item To capture images, light field and structured light cameras are used for objects for capturing dynamic 3D models and objects that change slow, respectively. \item The proposed framework can be easily implemented using the existing networks and devices. \end{itemize}& 3D holographic communication & Wireless for metaverse & Framework and experimental model & N.A\\ \midrule
    Plechata~\textit{et al.}~\cite{plechata2022can}& \begin{itemize} \item Proposed a theoretical framework for using metaverse in healthcare \item The proposed framework can be extended with modifications to many diseases     \end{itemize} & Healthcare & Metaverse for wireless & Framework & N.A \\ \midrule
    Wang~\textit{et al.}~\cite{wang2022development}& \begin{itemize} \item Proposed a MeTAI ecosystem for healthcare applications \item MeTAI has four applications: (a) virtual cooperative scanning, (b) raw data sharing, (c) augmented regulatory science, and (d) metaverse medical intervention.  \item MeTAI can be applied for many diseases with specific modifications.   \end{itemize}&  Healthcare & Metaverse for wireless & Framework & N.A \\ \midrule
    Lim~\textit{et al.}~ \cite{lim2022realizing}& \begin{itemize} \item The authors proposed an edge intelligence-based architecture for realizing metaverse. \item The authors also identified the key enablers. \item They also presented a case study showing the role of edge intelligence towards enabling metaverse \end{itemize} & Edge intelligence for metaverse & Wireless for metaverse & Framework & N.A \\ \midrule
    Zhou~\textit{et al.}~\cite{zhou2022vetaverse} & \begin{itemize} \item Presented a vetaverse architecture \item Identified artificial intelligence, speech understanding, humans motion detection, physiological parameters monitoring, and emotion recognition, as key enablers of vetaverse.     \end{itemize} & Intelligent transportation system & Wireless for metaverse & Framework & N.A \\ \midrule
    Alpala~\textit{et al.}~\cite{alpala2022smart} & \begin{itemize} \item  Presented an experimental framework for enabling communication between metaverse environments \item Presented a case study of smart factory \item Presented experimental results to show the validity of their proposal  \end{itemize} & Smart factory  &  Metaverse for wireless & Framework & N.A  \\  \midrule
Allam~\textit{et al.}~\cite{allam2022metaverse} & \begin{itemize} \item Presented various key enablers of metaverse architecture for enabling smart cities \item Identified use case of metaverse for metaverse in smart cities  \end{itemize} & Smart cities & Metaverse for wireless & N.A & This work discussed the key components of metaverse architecture without proposing a novel framework. \\ \midrule
 Du \textit{et al.}~\cite{du2022exploring} & \begin{itemize} \item Proposed an attention-aware network resource allocation scheme for a metaverse.  \item Their proposal allocates resources (i.e., edge devices rendering capacity) based on the predicted user object-attention values and shown promising results. \item Provided future research directions  \end{itemize} & Customized meta services & Wireless for metaverse & Architecture and experimental model & N.A \\ \bottomrule 
\end{tabular}
\end{center}

\end{table*}

\begin{figure}[!t]
	\centering
	\includegraphics[width=8cm, height=9cm]{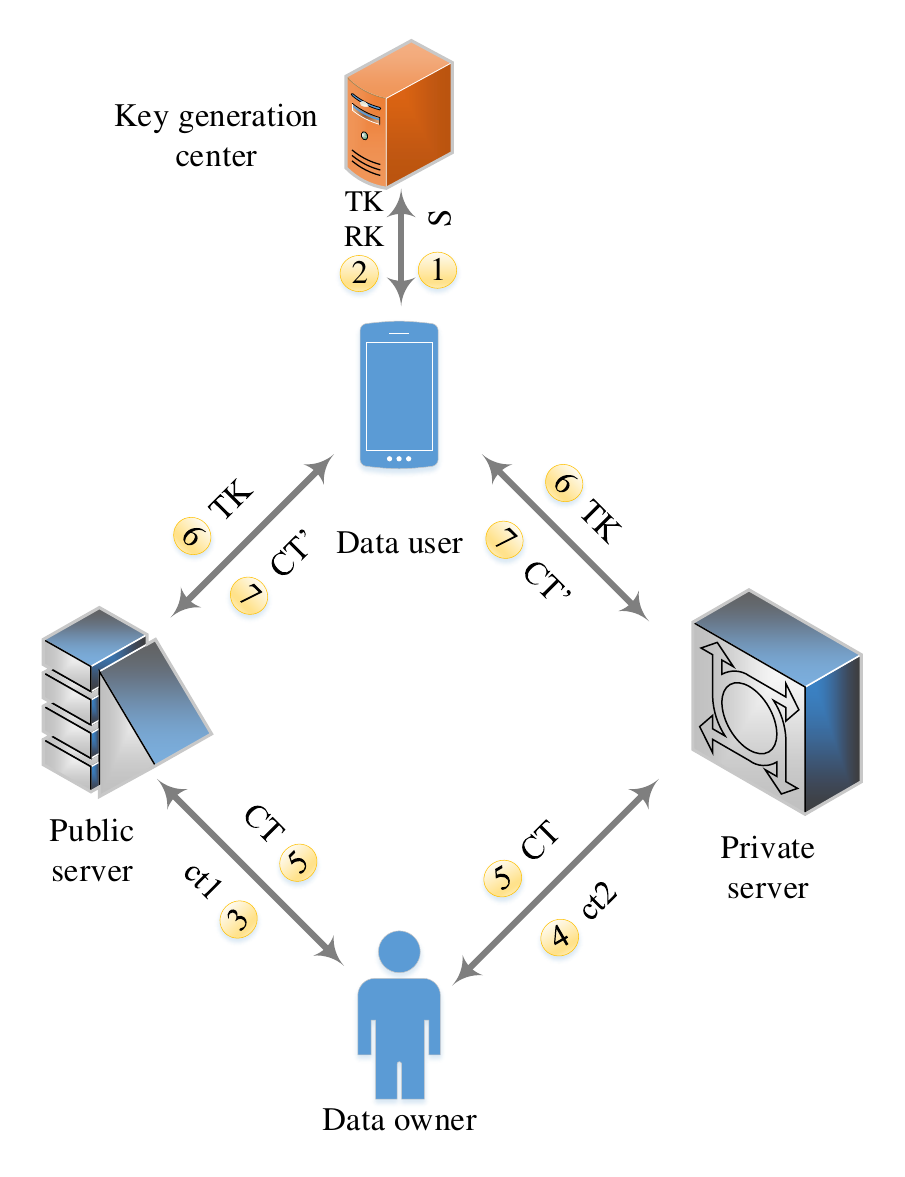}
	\caption{Encryption of data for metaverse \cite{zhang2022multi}.}
	\label{fig:ecnryption}
\end{figure}

\begin{figure}[!t]
	\centering
	\includegraphics[width=8cm, height=7cm]{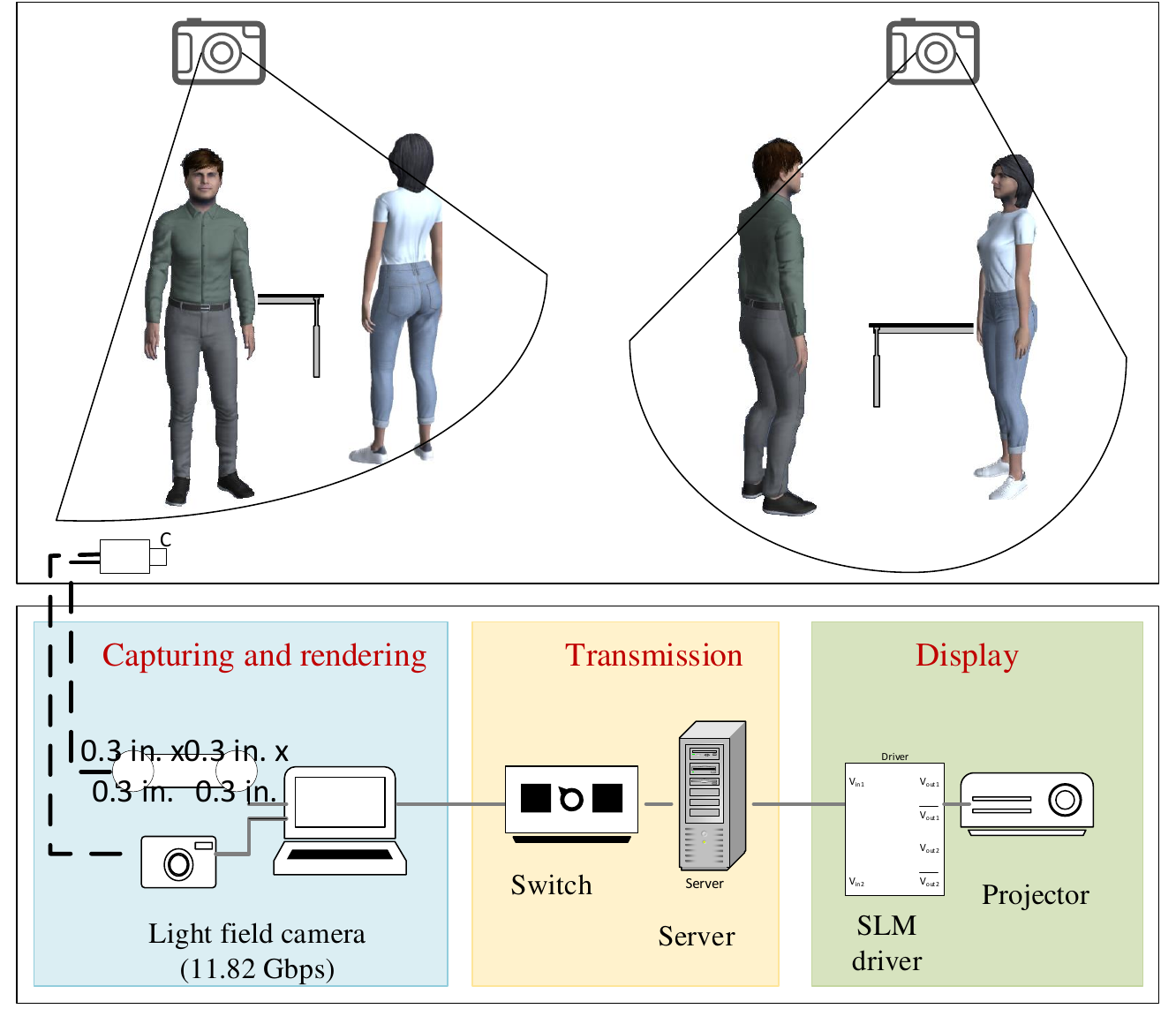}
	\caption{3D holographic communication framework \cite{he2023three}.}
	\label{fig:holo}
\end{figure}

\section{State-of-the-Art and Standardization}
\subsection{Advances}
In this section, we discuss various recent advances (i.e., summarized in Table~\ref{tab:advances1}) \cite{zhang2022multi,he2023three,plechata2022can,wang2022development,lim2022realizing,lim2022realizing,zhou2022vetaverse,alpala2022smart,allam2022metaverse,du2022exploring} towards enabling wireless system by a metaverse. As the metaverse is still in its infancy, only a few works presented architectures/frameworks for enabling emerging applications using the metaverse. In \cite{zhang2022multi}, the authors proposed a metaverse-enabled healthcare framework that diagnoses a patient. Meanwhile, there is healthcare data that is used by a metaverse and stored on edge servers. To ensure the privacy of such metaverse data, they propose the use of attribute-based encryption. The system consists of a data user, private server, public server, data owner, and key generation
center. The data user submits the attribution set for registration to the key generation center that issues reclaiming key and transformation key for the data owner. The intermediate cipher texts are given to the owner of data, as shown in Fig.~\ref{fig:ecnryption}. Then, the cipher texts are fed to the private and public servers. Finally, the transformation keys are shared with servers when the data is required to be downloaded by a user. The proposal of \cite{zhang2022multi} can be used for ensuring the privacy and security of data in a metaverse architecture presented in Section~\ref{High-Level Architecture} (i.e., Fig.~\ref{fig:architecture}). He~\textit{et al.} in \cite{he2023three} proposed a three-dimensional holographic communication system for the metaverse. Their system has four components, such as display, transmission, hologram generation, and capture, as shown in Fig.~\ref{fig:holo}. To support 3D communication, one must use 3D display and imaging technologies, such as light field (LF) display, volume display, and binocular vision display \cite{xu2019time,hiura2017measurement,north2016compact,su2020binocular,huang2017systematic}. To capture images, light field and structured light cameras are used for an object for capturing dynamic 3D models and objects that change slow, respectively. Next to capturing 3D images, computer-generated holograms (CGHs) are used to denote 3D intensity patterns in computer holography under coherent illumination. The phase-only CGHs are computed by the capturing and rendering part using the layer-based angular-spectrum method (ASM). The layer-based ASM used shading images and depth images. Next, the CGHs are transmitted over a wireless channel using some communication technology (e.g., $5$G). At the receiver side, a 3D video is generated using a holographic optical display system. Although the proposed 3D communication system offers many benefits, there are many challenges that need to be addressed. The first one is communication resource management. For a massive number of applications based on 3D holographic communication, we must propose efficient resource management schemes that increase the throughput of the overall system. Other than this issue, mobility management is necessary for such systems. For instance, a user might move outside the coverage area of one capturing device during the mid of capturing phase, and thus the capturing device will not get complete information. To resolve this, one can predict the mobility of the devices and based on the predicted mobility, one can better associate the user with a better image-capturing device. On the other hand, there must be novel encryption schemes for 3D holographic communication systems. A malicious user might access the wireless signal and thus, causes privacy leakage or alter important information. Therefore, there is a need for efficient and effective encryption schemes for 3D holographic communication systems. \\
\begin{figure}[!t]
	\centering
	\includegraphics[width=8cm, height=7cm]{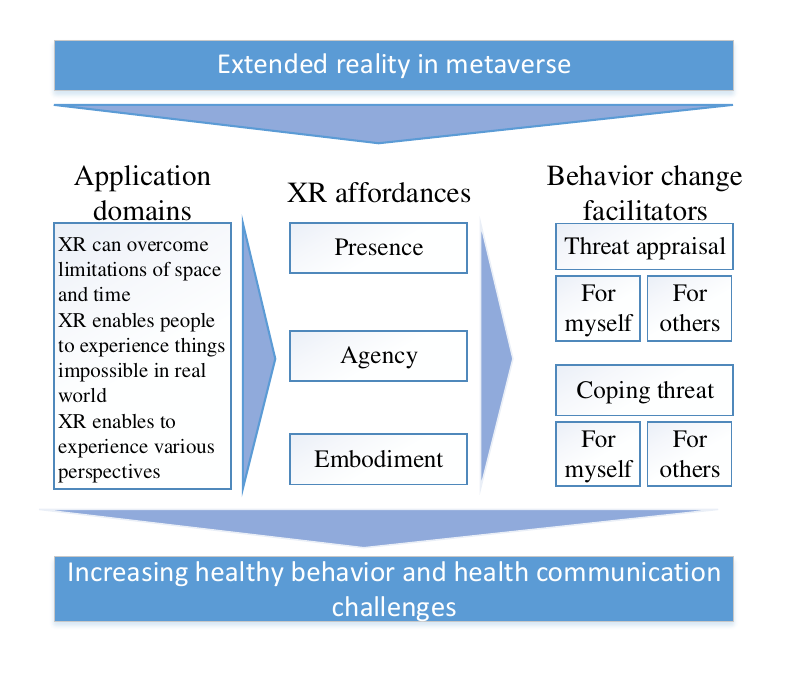}
	\caption{A theoretical framework for metaverse using extended reality \cite{plechata2022can}.}
	\label{fig:XR_app}
\end{figure}

   \begin{figure*}[!t]
	\centering
	\includegraphics[width=14cm, height=9cm]{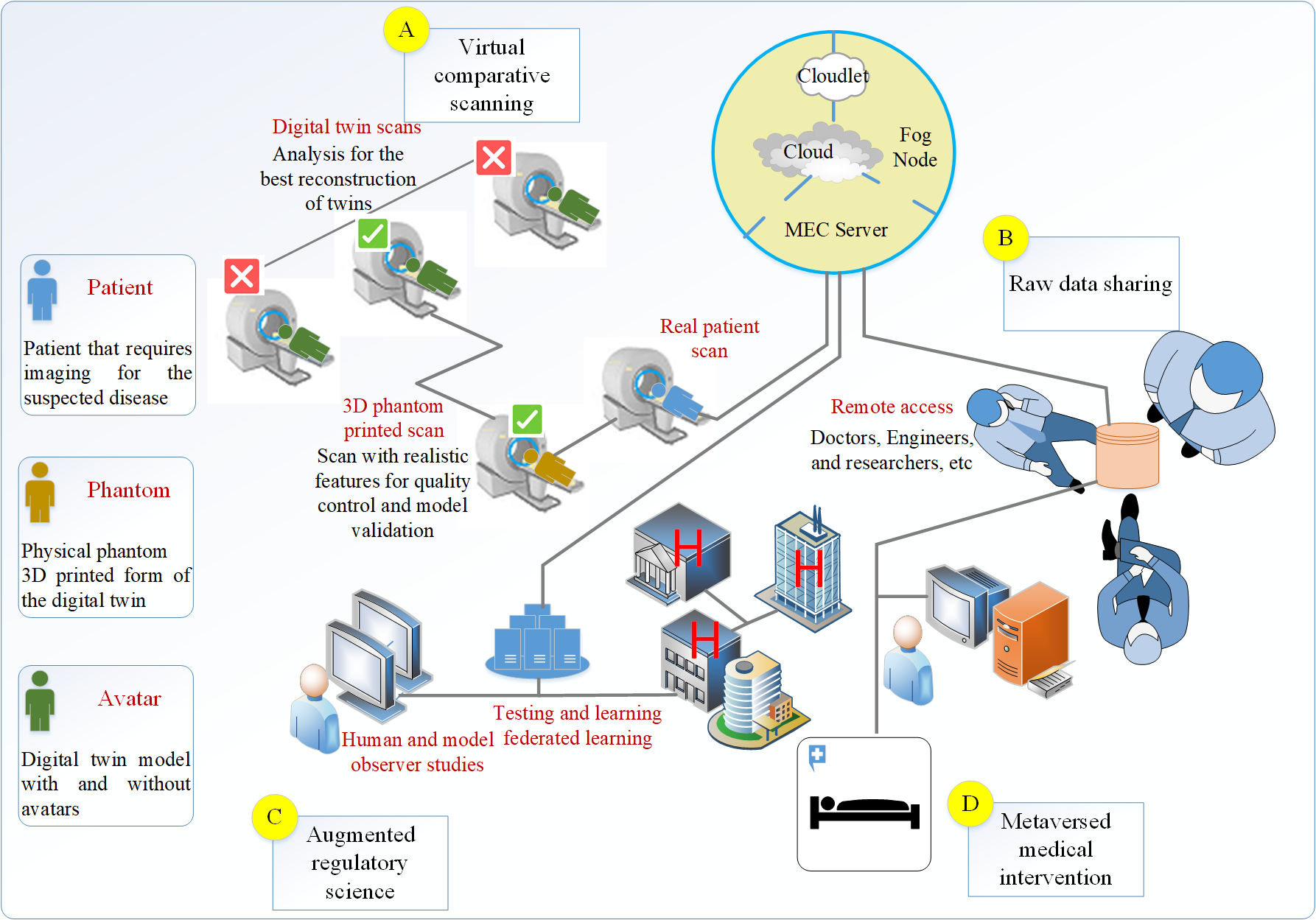}
	\caption{MeTAI ecosystem using AI and metaverse \cite{wang2022development}.}
	\label{fig:metai}
\end{figure*}

Plechata~\textit{et al.} in \cite{plechata2022can} highlighted the role of extended reality in enabling metaverse for healthcare applications. A metaverse-enabled architecture for disease prevention and health promotion was considered that is based mainly on two phases: threat appraisal and coping appraisal. Threat appraisal refers to vulnerability and threat severity (i.e., level of damage to health), whereas coping appraisal refers to self-efficacy and response efficacy. In response to efficacy, an individual belief's whether the measures of coping will minimize the health threat or not. On the other hand, self-efficacy refers to individual confidence in the ability for performing behavior recommended by the architecture. Similar to many existing applications, extended reality can play a crucial role in healthcare communications. To do so, one can use metaverse using extended reality to support patient support groups as well as expert-moderated health communities. Specifically, the metaverse using extended reality will provide presence, agency, and embodiment. Based on these extended reality affordances, the architecture can better enable the behavior change facilitators, as shown in Fig.~\ref{fig:XR_app}. The framework proposed by the authors can help in improving healthcare services using metaverse, but it needs further efforts. To implement the theoretical framework, in reality, there is a need t resolve many challenges. These challenges are sensing, adding healthcare annotations using extended reality, and communication of sensory data (e.g., human body temperature and 3D images of body parts). Therefore, there is a need for modifications in the framework of \cite{plechata2022can} to enable healthcare services. Wang~\textit{et al.} in \cite{wang2022development} proposed the use of metaverse for healthcare. They presented an architecture, namely, MeTAI ecosystem, for enabling intelligent healthcare based on the metaverse. The MeTAI ecosystem shown in Fig.~\ref{fig:metai}, has four applications: (a) virtual cooperative scanning, (b) raw data sharing, (c) augmented regulatory science and (d) metaverse medical intervention. The purpose of virtual cooperative scanning is to find suitable scanning technology for healthcare diseases. The digital twin scanners are installed to take scans of digital avatars. The architecture also provides ubiquitous and secure medical data access to various patients for using it by healthcare personnel and experts. Although the framework presented in \cite{wang2022development} can be applied to many healthcare applications, it needs much effort to apply in reality. For instance, to immerse interactive experience technologies with medical imaging, there is a need to design various schemes depending on the nature of the disease. Additionally, there is a need for effective three-dimensional (3D) computed tomography (CT) of human models for use in the analysis. On the other hand, there are many challenges that need to be resolved prior to using MeTAI system. These challenges are privacy, security, management, and disparity reduction. As MeTAI system can be deployed commercially on a large scale, therefore, there must be some set of laws to ensure the privacy of users (e.g., the Health Insurance Portability and Accountability Act (HIPAA) in the United States). In addition to laws, one must use modern security-related technologies, such as blockchain and privacy-aware distributed learning. Other than security and privacy, there must be an efficient mechanism for the management of such a complex system. \\
\begin{figure*}[!t]
	\centering
	\includegraphics[width=16cm, height=11cm]{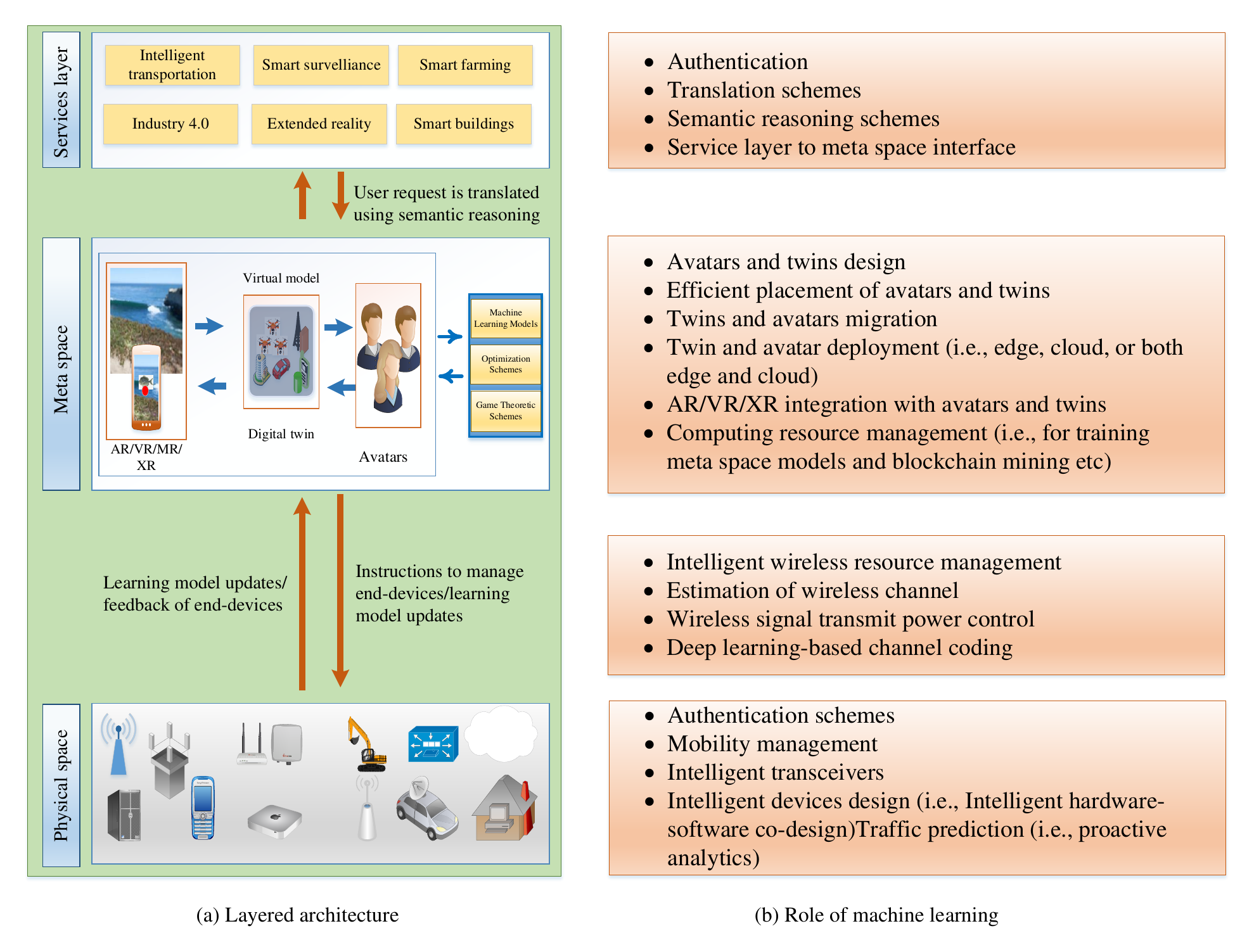}
	\caption{Role of ML for the metaverse.}
	\label{fig:ML}
\end{figure*}

\begin{figure*}[!t]
	\centering
	\includegraphics[width=14cm, height=10cm]{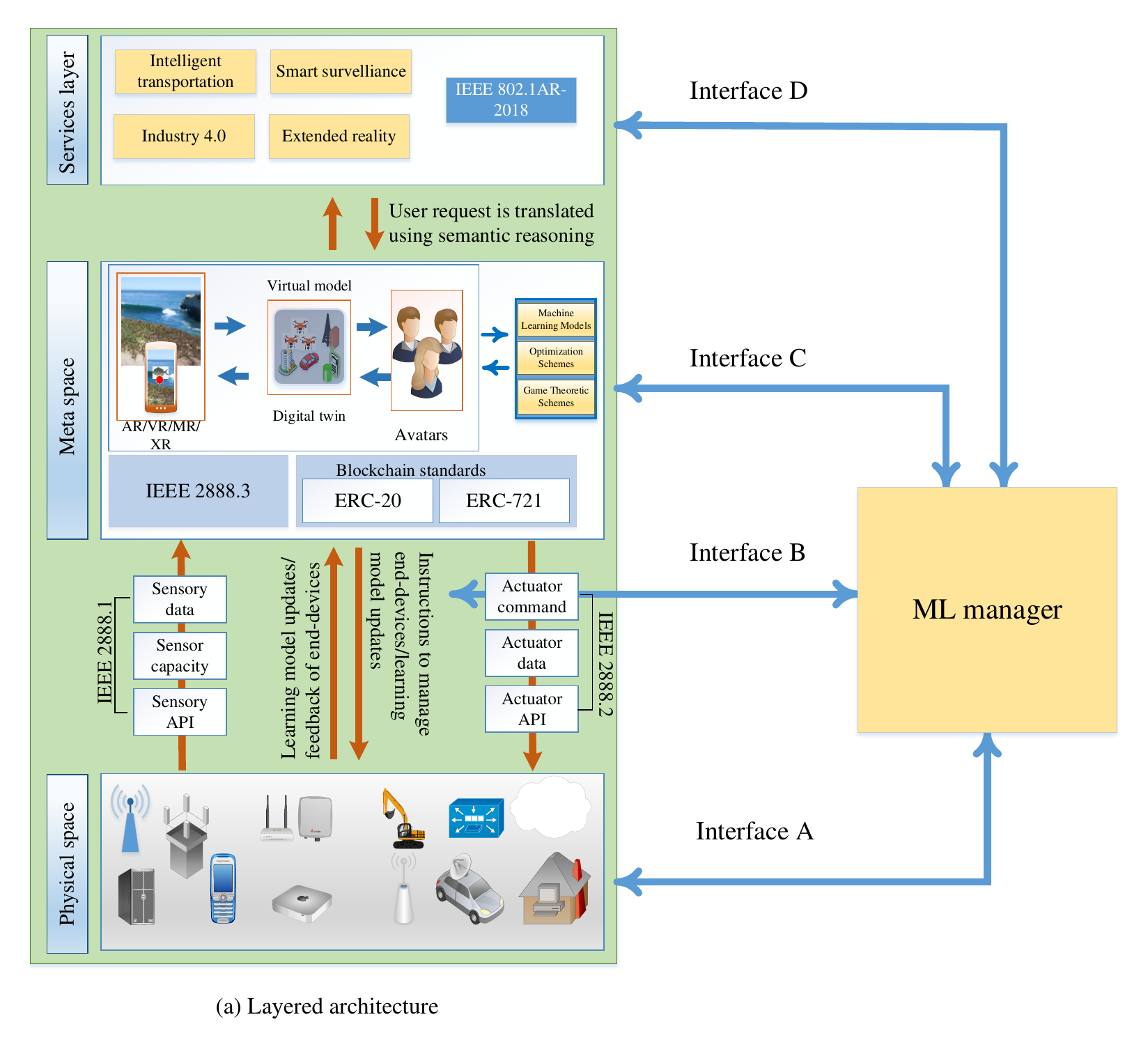}
	\caption{Standardization of ML-enabled metaverse.}
	\label{fig:stnadardization}
\end{figure*}

Lim \textit{et al.} in \cite{lim2022realizing} proposed an edge intelligence-based architecture for realizing the metaverse. They focused on infrastructure, the metaverse engine, the virtual world, and the physical world. They identified the key requirements for enabling metaverse. Additionally, they discussed various interfaces for communication among various players of the metaverse architecture. They also presented a case study of the edge-based metaverse and finally, they presented open research challenges. The authors in \cite{lim2022realizing} considered the aspect wireless for the metaverse. On the other hand, one can use metaverse to fulfill the diverse requirements of various applications (e.g., intelligent transportation systems). For doing so, one can deploy a metaverse that uses digital twins, avatars, and other schemes for efficient and effective management of various resources. In another work \cite{zhou2022vetaverse}, the authors presented vision, applications, and technologies of enabling vehicular networks by metaverse and they named it \emph{vetaverse}. Their identified key technologies are artificial intelligence, speech understanding, human motion detection, physiological parameters monitoring, and emotion recognition. They also gave an architecture for vetaverse. Finally, they presented open challenges with suggestions. In \cite{alpala2022smart}, Alpala \textit{et al.} presented an experimental framework for enabling collaboration between virtual environments using a virtual reality-based metaverse. Their system consists of an online multi-user system, interfaces, object-oriented configurations, and other functional components. As a case study, they presented a metaverse-based digital factory and presented experimental results. Although the framework of \cite{alpala2022smart} can be used for smart factories, the authors did not consider a few important aspects, such as security, privacy, and resource management. Another work \cite{allam2022metaverse} proposed the use of metaverse to enable smart cities. Specifically, the authors presented a high-level architecture element of a metaverse to enable smart applications. Additionally, it helps in providing guidelines for using emerging technologies in the metaverse. They also presented the key projects for real-time implementation of the metaverse. Although the authors discussed various key enablers and use cases of the metaverse, they did not provide a more concrete implementation of the use cases of the metaverse. In \cite{du2022exploring}, Du \textit{et al.} proposed an attention-aware network resource allocation scheme for a metaverse. The key idea is to allocate more resources to the virtual objects that are more important to users. Specifically, they discussed the key requirements (i.e., remote rendering, eye-tracking,  and QoE analysis) of enabling of the metaverse. Then, using the existing user-object-attention level, an attention-aware network resource allocation algorithm that has two steps (i.e., QoE maximization and attention prediction is proposed. Their proposal allocates resources (i.e., edge devices rendering capacity) based on the predicted user object-attention values and shows promising results. Finally, the authors provided future directions. \par



\subsection{Standardization}
In this section, we will discuss the standardization of a metaverse-based wireless system. Prior to discussing the standardization of the metaverse, there is a need to highlight the role of many emerging technologies in enabling the metaverse. First of all, we will highlight the role of machine learning in enabling metaverse for wireless systems. At the physical layer, one can use machine learning to enable efficient authentication schemes to avoid unauthorized users to access distributed devices. As the devices are distributed in physical space, therefore, enabling security to avoid unauthorized access can be performed using effective authentication schemes. Such authentication schemes can be based on machine learning \cite{kung2005biometric,punithavathi2019lightweight,xun2019automobile,hazratifard2022using}. Other than authentication, one can use machine learning for the mobility of the management of devices. Such mobility management schemes can use prediction based on various machine learning schemes (e.g., convolutional neural networks). Based on the predicted outcomes, one can better manage the mobility of devices \cite{simsek2015context,zhang2021mobility,chowdhury2020mobility,liu2019green}. Furthermore, one can use machine learning for enabling intelligent transceivers. These transceivers will use machine learning for intelligent resource allocation, intelligent channel estimation, and intelligent transmit power control, among others \cite{zheng2022survey,zheng2019intelligent,wang2020channel,nadeem2019intelligent}. Also, one can use machine learning to design efficient devices using hardware-software co-design \cite{khan2020federated,khan2022digital}. Generally, training of local models for training a distributed learning meta space model consumes a significant amount of computation resources which in turn will consume significant power/energy. Therefore, one must use neural architecture search (NAS) that tries various architectures of machine learning models in order to select optimal architecture for a particular dataset and task \cite{zoph2016neural,elsken2019neural,jaafra2019reinforcement}. NAS is a sub-field of automated machine learning that enables one to find a suitable design for a given design. Although NAS enables efficient software design, there is a need for software-hardware co-design that consider both hardware and software during the design of end-devices. Such designs can be based on machine learning \cite{khan2020federated}. Therefore, there is a need to propose standardization schemes for machine learning-enabled metaverse, as shown in Fig.~\ref{fig:ML}. \par

In 2019, IEEE 2888 project was launched to standardize interfaces between the cyber and physical worlds, as shown in Fig.~\ref{fig:stnadardization}. One can use these interfaces along with other interfaces in the metaverse. IEEE 2888.1 and IEEE 2888.2 interfaces can be used for moving sensory information from physical space to meta space and actuator controls from meta space to physical space, respectively. On the other hand, IEEE 2888.3 standard can be used for the definition of digital things \cite{wang2022survey}. Additionally, for efficient communication between meta space entities (e.g., avatars and twins), there is a need for novel interfaces based on novel standards. Due to the important role of machine learning in enabling metaverse systems, there is a need for a standardized ML manager that can control various interfaces, such as interface A, interface B, interface C, and interface D. Interface A will deal with the efficient deployment and resources (i.e., computing and communication) management of the physical space using machine learning. Interface B will control communication between the meta space and physical space by modifying/assisting the existing IEEE 2888.1, IEEE 2888.2, and IEEE 2888.3 standardized interfaces using machine learning schemes. Interface C will use machine learning to deploy meta space using the physical space infrastructure (e.g., edge/cloud servers and unmanned aerial vehicles). Such a deployment will include virtual machines/containers-based design. Additionally, the deployment of these containers/virtual machines on single/multiple hardware devices. Such kind of operations/functions will be performed by interface C. To handle the meta space data, one can use blockchain. ERC-20 helps in the implementation of standard APIs tokens. Additionally, ERC-20 supports basic functionality for transferring tokens \cite{casale2021networks}. On the other hand, ERC-721 helps in the implementation of a standard programming interface for non-fungible tokens within a smart contract \cite {christodoulou2020decentralized}. Other than ERC-20 and ERC-721, there is a need for other standards using machine learning to enable consensus among blockchain nodes with less latency and energy consumption. Finally, interface D will perform secure authentication using existing/modified schemes. IEEE 802.1AR-2018 (IEEE 802.1AR-2009 suspended) can provide unique per-device identifiers (DevID) as well as cryptographic binding of identifiers with devices \cite{8423794}.

\subsection{Summary: Lessons Learned and Insights}
In this section, we discussed recent advances in the metaverse. Moreover, we identified their design aspect along with their primary focus. Several lessons learned from this section are as follows:
\begin{itemize}
    \item There is a need for wireless channel models for holographic communications. One can use 3D holographic communication for the transmission of 3D human images. To efficiently perform this communication, there is a need for wireless channel models similar to existing channel models (e.g., Stanford University interim (SUI) Channel models, such as SUI-1, SUI-2, SUI-3, SUI-4, SUI-5, and SUI-6) \cite{khan2013comparison,khan2014novel}. Such a specialized channel model will be used for the analysis of holographic communication systems \cite{wei2022multi}. Additionally, to cope with fading effects of a wireless channel, one should design an efficient and effective channel estimator. To do so, the channel model can help to analyze the performance of the various channel estimators prior to actual implementation for a metaverse-based wireless system. 
    \item To enable various emerging applications using metaverse, there is a need to resolve the issue of interoperability due to the presence of many players (e.g., edge/cloud servers, devices, blockchain miners, and unmanned aerial vehicles). Therefore, enabling a seamless interaction among these players is challenging due to their different underlying technologies. To do so, one can propose a general interface that will allow us to efficiently and seamlessly communicate. 
    \item Most of the existing works presented theoretical frameworks for metaverse-based wireless system applications. These theoretical frameworks (e.g., autonomous driving cars) can be extended to many specific applications (e.g., lane change assistance) by medications. Although theoretical frameworks offer many benefits, there is a need for mathematical models related to specific applications.  
    \item Due to the important role of machine learning in enabling a metaverse-based wireless system, there is a need for effective standardization of machine learning for a metaverse-enabled wireless system. To do so, one can propose a machine learning manager that can control various, diverse players of the metaverse using different interfaces. Meanwhile, the machine learning-based metaverse system can use existing standards in addition to novel standards for effectively enabling a metaverse-enabled wireless system.  
\end{itemize}

\begin{table}[]
\fontsize{7}{6}\selectfont
\caption {Summary of challenges listed in existing surveys and tutorials} \label{tab:challenges} 
\begin{center}
\begin{tabular}{p{2cm}p{5cm}}
\toprule
    \textbf{Reference}   & \textbf{Challenges} \\ \midrule
    Ning~\textit{et al.}~\cite{ning2021survey} & \begin{itemize} \item Interaction problem \item Computation issues \item Ethical issues \item Privacy issues \item Compatibility  \end{itemize} \\ \midrule
    Wang~\textit{et al.}~\cite{wang2022survey} & \begin{itemize} \item  Endogenous security empowered metaverse \item Cloud-edge-end orchestrated secure metaverse \item Cross-chain interoperable and regulatory metaverse  \item Energy-efficient and green metaverse \item  Content-centric and human-centric metaverse \end{itemize} \\ \midrule
    Gadekallu~\textit{et al.}~\cite{gadekallu2022blockchain} & The authors summarized and discussed about blockchain to enable metaverse. \\ \midrule
    Khan~\textit{et al.}~\cite{khan2022metaverse} & \begin{itemize} \item Resource optimization \item Blockchain-based data management \item Incentive mechanism \item Prototyping \end{itemize}\\ \midrule
    Khan~\textit{et al.}~\cite{khan2022machine} & \begin{itemize} \item  Training fashion \item Standardization of ML-based metaverse \item Blockchain for secure ML-enabled metaverse-based wireless systems \end{itemize}\\ \midrule
    Xu~\textit{et al.}~\cite{xu2022full} &  \begin{itemize} \item Advanced multiple access for immersive streaming \item Multi-sensory multimedia networks  \item Multimodal semantic/goal-aware Communication \item Integrated sensing and communication \item Digital edge twin networks \item Edge intelligence and intelligent edge \item Sustainable resource allocation \item Avatars (Digital Humans) \item The industrial/vehicular metaverse \item Quality of experience \item Market and mechanism design for metaverse services \end{itemize}\\ \midrule
    Our Tutorial &  Interoperable meta spaces, Non-fungible tokens for metaverse trading, Personalized distributed learning-based avatars modeling, Isolation of meta spaces, Mobility management, Intelligent interfaces, Zero-touch networking for metaverse, Machine Learning-enabled semantic communication for metaverse
    \\ 
\bottomrule
\end{tabular}
\end{center}
\end{table}

\begin{table*}[htp!]
\fontsize{8}{7}\selectfont
\caption {Summary of the research challenges and their guidelines.} \label{tab:challenges} 
 \begin{center}
  \begin{tabular}{p{3cm}p{3.5cm}p{4.5cm}p{4.5cm}}
    \toprule
   
     \textbf{Challenges} & \textbf{Design aspect} &  \textbf{Causes} & \textbf{Guidelines} \\
     \midrule
     \textbf{Interoperable meta spaces} & Wireless for metaverse &\begin{itemize} \item  Wide variety of players in a metaverse-enabled wireless system \item Different computing hardware (e.g., edge servers)   \end{itemize} & \begin{itemize} \item  Virtual machine and containers-based meta space implementation \item General wireless interfaces for devices \end{itemize}\\
     \midrule

     \textbf{Non-fungible tokens for metaverse trading}  & Wireless for metaverse (This challenge focused mainly on business model)  & \begin{itemize} \item How to trade multiple players in a metaverse-enabled wireless system \item How to represent metaverse assets \end{itemize} & \begin{itemize} \item Non-fungible tokens for the trading of metaverse assets \item Novel unique numbering for non-fungible tokens  \end{itemize} \\
    \midrule
    
     \textbf{Personalized distributed learning-based avatars modeling} &  Wireless for metaverse & \begin{itemize} \item Personalized local datasets associated with mobile devices/users in a physical space \item Privacy leakage in centralized learning \end{itemize} & \begin{itemize} \item Noise-less local datasets \item Clustering-based personalized distributed meta space models \end{itemize} \\
    \midrule
    \textbf{Isolation of meta spaces} & Wireless for metaverse & \begin{itemize} \item Wireless and computing resources constraints \item Efficient use of physical space resources \end{itemize} & \begin{itemize} \item Matching-theory enabled access network isolation \item Optimization theory-based isolation schemes  \end{itemize} \\
    \midrule
    \textbf{Mobility Management}  &  Metaverse for wireless and wireless for metaverse & \begin{itemize} \item Effect of users mobility on meta space modeling \item Devices/users mobility in physical space during service provided by meta space \end{itemize} & \begin{itemize} \item Meta space migration \item Deep learning-enabled prediction for mobility management \end{itemize} \\
    \midrule
    \textbf{Intelligent Interfaces} & Wireless for metaverse and metaverse for wireless & \begin{itemize} \item Communication resources constraints \item Wireless channel uncertainties \end{itemize} & \begin{itemize} \item Centralized learning-enabled intelligent interfaces \item Distributed learning-enabled interfaces \end{itemize} \\
  \midrule
        \textbf{Zero-touch networking for metaverse} & Wireless for metaverse and metaverse for wireless & \begin{itemize} \item Resource constraints \item Robustness issues \item High network complexity in serving massive number of users \end{itemize} & \begin{itemize} \item Machine learning-enabled resource optimization \item Network slicing \item Machine learning-enabled fault tolerance and security schemes \end{itemize} \\
       \midrule  
          \textbf{Machine Learning-Enabled semantic communication for metaverse} & Wireless for metaverse & \begin{itemize} \item Context-awareness requirement \item High communication resources requirements due to a massive number of devices \item Strict latency requirements of various (e.g., healthcare) applications based on metaverse \end{itemize} & \begin{itemize} \item Auto-encoder based semantic encoders and decoders  \item Deep learning-enabled semantic communication \item Distributed learning-enabled privacy-aware semantic communication systems \end{itemize} \\ \midrule
          \textbf{Hybrid Modeling of Twins and Avatars} & Metaverse for wireless & \begin{itemize} \item Diverse players in meta space \item Limitations of mathematical, experimental, and machine learning models  \end{itemize} & \begin{itemize} \item Mathematical optimization-enabled modeling  \item Joint mathematical, machine learning, and experimental modeling.  \end{itemize} \\ 
\bottomrule  
\end{tabular}
\end{center}
\end{table*}

\section{Open Challenges}
In this section, we present open research challenges. Existing tutorials and surveys on metaverse considered interaction problem, computation issues, ethical issues, privacy issues, compatibility, endogenous security, empowered metaverse, cloud-edge-end orchestrated secure metaverse, cross-chain interoperable and regulatory metaverse, energy-efficient and green metaverse, content-centric and human-centric metaverse, resource optimization, blockchain-based data management, incentive mechanism, prototyping, training fashion, standardization of ML-based metaverse, blockchain for secure ML-enabled metaverse-based wireless systems, advanced multiple access for immersive streaming, multi-sensory multimedia
networks, multimodal semantic/goal-aware communication, integrated sensing and communication, digital edge twin networks, edge intelligence and intelligent edge, sustainable resource allocation, avatars (Digital Humans), the industrial/vehicular metaverse, quality of experience, market and mechanism design for metaverse services, as challenges as given in Table~\ref{tab:challenges}. In contrast, we consider interoperable meta spaces, non-fungible tokens for metaverse trading, personalized distributed learning-based avatars modeling, isolation of meta spaces, mobility management, and intelligent interfaces.

\subsection{Interoperable Meta Spaces}
{\em How do we enable seamless interaction between the avatars and twins modeled for different meta spaces?} In a metaverse, the concept of interoperability is different compared to existing wireless systems. In a traditional wireless system, the goal of interoperability is to enable seamless interaction between a wide variety of players (e.g., devices and edge/cloud servers). In contrast, here, the metaverse has two main aspects: (a) wireless devices and (b) meta spaces. To enable interoperability between various wireless devices, there is a need for the design of general interfaces that can enable seamless communication. However, different devices have different structures. Therefore, we must define novel interfaces for a metaverse-based wireless system. On the other hand, meta space mainly constituted by digital avatars and twins must be interoperable (i.e., one virtual machine-based meta space (as explained already in Section~\ref{Meta Space}) must work on the new edge/cloud servers as well due to meta space migration). For instance, meta space based on a virtual machine might not work on containers deployed at the network edge. Additionally, within a single design (i.e., container-based or virtual machine-based), the meta space might not be compatible. Therefore, for efficient deployment of wireless systems, one must propose interoperable meta spaces. For virtual machines-based meta space, one can have three levels of interoperability \cite{lenk2014tiosa}. The first one (i.e., level 1) involves running of virtual machine-based meta space on a virtual hardware selection/ CPU architecture, and or particular virtualization product. Level-1 migration is equivalent to suspending at the source and resuming at the destination. Additionally, one can live to migrate meta space based on level-1, it faces some limitations. The prominent one is the preservation of IP addresses. In other level-2, virtual machine-based meta space will run on a specific family of hardware and works by shutting down in the current edge and rebooting at the destination edge. On the other hand, level-3 has more freedom of running meta space on multiple hardware and thus gives more flexible operation with better interoperability. \par

\subsection{Non-Fungible Tokens for Metaverse Trading}
{\em How does one use non-fungible tokens for the trading of metaverse entities (e.g., digital avatars and twins) among various players?} Enabling ownership of digital items (e.g., in-game items, collectibles, videos, art, and music) in a metaverse is challenging and needs careful design. In a metaverse, to uniquely represent the digital assets, non-fungible tokens are used that are a unit of data stored on a blockchain. Alternately, non-fungible tokens serve as a certificate of authenticity in a metaverse-enabled wireless system. One can also say that non-fungible tokens form a link between physical world items and metaverse virtual items. A unique value is associated with a non-fungible token in the metaverse that is used for permanently storing them in a blockchain network. One of the recent events of non-fungible tokens was the selling of digital work created by Beeple \cite{xu2022full}. Although non-fungible tokens can be effectively used for representing digital assets in the metaverse, their many challenges that must be resolved. The first one is how to use non-fungible tokens for the representation of digital assets in a wireless system. For instance, in a metaverse, how do we use a non-fungible token to define an entity? The entity can be an end-device, a system made of many devices, or a complete application (e.g., autonomous cars) made of many systems. There should be a proper and worldwide acceptable framework for assigning non-fungible tokens to wireless systems. Also, the unique numbering of non-fungible tokens must be done in an efficient way to effectively cover all massive numbers of entities in a metaverse.

\subsection{Personalized Distributed Learning-based Avatars Modeling}
{\em How do we enable efficient modeling of avatars using personalized, privacy-aware distributed learning schemes?} To model avatars, one can use distributed learning. However, getting a generalized global model using distributed for modeling avatars might not effectively model them. Therefore, there is a need for modified distributed learning modeling. One can use personalized distributed learning to model avatars. For instance, consider a metaverse-based vehicular network, there is a wide variety of vehicles, such as cars, trucks, and bikes, among others. If we want to model mobility and driving assistance using distributed learning in a metaverse-based intelligent transportation system, there is a need for more personalized models. Such modeling will be based on the training of a general global model and then further training of local data to make it more personalized. Although such an approach will enable efficient modeling, it may face challenges. The local data might not be sufficient to well train the personalized model. Additionally, the local data may have noise. Therefore, we must effectively take into account all the factors while using personalized distributed learning for the modeling of avatars in a metaverse. On the other hand, there might be very less local data associated with some of the devices. To address this challenge, one can use a clustering approach that will be based on the clustering of devices with similar data distribution. In each cluster, after getting a global model, a local model will be trained that will be used by the associated devices.

\subsection{Isolation of Meta Spaces}
{\em How does one enable isolated operation of meta spaces using shared hardware without affecting the performance of each other?} To deploy meta spaces (i.e., twins and avatars along with computing storage), there are two main ways: dedicated hardware and shared hardware. Dedicated hardware will result in good performance, but it comes with a high cost that is not practically feasible. There is a need for shared hardware usage for various meta spaces associated with various applications/functions. To do so, there is a need for isolation at various levels (e.g., access network and core network). For an access network, one can use the concept of virtualization which will consist of buying network resources from the operators and selling them to metaverse users. For such a design, one can define a utility that will jointly maximize the profit of network operators and metaverse users. For such maximization, one can use mathematical optimization, matching theory, and game theory, among others. On the other hand, computing resources must be efficiently managed in such a way as to run multiple meta spaces on computing hardware (i.e., edge/cloud server) without affecting the performance of other metaverse users.      

\subsection{Mobility Management}
{\em How does one efficiently model and manage the mobility of avatars and mobile users, respectively?} Regarding mobility in a metaverse-enabled wireless system, there are two aspects, such as mobility modeling for avatars in meta space and mobility of users during run time. To analyze the wireless system prior to deployment for emerging applications, there is a need to effectively model the mobility of devices. For instance, wireless systems require accurate modeling of devices/users' mobility. Such mobility modeling will enable efficient and accurate analysis of wireless systems (e.g., channel estimation design and deployment of equipment/devices for existing/new applications). On the other hand, the mobility of devices served by meta space must also be handled effectively. To do so, there is a need to propose novel migration of meta space. Such migration schemes can be either live or non-live. Live migration schemes are preferable for real-time applications, whereas non-live migration schemes can be used for non-real-time applications. For both migration schemes, one can use deep learning-based mobility prediction.

\subsection{Intelligent Interfaces}
{\em How do we enable efficient management of wireless resources using metaverse for various applications?} In a metaverse-enabled wireless system, the meta space will perform efficient management of resources for running the physical space devices to serve the end-users. There is a need for efficient intelligent interfaces. Such intelligent interfaces will enable efficient resource management as well as effective channel estimation. For such intelligent interfaces, one can use centralized and distributed learning schemes \cite{ullah2020applications,yang2020machine,yang2020machine}. For estimation and resource allocation schemes, in the case of centralized learning, learning takes place at a centralized location. This centralized learning will take place in a meta space and thus might suffer from privacy issues in case of malicious security attacks. In contrast, distributed learning takes place in a distributed manner at devices and aggregation takes place at a meta space, and thus better preserves privacy compared to centralized learning. However, distributed learning models for intelligent interfaces will have a large convergence time due to heterogeneity in data and systems. Therefore, there is a need to propose novel distributed learning algorithms for intelligent interfaces in a metaverse-enabled wireless system.     

\subsection{Zero-Touch Networking for Metaverse}
{\em How does one use zero-touch networking to enable effective self-sustaining metaverse-enabled wireless applications?} Deploying metaverse for emerging applications to service a massive number of users requires seamless metaverse signaling. Such signaling must be done in a way that requires less intervention from end users and operators. To do so, one can use zero-touch networking (i.e., autonomous networking) for metaverse signaling. For the efficient realization of zero-touch networking for the metaverse, one can use various schemes/technologies, such as network slicing, machine learning, and optimization theory. Note that there are two aspects: zero-touch networking for metaverse and metaverse for zero-touch networking. Metaverse for zero-touch networking requires training of meta space models using emerging machine learning schemes and mathematical tools that can assist the network operation with the lowest possible intervention of network operators. On the other hand, zero-touch networking for the metaverse deals with the efficient signaling of a metaverse using emerging schemes to enable various metaverse-based applications. One can use machine learning schemes (specifically distributed learning) to train various models for performing metaverse signaling. Such models will perform optimization of computing and communication resources for performing signaling. Additionally, the interruption in metaverse services due to faults or security attacks must be addressed using zero-touch networking models.   

\subsection{Machine Learning-enabled Semantic Communication for Metaverse}
{\em How do we enable applications using metaverse and machine learning while performing service-level optimization and service diversity?} Enabling the metaverse for a massive number of devices requires service-level optimization and service diversity for cost-efficient operation. In contrast to traditional data-oriented wireless systems that require a channel with an infinite capacity for real-time applications, there is a need to combine reasoning tools and knowledge representation in training machine learning tools for the metaverse. Traditional data-oriented wireless systems represent information simply as bits that are not sufficient. Therefore, semantic communication combines reasoning tools and knowledge representation along with machine learning tools for communication in a metaverse. Semantic communication only sends important information in contrast to traditional data-oriented communication systems and thus improves the system's efficiency. The key components of semantic communication in a metaverse will be a semantic encoder, semantic decoder, and semantic noise interference. The purpose of a semantic encoder is to detect semantic information out of all available information. On the receiving end, the semantic decoder decodes the relevant information from the received information. In \cite{luo2022autoencoder}, a semantic communication scheme based on auto-encoder over a Rayleigh channel was proposed. The purpose of the auto-encoder is to encode and decode the information in semantic communication. Another work \cite{xie2021deep} proposed a deep learning-enabled semantic communication. Specifically, a DeepSC, using a transformer encoder and decoder for text transmission was proposed. Based on the aforementioned facts, there is a need to propose a novel distributed learning scheme for a privacy-aware semantic communication system.          

\subsection{Hybrid Modeling for Meta Space}
{\em How do we effectively model meta space that truly reflects the actual entities in the physical space?} Modeling of twins and avatars can be performed using various techniques, such as machine learning, experimental, and mathematical. Every technique has pros and cons, therefore, it might not be more suitable to model twins and avatars using a single technique. For instance, machine learning-enabled might not converge well, and thus fail to effectively model twins and avatars. Similarly, mathematical modeling also has limitations due to the various assumptions required for modeling. Moreover, experimental modeling also has experimental errors. Keeping in view the aforementioned facts, one can conclude that there is a need for hybrid modeling based simultaneously on different techniques. For instance, consider a wireless system that has a variety of players. For mobility modeling, one can use deep learning (i.e., machine learning-enabled modeling). For some entities (e.g., resource block allocation), one can use mathematical modeling (e.g., optimization theory, game theory, and graph theory). For 3D modeling of mobile devices/humans, one can use experimental modeling to effectively model the effect on wireless communication (e.g., the effect on THz communication). Therefore, there is a need to propose hybrid models for the effective modeling of meta space.      




\section{Conclusions}
In this tutorial, we have presented a detailed overview of the fundamentals of the metaverse for wireless systems. Specifically, we presented design aspects, key enablers, general architecture, and practical use cases. As a part of the general architecture, we studied the network management, reliability, and security of both meta space and physical space. We also outlined the recent advances and evaluated them. Furthermore, we presented the standardization of the machine learning-enabled metaverse. Finally, open challenges are presented with possible guidelines.





\bibliographystyle{IEEEtran}
\bibliography{Database}

\begin{IEEEbiography}[{\includegraphics[width=1in,height=1.25in,clip,keepaspectratio]{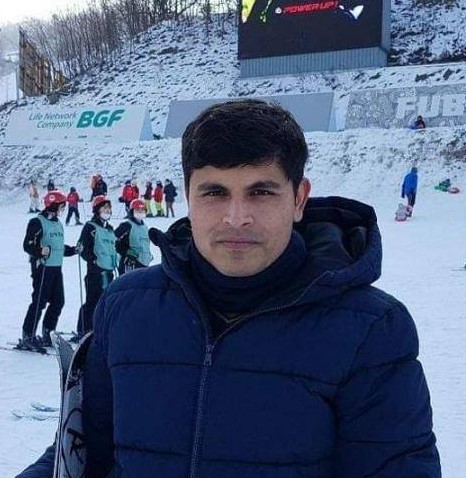}}]{Latif U. Khan} received his Ph.D. degree in Computer Engineering and M.S. degree in Electrical Engineering with distinction from Kyung Hee University (KHU), South Korea in 2021 and UET Peshawar in 2017, respectively. He worked as a leading researcher in the intelligent Networking Laboratory under a project jointly funded by the prestigious Brain Korea 21st Century Plus and Ministry of Science and ICT, South Korea. Prior to joining the KHU, he has served as a faculty member and research associate in the UET, Peshawar, Pakistan. He has published his works in highly reputable conferences and journals. He is the recipient of KHU best thesis award. He is the author/co-author of two conference best paper awards. He is also author of two books titled "Network Slicing for $5$G and Beyond Networks" and "Federated Learning for Wireless Networks". His research interests include analytical techniques of optimization and game theory to edge computing, end-to-end network slicing, digital twins, and federated learning for wireless networks. 
\end{IEEEbiography}

\begin{IEEEbiography}[{\includegraphics[width=1.5in,height=1.25in,clip,keepaspectratio]{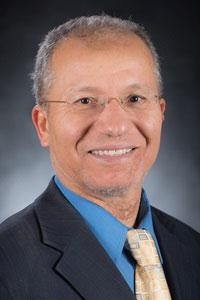}}]{Mohsen Guizani} (S’85, M'89, SM'99, F’09) received his B.S. (with distinction) and M.S. degrees in electrical engineering, and M.S. and Ph.D. degrees in computer engineering from Syracuse University, New York, in 1984, 1986, 1987, and 1990, respectively. He is currently a Professor with the Machine Learning Department, Mohamed Bin Zayed University of Artificial Intelligence (MBZUAI), Abu Dhabi, UAE. Previously, he served in different academic and administrative positions at the University of Idaho, Western Michigan University, the University of West Florida, the University of Missouri-Kansas City, the University of Colorado-Boulder, and Syracuse University. His research interests include wireless communications and mobile computing, computer networks, mobile cloud computing, security, and smart grid. He was the Editor-in-Chief of IEEE Network. He serves on the Editorial Boards of several international technical journals, and is the Founder and Editor-in-Chief of the Wireless Communications and Mobile Computing journal (Wiley). He is the author of nine books and more than 500 publications in refereed journals and conferences. He has guest edited a number of Special Issues in IEEE journals and magazines. He has also served as a TPC member, Chair, and General Chair of a number of international conferences. Throughout his career, he received three teaching awards and four research awards. He also received the 2017 IEEE Communications Society WTC Recognition Award as well as the 2018 AdHoc Technical Committee Recognition Award for his contribution to outstanding research in wireless communications and ad hoc sensor networks. He was the Chair of the IEEE Communications Society Wireless Technical Committee and the Chair of the TAOS Technical Committee. He served as a IEEE Computer Society Distinguished Speaker and is currently an IEEE ComSoc Distinguished Lecturer. He is a Senior Member of ACM.
\end{IEEEbiography}

\begin{IEEEbiography}[{\includegraphics[width=1in,height=1.25in,clip,keepaspectratio]{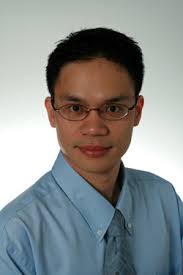}}]{Dusit Niyato}(M’09–SM’15–F’17) received the Ph.D. degree in electrical and computer engineering from the University of Manitoba, Winnipeg, MB, Canada, in 2008. He is currently a Professor with
	the School of Computer Science and Engineering, Nanyang Technological University, Singapore. He has published more than 400 technical articles in
	the area of wireless and mobile computing. He received the Best Young Researcher Award of the IEEE Communications Society Asia Pacifica and
	the 2011 IEEE Communications Society Fred W. Ellersick Prize Paper Award. He is also serving as a Senior Editor of the IEEE Wireless Communication Letters, an Area Editor of the IEEE Transactions on wireless Communications and the IEEE Communications Surveys and Tutorials, an Editor of the IEEE Transactions on Communications, and an Associate Editor of the IEEE Transactions on Mobile Computing, the IEEE Transactions on Vehicular Technology, and the IEEE Transactions on Cognitive Communications and Networking. He was a Distinguished Lecturer of the IEEE Communications Society from 2016 to 2017. He was named a highly cited researcher in computer science.\end{IEEEbiography}

\begin{IEEEbiography}[{\includegraphics[width=1in,height=1.25in,clip,keepaspectratio]{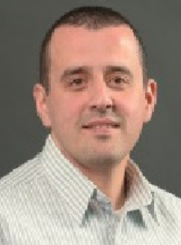}}]{Ala Al-Fuqaha} (Senior Member, IEEE) received the
	Ph.D. degree in computer engineering and networking from the University of Missouri-Kansas City,
	Kansas City, MO, USA, in 2004.
	He is currently a Professor at the Information and
	Computing Technology Division, College of Science
	and Engineering, Hamad Bin Khalifa University
	(HBKU), Doha, Qatar. His research interests include
	the use of machine learning in general and deep
	learning in particular in support of the data-driven
	and self-driven management of large-scale deployments of the Internet of Things (IoT) and smart city infrastructure and
	services, wireless vehicular networks (VANETs), cooperation and spectrum
	access etiquette in cognitive radio networks, and management and planning
	of software-defined networks (SDNs).
\end{IEEEbiography}

\begin{IEEEbiography}[{\includegraphics[width=1.5in,height=1.25in,clip,keepaspectratio]{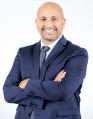}}]{Merouane Debbah} is Chief Researcher at the Technology Innovation Institute in Abu Dhabi. He is a Professor at Centralesupelec and an Adjunct Professor with the Department of Machine Learning at the Mohamed Bin Zayed University of Artificial Intelligence. He received the M.Sc. and Ph.D. degrees from the Ecole Normale Supérieure Paris-Saclay, France. He was with Motorola Labs, Saclay, France, from 1999 to 2002, and also with the Vienna Research Center for Telecommunications, Vienna, Austria, until 2003. From 2003 to 2007, he was an Assistant Professor with the Mobile Communications Department, Institut Eurecom, Sophia Antipolis, France. In 2007, he was appointed Full Professor at CentraleSupelec, Gif-sur-Yvette, France. From 2007 to 2014, he was the Director of the Alcatel-Lucent Chair on Flexible Radio. From 2014 to 2021, he was Vice-President of the Huawei France Research Center. He was jointly the director of the Mathematical and Algorithmic Sciences Lab as well as the director of the Lagrange Mathematical and Computing Research Center. Since 2021, he is leading the AI \& Digital Science Research centers at the Technology Innovation Institute. He has managed 8 EU projects and more than 24 national and international projects. His research interests lie in fundamental mathematics, algorithms, statistics, information, and communication sciences research. He is an IEEE Fellow, a WWRF Fellow, a Eurasip Fellow, an AAIA Fellow, an Institut Louis Bachelier Fellow and a Membre emerite SEE. He was a recipient of the ERC Grant MORE (Advanced Mathematical Tools for Complex Network Engineering) from 2012 to 2017. He was a recipient of the Mario Boella Award in 2005, the IEEE Glavieux Prize Award in 2011, the Qualcomm Innovation Prize Award in 2012, the 2019 IEEE Radio Communications Committee Technical Recognition Award and the 2020 SEE Blondel Medal. He received more than 20 best paper awards, among which the 2007 IEEE GLOBECOM Best Paper Award, the Wi-Opt 2009 Best Paper Award, the 2010 Newcom++ Best Paper Award, the WUN CogCom Best Paper 2012 and 2013 Award, the 2014 WCNC Best Paper Award, the 2015 ICC Best Paper Award, the 2015 IEEE Communications Society Leonard G. Abraham Prize, the 2015 IEEE Communications Society Fred W. Ellersick Prize, the 2016 IEEE Communications Society Best Tutorial Paper Award, the 2016 European Wireless Best Paper Award, the 2017 Eurasip Best Paper Award, the 2018 IEEE Marconi Prize Paper Award, the 2019 IEEE Communications Society Young Author Best Paper Award, the 2021 Eurasip Best Paper Award, the 2021 IEEE Marconi Prize Paper Award, the 2022 IEEE Communications Society Outstanding Paper Award, the 2022 ICC Best paper Award as well as the Valuetools 2007, Valuetools 2008, CrownCom 2009, Valuetools 2012, SAM 2014, and 2017 IEEE Sweden VT-COM-IT Joint Chapter best student paper awards. He is an Associate Editor-in-Chief of the journal Random Matrix: Theory and Applications. He was an Associate Area Editor and Senior Area Editor of the IEEE TRANSACTIONS ON SIGNAL PROCESSING from 2011 to 2013 and from 2013 to 2014, respectively. From 2021 to 2022, he serves as an IEEE Signal Processing Society Distinguished Industry Speaker.
\end{IEEEbiography}

\end{document}